\documentclass[oneside,english]{amsart}
\usepackage{lmodern}
\usepackage[T1]{fontenc}
\usepackage[latin9]{inputenc}
\usepackage[letterpaper]{geometry}
\usepackage{babel}
\usepackage{amsbsy}
\usepackage{amstext}
\usepackage{amsthm}
\usepackage{amssymb}
\usepackage{mathdots}
\usepackage{graphicx}
\usepackage{booktabs}
\usepackage{xifthen,xcolor,tikz,setspace}
\usetikzlibrary{decorations.pathmorphing,patterns,shapes}

\usepackage[colorlinks=true]{hyperref}

\makeatletter
\numberwithin{equation}{section}
\numberwithin{figure}{section}
 \theoremstyle{definition}
 \newtheorem*{defn*}{\protect\definitionname}
  \theoremstyle{plain}
  \newtheorem*{conjecture*}{\protect\conjecturename}
  \theoremstyle{plain}
  \newtheorem*{thm*}{\protect\theoremname}
\theoremstyle{plain}
\newtheorem{thm}{\protect\theoremname}[section]
  \theoremstyle{remark}
  \newtheorem*{rem*}{\protect\remarkname}
  \theoremstyle{plain}
  
  \theoremstyle{definition}
  
  \theoremstyle{remark}
  
  \theoremstyle{plain}
  
  \theoremstyle{plain}

\newtheorem{theorem}{\protect\theoremname}[section]
  \theoremstyle{remark}
  \newtheorem{remark}{\protect\remarkname}
  \theoremstyle{plain}
  \newtheorem{lemma}[thm]{\protect\lemmaname}
  \theoremstyle{plain}
  \newtheorem{proposition}[thm]{\protect\propositionname}
  \theoremstyle{definition}
  \newtheorem{definition}{\protect\definitionname}
  \theoremstyle{definition}
  \newtheorem*{definition*}{\protect\definitionname}
  \theoremstyle{plain}
  \newtheorem{corollary}[thm]{\protect\corollaryname}
  \theoremstyle{remark}
  \newtheorem*{remark*}{\protect\remarkname}
  \theoremstyle{plain}
  \newtheorem*{lemma*}{\protect\lemmaname}
  \theoremstyle{plain}
  \newtheorem*{theorem*}{\protect\theoremname}

\newcommand{\Hm}{\mbox{hm }}
\newcommand{\pf}{\mbox{Pf }}
\newcommand{\nor}{\frac{1}{2\sqrt{2}}}
\newcommand{\norin}{2\sqrt{2}}
\newcommand{\nori}{\frac{i}{2\sqrt{2}}}
\newcommand{\norsquare}{\frac{1}{8}}
\newcommand{\re}{\mbox{Re }}
\newcommand{\im}{\mbox{Im }}
\usepackage{hyperref}
\hypersetup{
    colorlinks,
    citecolor=black,
    filecolor=black,
    linkcolor=black,
    urlcolor=black
}

\makeatother

  \providecommand{\conjecturename}{Conjecture}
  \providecommand{\corollaryname}{Corollary}
  \providecommand{\definitionname}{Definition}
  \providecommand{\lemmaname}{Lemma}
  \providecommand{\propositionname}{Proposition}
  \providecommand{\remarkname}{Remark}
  \providecommand{\theoremname}{Theorem}
\providecommand{\theoremname}{Theorem}

\begin{document}

\title{Ising Model: Local Spin Correlations and Conformal Invariance}

\author{Reza Gheissari, Cl\'ement Hongler, and S.\ C.\ Park}

\address{Courant Institute of Mathematical Sciences, New York University.
251 Mercer St, New York, NY 10012, USA. }

\email{\textsf{reza@cims.nyu.edu}}

\address{Ecole Polytechnique F\'ed\'erale de Lausanne, EPFL SB MATHAA CSFT, CH-1015
Lausanne, Switzerland}

\email{\textsf{clement.hongler@epfl.ch}}

\address{Ecole Polytechnique F\'ed\'erale de Lausanne, EPFL SB MATHAA CSFT, CH-1015
Lausanne, Switzerland}

\email{\textsf{sungchul.park@epfl.ch}}

\begin{abstract}
We study the 2-dimensional Ising model at critical temperature on
a simply connected subset $\Omega_{\delta}$ of the square grid $\delta\mathbb{Z}^{2}$.
The scaling limit of the critical Ising model is conjectured to be
described by Conformal Field Theory; in particular, there is expected to be a precise correspondence between local lattice fields of the Ising model and the
local fields of Conformal Field Theory.

Towards the proof of this correspondence, we analyze arbitrary spin
pattern probabilities (probabilities of finite spin configurations occurring at the origin), explicitly obtain their infinite-volume limits, and prove
their conformal covariance at the first (non-trivial) order. We formulate
these probabilities in terms of discrete fermionic observables, enabling
the study of their scaling limits. This generalizes results of \cite{hon2010,hosm2013}
and \cite{chelkak-hongler} to one-point functions of any local spin correlations. 

We introduce a collection of tools which allow one to exactly and
explicitly translate any spin pattern probability (and hence
any lattice local field correlation) in terms of discrete complex
analysis quantities. The proof requires working with multipoint
lattice spinors with monodromy (including construction of explicit formulae
in the full plane), and refined analysis near their source points to prove convergence to the appropriate continuous conformally covariant functions. 
\end{abstract}

\maketitle
\tableofcontents{}

\section{Introduction}

The 2D Ising model is one of the most studied models of statistical
mechanics. In its simplest formulation it consists of a random assignment
of $\pm1$ spins $\sigma_{x}$ to the faces of (subgraphs of) the
square grid $\mathbb{Z}^{2}$; the spins tend to align with their
neighbors; the probability of a configuration is proportional to $e^{-\beta H\left(\sigma\right)}$
where the energy $H(\sigma)=-\sum_{i\sim j}\sigma_{i}\sigma_{j}$
sums over pairs of adjacent faces; alignment strength is controlled
by the parameter $\beta>0$, usually identified with the inverse temperature. 

The 2D Ising model has found applications in many areas of science,
from description of magnets to ecology and image processing. Due to
its simplicity and emergent features, it is interesting both as a
discrete probability and statistical field theory model. Of particular
physical interest is the phase transition at the critical point $\beta_{c}$:
for $\beta<\beta_{c}$ the system is disordered at large scales while
for $\beta>\beta_{c}$ a long-range ferromagnetic order arises. Classically,
the phase transition can be described in terms of the infinite-volume
limit: in the disordered phase $\beta<\beta_{c}$ there is a unique
Gibbs measure, while in the ordered phase $\beta>\beta_{c}$ infinite-volume
measures are convex combinations of two extremal measures. It has
a continuous phase transition: only one Gibbs measure exists at $\beta=\beta_{c}$. 

Critical lattice models at continuous phase transition points are
widely expected to have universal scaling limits (independent of the
choice of lattice and other details). In 2D, such scaling limits are
expected to exhibit conformal symmetry. This can be loosely formulated
as follows: for a conformal mapping $\varphi:\Omega\to\tilde{\Omega}$,
 
\[
\varphi\left(\mbox{scaling limit on }\Omega\right)=\mbox{scaling limit on }\tilde{\Omega}\,.
\]

There are two main tools used to describe the scaling limits of planar
lattice models: curves and fields. Schramm-Loewner Evolution (SLE)
curves naturally arise in conformally invariant setups: for the Ising
model, they describe the scaling limit of interfaces between opposite
spins (\cite{chelkak-duminil-copin-hongler-kemppainen-smirnov,benoist-hongler-cle}).
The fields on a discrete level, such as the $\pm1$-valued \emph{spin
field} $\sigma_{i}$, can be described by Conformal Field Theory (CFT):
their correlations, in principle, are conjecturally described using
representation-theoretic methods. Such conjectures have been proved
for a number of natural fields (\cite{hon2010,hosm2013,chelkak-hongler});
the present paper is part of this program.

What makes it possible to mathematically analyze the 2D Ising model
with great precision is its exactly solvable structure, first revealed
by Onsager \cite{onsager}. The exact solvability can be formulated
in many different ways; in recent years, the formulation in terms
of discrete complex analysis has emerged as one the most powerful
ways to understand the scaling limit of the model rigorously. In particular,
the model's conformal symmetry becomes much more transparent in this
context.

The results of \cite{hon2010} and \cite{chelkak-hongler} on (asymptotic)
conformal invariance of spin and energy fields can be formulated,
in their simplest cases, as follows: consider the critical Ising model
with plus boundary conditions on the discretization $\Omega_{\delta}$
by a square grid of mesh size $\delta>0$ of a simply-connected domain
$\Omega$ around the origin. Take the origin $0\in\Omega$, identify
it with the closest face of $\Omega_{\delta}$ and let $\delta$ be
the face to the right of $0$. Then, as $\delta\to0$, we have the asymptotic
expansions,

\begin{align*}
[\mbox{spin field}] &  & \mathbb{E}_{\Omega_{\delta}}[\sigma_{0}] & =0+C_{\sigma}\left|\varphi'\left(0\right)\right|^{-\frac{1}{8}}\delta^{\frac{1}{8}}+o\left(\delta^{\frac{1}{8}}\right)\,,\\{}
[\mbox{energy density field}] &  & \mathbb{E}_{\Omega_{\delta}}[\sigma_{0}\sigma_{\delta}] & =\frac{\sqrt{2}}{2}+C_{\epsilon}\left|\varphi'\left(0\right)\right|^{-1}\delta+o\left(\delta\right),
\end{align*}
where $C_{\sigma},C_{\epsilon}>0$ are explicit (lattice-dependent)
constants and $\varphi$ is any conformal map from the unit disk $\mathbb{D}$
to $\Omega$ fixing the origin. The first terms in the respective
expansions are the infinite-volume limits of the left-hand side quantities.
These results illustrate the following: for any local field one-point function,
the correction to its infinite-volume expectation is described by
Conformal Field Theory (CFT) quantities. 

\begin{figure}
\begin{center}
\includegraphics[scale=0.58]{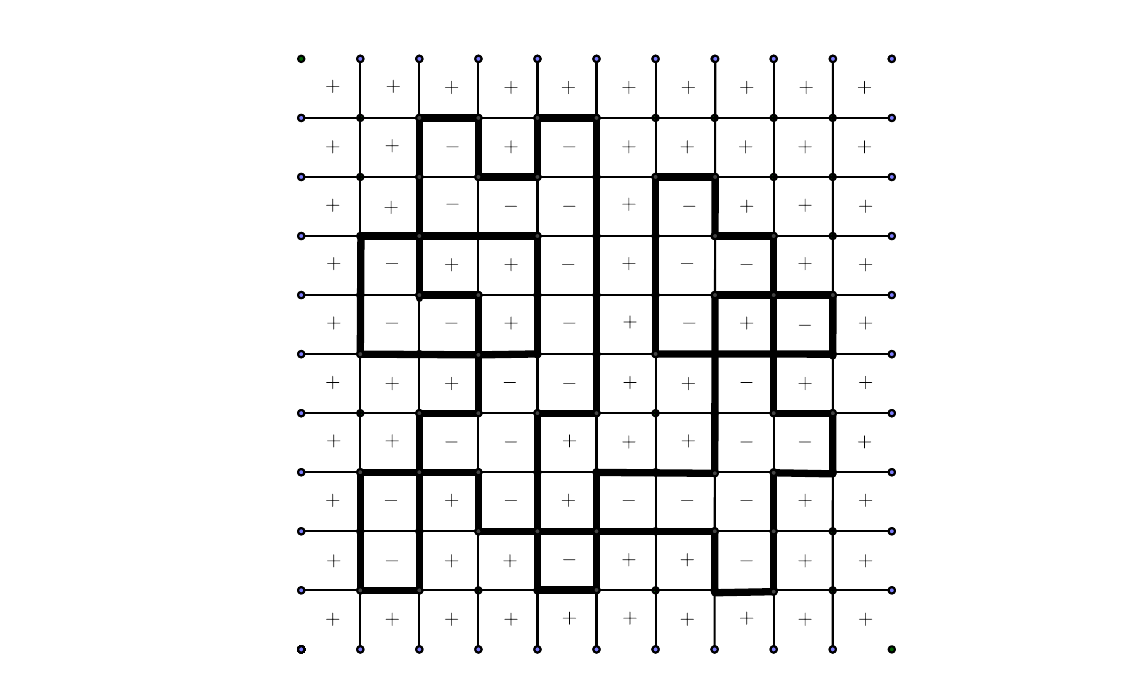}
\end{center}
\caption{An Ising model configuration on the faces of a square subset of $\mathbb Z^2$ with all-plus boundary conditions,
along with its ``low-temperature expansion", indicating interfaces separating plus and minus spins.}\label{fig:ising-grid}
\end{figure}

The Ising Model is conjectured to be described by the unitary CFT
minimal model $\mathcal{M}_{3,4}$ (see e.g. \cite{BPZ}, \cite{di-francesco-mathieu-senechal}),
also known as the Ising CFT. The Ising CFT consists of three primary
fields: those of respective scaling dimensions $0$ (the identity
field \textendash{} a constant field identically equal to one), $\frac{1}{8}$ (the spin field)
and $1$ (the energy field). Each of these primary fields generates
an infinite-dimensional tower of fields called its \emph{descendants}. 

We conjecture that the space of the Ising CFT fields describes the
limits of Ising \emph{lattice local fields}: 
\begin{definition*}
Let $\mathcal{F}$ be a finite connected collection of faces of $\mathbb{Z}^{2}$
including $0$. For any $F:\{\pm1\}^{\mathcal{F}}\to\mathbb{C}$,
a \emph{lattice local field} $\Phi_{\delta}^{F}$ is a random field
on the faces of $\Omega_{\delta}$ whose values are given by $\Phi_{\delta}^{F}(x)=F(\sigma\text{|}{}_{x+\delta\mathcal{F}})$.
We call a local field \emph{spin-antisymmetric} if $F(-\sigma)=-F(\sigma)$
and \emph{spin-symmetric} if $F(-\sigma)=F(\sigma)$.
\end{definition*}
\begin{conjecture*}
For any nonzero lattice field (i.e. whose correlations do not vanish generically; see~\cite{lattice-virasoro}) $\Phi_{\delta}^{F}$, there exists
$D\in\mathbb{N}\cup\left(\mathbb{N}+\frac{1}{8}\right)$ such that
\begin{align*}
\delta^{-D}\Phi_{\delta}^{F}\to  \Phi
\end{align*}
in the sense of correlations (meaning that the $n$ point functions converge), where $\Phi$ is a nonzero primary or
descendant CFT field. If $\Phi$ is spin-antisymmetric then $D\in\mathbb{N}+\frac{1}{8}$;
if $\Phi$ is spin-symmetric then $D\in\mathbb{N}$. Moreover every
Ising CFT field can be obtained in such a manner. 
\end{conjecture*}

\subsubsection*{Spin Pattern Probabilities}

Correlations of any lattice local field at a point $x$ can be rewritten
in terms of probabilities of observing certain \emph{spin patterns}
centered at $x$, i.e.\ probabilities of spin configurations in a microscopic
neighborhood of $x$. The main objective of this paper is to obtain
explicit representations for probabilities of spin pattern events,
which are the most general local quantities describing the model:
we obtain infinite-volume limits for arbitrary local pattern probabilities
and give their first-order corrections corresponding to what can be
expected from the Ising-CFT correspondence (see Theorems~\ref{thm:spin-sym-energy}--\ref{thm:spin-sens-energy} and Corollary~\ref{thm:spin-pattern}). 

More precisely, Let $\mathcal{F}$ be a finite connected set of faces
in $\mathbb{Z}^{2}$ and fix a configuration $\rho\in\{\pm1\}^{\mathcal{F}}$.
We look at two types of lattice local fields: 
\begin{itemize}
\item \emph{spin-antisymmetric pattern fields} $\Phi_{\delta}^{F_{\rho}}(x)$
where $F_{\rho}=\boldsymbol{1}\{\sigma|_{\mathcal F}=\rho\}$, whose expectation
gives the probability of the \emph{spin-antisymmetric pattern} $\rho$
on $\mathcal{F}$,
\item \emph{spin-symmetric pattern fields} $\Phi_{\delta}^{F_{\rho}^{\pm}}(x)$
where $F_{\rho}^{\pm}=\boldsymbol{1}\{\sigma|_{\mathcal F}\in\{\pm\rho\}\}$, whose
expectation gives the probability of the \emph{spin-symmetric pattern}
$\pm\rho$ on $\mathcal{F}$.
\end{itemize}
Every Ising lattice local field can easily be seen to be a finite
linear combination of such fields. The main result of the paper is
the following. 
\begin{theorem*}[see Corollary \ref{thm:spin-pattern}]
Let $\mathcal{F}$ and $\rho$ be as above and let $\mathbb{P}_{\mathbb{Z}^{2}}$
be the infinite-volume measure of the critical Ising model. Consider
the critical Ising model on $\Omega_{\delta}$ with plus boundary
conditions, and denote it by $\mathbb P_{\Omega_\delta}$. Then as $\delta\to0$, we have
\begin{eqnarray*}
\mathbb{P}_{\Omega_{\delta}}\left[\sigma|_{\delta\mathcal{F}}=\rho\right] & = & \mathbb{P}_{\mathbb{Z}^{2}}\left[\sigma|_{\mathcal{F}}=\rho\right]+\delta^{\frac{1}{8}}\cdot\mathrm{geometric}\;\mathrm{effect}(\rho,\Omega)+o\left(\delta^{\frac{1}{8}}\right),\\
\mathbb{P}_{\Omega_{\delta}}\left[\sigma|_{\delta\mathcal{F}}\in\{\pm\rho\}\right] & = & \mathbb{P}_{\mathbb{Z}^{2}}\left[\sigma|_{\mathcal{F}}\in\{\pm\rho\}\right]+\delta\cdot\mathrm{\mathrm{geometric}\;\mathrm{effect}}(\pm\rho,\Omega)+o\left(\delta\right).
\end{eqnarray*}
The infinite-volume probability and the geometric effects are given in terms of explicit Pfaffian formulae.
\end{theorem*}

The distinction between spin-symmetric and spin-antisymmetric pattern
fields is both natural and important in the CFT framework: the spin
and energy fields are the most elementary instances of spin-antisymmetric
and spin-symmetric local fields respectively. The space of lattice
local fields is a vector space that can be decomposed into the direct
sum of fields that are symmetric and antisymmetric under spin flip. 

The above theorem proves the aforementioned conjecture
in the following specific case: it allows one to study the scaling
limit of the one-point function of $\delta^{-D}\Phi_{\delta}^{F}$
for $D\leq1$ in simply-connected domains. Following~\cite{hon2010,hosm2013,chelkak-hongler}, we expect the proof naturally extends to multi-field correlations, in order to prove\textemdash in full\textemdash the
conjecture for fields of scaling dimension $D\in \{0,1/8,1\}$.
Beyond that, the method we
use provides a general toolbox to express multipoint correlations
of any local lattice field in terms of discrete fermionic observables
in discrete domains, and hence to give explicit infinite-volume limits
and first order corrections, and reduce the calculation of all subsequent CFT terms to questions in discrete complex analysis (see Applications
\ref{subsec:Lattice-local-fields}). 

Moreover, the results allow one to study new interesting quantities:
for instance, one can estimate spin flip rates for critical Ising
Glauber dynamics, including the geometric effects on them up to first order in the mesh size (see Applications
\ref{subsec:Local-Markov-chain}).

The results and approach of this paper, as well as the conjectured
connection between Ising lattice local fields and CFT, are expected
to straightforwardly generalize in two directions. First, the approach can be generalized
to arbitrary combinations of $+,-$ and free boundary conditions (the
three conformally invariant boundary conditions according to CFT \cite{cardy-i}).
Second, the results should extend to more general planar graphs, including,
in particular, isoradial graphs.

Let us also point out similar connections between pattern probabilities
and conformal invariance obtained by Boutillier in \cite{boutillier}
in the context of the dimer model. 

The proof relies mainly on discrete complex analytic methods: we use
lattice observables, modifying the objects introduced in \cite{hon2010}
and \cite{chelkak-hongler}, to connect pattern probabilities with
solutions of discrete boundary value problems. This requires precise
treatment of multipoint observables on a topological double cover
of the lattice with microscopically separated source points at their
singularities. We then study the scaling limits of such solutions
using discrete complex analysis technique, where, in particular, the
neighborhood of the monodromy of the double cover needs to be analyzed
delicately. The new techniques introduced for this purpose are: refined
analysis of convergence of observables and constructions and characterizations
of lattice spinor observables on the slit plane $(\mathbb{C}\backslash\mathbb{R}_{>0})_{\delta>0}$,
both as limits of finite-volume ones and in terms of discrete harmonic
measures (explicitly computed with Fourier techniques). 

\subsubsection*{Acknowledgements}
Firstly, the authors thank the anonymous referees for their many helpful suggestions and comments. 

This research was initiated during the Research Experience for Undergraduates
program at Mathematics Department of Columbia University, funded by
the NSF under grant DMS-0739392. We would like to thank the T.A.,
Krzysztof Putyra, for his help during this program and the program
coordinator Robert Lipshitz, as well as all the participants, in particular
Adrien Brochard and Woo Chang Chung. 

R.G. would like to thank Chuck Newman and Eyal Lubetzky for interesting
discussions. 
C.H. would like to thank Dmitry Chelkak and Stanislav Smirnov, for
sharing many ideas and insights about the Ising model and conformal
invariance; Itai Benjamini and Curtis McMullen for asking questions
that suggested we look at this problem; St\'ephane Benoist, John Cardy,
Julien Dub\'edat, Hugo Duminil-Copin, Konstantin Izyurov, Kalle Kyt\"ol\"a
and Wendelin Werner for interesting discussions, the NSF under grant
DMS-1106588, the ERC under grant SG CONSTAMIS, the NCCR Swissmap, the Blavatnik family foundation, the Latsis family foundation, and the Minerva Foundation
for financial support.
S.P. would like to thank C. M. Munteanu for interesting discussions.
\subsection{Notation}

We begin by defining the most important notation, that will be necessary for the statements of the main theorems. We defer a more extensive discussion of the notation used in the proofs to \S\ref{subsec:Extra-Notation}.

In this paper, we consider the Ising model with spins on the faces
of the graph $\Omega_{\delta}$, a discretization of $\Omega$ of
mesh size $\delta>0$. More precisely:

Identify $\mathbb{Z}^{2}$ with the square lattice with vertex set at
$\mathbb{Z}+i\mathbb{Z}\subset\mathbb{C}$ and nearest-neighbor edges. Let 
\begin{align*}
\mathbb{C}_{1}:=(1+i)\mathbb{Z}^{2}+1\qquad \mbox{ and } \qquad \mathbb C_\delta = \delta \mathbb C_1
\end{align*}
be the rescaled, rotated and shifted lattice, and its
rescaling by a \emph{mesh size} $\delta>0$, respectively.

For a simply connected open domain $\Omega\subset\mathbb{C}$ bounded
by a smooth curve containing $0$ (this is easily relaxed to arbitrary simply connected
domains~\cite[Remark 2.10]{chelkak-hongler}), define $\Omega_{\delta}$
as the largest connected component of the graph $\Omega\cap\mathbb{C}_{\delta}$.
Denote by $\mathcal{V}_{\Omega_{\delta}}$ the set of vertices of
$\Omega_{\delta}$, and by $\mathcal{E}_{\Omega_{\delta}}$ the set
of edges\emph{ }in $\Omega_{\delta}$. We denote the set of faces\emph{
}of the graph by $\mathcal{F}_{\Omega_{\delta}}$. Whenever needed,
we identify the edges in $\mathcal{E}_{\Omega_{\delta}}$ with their
midpoints, and the faces in $\mathcal{F}_{\Omega_{\delta}}$ with
their centers, such that the origin is identified with a face. 

\subsubsection{Ising Model}\label{subsubsec:basic-ising-def}
An Ising \emph{configuration} $\sigma$ is an assignment of $\pm1$
spins to the faces in $\mathcal{F}_{\Omega_{\delta}}$. We consider
the \emph{critical Ising model} on $\mathcal{F}_{\Omega_{\delta}}$
with \emph{plus boundary conditions}, given by,
\[
\mathbb{P}_{\Omega_{\delta}}(\sigma)=\mathbb P^+_{\Omega_\delta}(\sigma)\propto e^{-\beta_{c}H(\sigma)}\,,
\]
where $\beta_{c}=\frac{1}{2}\ln(\sqrt{2}+1)$ is the critical inverse
temperature and $H\left(\sigma\right)=-\sum_{x\sim y}\sigma_{x}\sigma_{y}$
with boundary faces fixed to have $+1$ spin. Let $\mathbb{E}_{\Omega_{\delta}}= \mathbb E^+_{\Omega_\delta}$
be its corresponding expectation.

Define the \emph{energy density} \emph{field} $(\epsilon(\delta e))_{e\in \mathbb C_1}$ as follows: for $\delta e\in \mathcal E_{\Omega_{\delta}}$ separating faces $\delta f_1\sim \delta f_2$, 
\begin{align*}
\epsilon\left(\delta e\right)=\mu-\sigma_{\delta f_1}\sigma_{\delta f_2}\,, \qquad \mbox{where} \qquad \mu := \frac{\sqrt 2}{2}= \mathbb E_{\mathbb C_\delta}[\sigma_{\delta f_1} \sigma_{\delta f_2}]=\mathbb E_{\mathbb C_1}[\sigma_{f_1} \sigma_{f_2}]\,.
\end{align*}

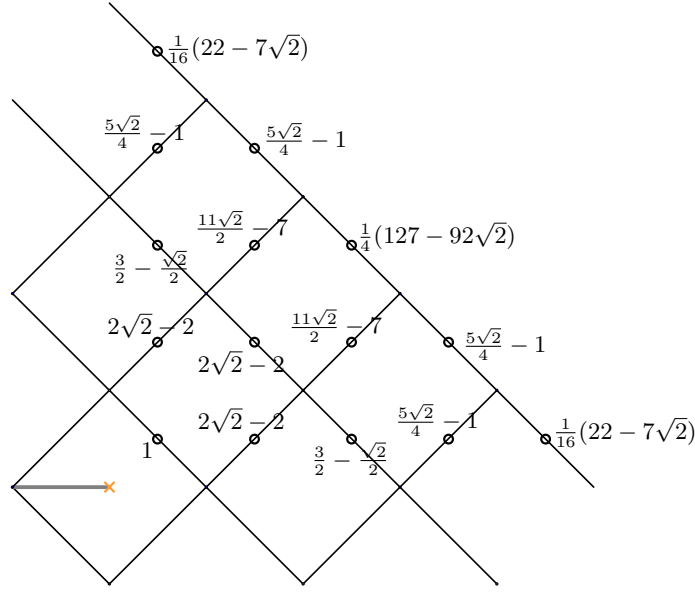
\begin{figure}
\begin{center}
\begin{tikzpicture}
\clip (-5cm,-4cm) rectangle (5cm,4cm);
\node{
\includegraphics[scale=1.6]{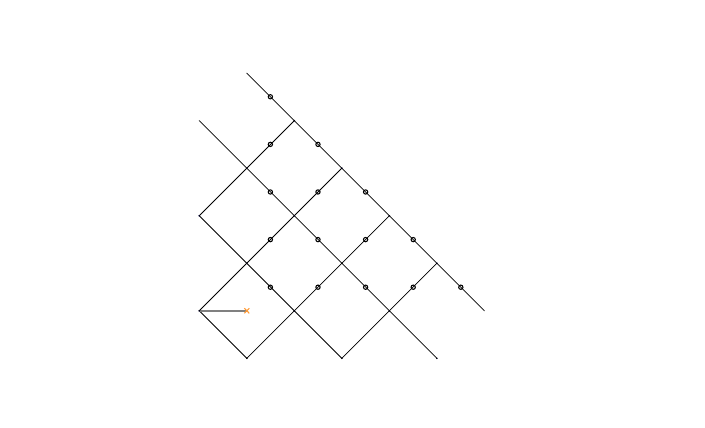}};
\node[font=\small] at (-5.65,-1.38) {$1-\frac {\sqrt 2}{2}$};

\node[font=\small] at (-1.1,-.98) {$2\sqrt 2 - 2$};
\node[font=\small] at (-1.1,-1.74) {$2\sqrt 2 - 2$};
\node[font=\small] at (-2.3,-.48) {$2\sqrt 2 -2$};
\node[font=\small] at (-2.35,-2.15) {$1$};
\node[font=\small] at (-2.3,.3) {$\frac{3}{2}-\frac{\sqrt 2}{2}$};

\node[font=\small] at (.35,-2.25) {$\frac{3}{2}-\frac{\sqrt 2}{2}$};
\node[font=\small] at (2.4,-.73) {$\frac{5\sqrt 2}{4}-1$};
\node[font=\small] at (.15,-.48) {$\frac{11\sqrt 2}{2}-7$};
\node[font=\small] at (1.5,-1.74) {$\frac{5\sqrt 2}{4}-1$};

\node[font=\small] at (-1.1,.83) {$\frac{11\sqrt 2}{2}-7$};
\node[font=\small] at (-.25,2) {$\frac{5\sqrt 2}{4}-1$};
\node[font=\small] at (-2.4,2.1) {$\frac{5\sqrt 2}{4}-1$};

\node[font=\small] at (1.5,.7) {$\frac{1}{4}(127-92\sqrt 2)$};
\node[font=\small] at (4,-1.9) {$\frac{1}{16}(22-7\sqrt 2)$};

\node[font=\small] at (-1.15,3.2) {$\frac{1}{16}(22-7\sqrt 2)$};

\end{tikzpicture}
\caption{Some explicit values of $(\mu_e)_{e\in \mathcal E_{\mathbb C_1}}$. This represents the mean of the physical quantity, the \emph{spin-weighted energy field}. The orange $\times$ marks the identified origin face, whose spin $\mu_e$ is weighted by; as $|e|\to \infty$, we have $\mu_e\to \mu$.}\label{fig:mu-values}
\end{center}
\end{figure}

Define the \emph{spin-weighted energy density field} $(\epsilon_{[0]}(\delta e))_{e\in \mathbb C_1}$ on $\mathcal E_{\Omega_\delta}$ by 
\begin{align}\label{eq:mu-e}
\epsilon_{[0]}(\delta e)= \mu_{ e} - \sigma_{\delta f_1} \sigma_{\delta f_2}\,, \qquad \mbox{where} \qquad \mu_{ e} := \lim_{\Omega\to\mathbb C} \frac {\mathbb E_{\Omega_\delta} [\sigma_0 \sigma_{\delta f_1} \sigma_{\delta f_2}]}{\mathbb E_{\Omega_\delta }[\sigma_0]}=\lim_{\Omega\to\mathbb C} \frac {\mathbb E_{\Omega_1} [\sigma_0 \sigma_{ f_1} \sigma_{ f_2}]}{\mathbb E_{\Omega_1 }[\sigma_0]}\,
\end{align}
where the limit is trivially independent of $\delta$ and exists for every $e\in \mathbb C_1$ by Theorem~\ref{thm:spin-sens-energy} (see Figure~\ref{fig:mu-values} for some exact values).
Given a set of edges $ \mathcal{B}\subset\mathcal{E}_{\mathbb C_1}$,
we write 
\begin{align*}
\epsilon(\delta \mathcal{B}):=\prod_{e\in \mathcal B}\epsilon(\delta e)\qquad \mbox{and analogously} \qquad \epsilon_{[0]}(\delta \mathcal B) := \prod_{e\in \mathcal B} \epsilon_{[0]}(\delta e)\,.
\end{align*}

\subsection{Main Results}
In this section we present the main results. By translation, it suffices to consider the statistics of fields centered at $x=0$.
For a collection $\mathcal{B}=\{e_{1},\ldots,e_{n}\}\subset\mathcal{E}_{\mathbb{C}_{1}}$,
consider the \emph{spin-symmetric field} $\epsilon(\delta\mathcal{B})=\epsilon({\delta e_{1}})\cdots\epsilon({\delta e_{n}})$,
i.e., the product of energy densities on a collection of edges around
$x$, and the \emph{spin-antisymmetric field} $\sigma_{0}\epsilon_{[0]}(\delta\mathcal{B})=\sigma_{0}\epsilon_{[0]}(\delta e_{1})\cdots\epsilon_{[0]}(\delta e_{n})$.
If $\mathbf A, \mathbf B$ are anti-symmetric square matrices of the same dimensions, define the directional derivative of the Pfaffian $\mathrm {Pf}(\mathbf B)$ (defined in~\eqref{eq:pfaffian}) by
\begin{align*}
D_{\mathbf A} \mathrm {Pf}(\mathbf B)= \lim_{t\downarrow 0}\frac {\mathrm {Pf}(\mathbf B+t\mathbf A)-\mathrm{Pf}(\mathbf B)}{t}\,.
\end{align*} 

For spin-symmetric and spin-antisymmetric fields, we obtain the following two convergence results:
\begin{theorem}[Spin-symmetric correlations]
\label{thm:spin-sym-energy}Let $\mathcal{B}=\left\{ e_{1},...,e_{n}\right\} \subset\mathcal{E}_{\mathbb{C}_{1}}$.
There exist explicit, real-valued anti-symmetric $2n \times 2n$  matrices $\mathbf{F}^{\mathcal{B}}$ and $\mathbf{E}^{\mathcal{B}}$,
such that as $\delta\to0$,

\begin{align*}
\mathbb{E}_{\Omega_{\delta}}\left[\epsilon(\delta\mathcal{B})\right]=\; & (-2)^{n}\cdot\mathrm{Pf}(\mathbf{F}^{\mathcal{B}})+(-2)^{n}\cdot\delta\cdot r_{\Omega}^{-1}(0)\cdot D_{\mathbf{E}^{\mathcal{B}}}\mathrm{Pf}(\mathbf{F}^{\mathcal{B}})+o(\delta)\,,
\end{align*}
where $r_{\Omega}(z)$ is the conformal radius of $\Omega$ seen from
$z\in\Omega$ (i.e. $r_{\Omega}(z)=\left|\varphi'(0)\right|$ where
$\varphi:\mathbb{D}\to\Omega$ is the conformal map such that $\varphi(0)=z$).
\end{theorem}
\begin{theorem}[Spin-antisymmetric correlations]
\label{thm:spin-sens-energy}Let $\mathcal{B}=\left\{ e_{1},...,e_{n}\right\} \subset\mathcal{E}_{\mathbb{C}_{1}}$.
The limits $\mu_{ e}$ defined in~\eqref{eq:mu-e} exist for every $e\in \mathcal E_{\mathbb C_1}$ and are given explicitly.  There exist explicit anti-symmetric $2n \times 2n$ matrices $\mathbf{F}_{[0]}^{\mathcal{B}}$ and
$\mathbf{E}_{[0]}^{\mathcal{B}}$, the former being real-valued, such that as $\delta\to0$,

\begin{align*}
\frac{\mathbb{E}_{\Omega_{\delta}}\left[\sigma_{0}\epsilon_{[0]}(\delta\mathcal{B})\right]}{\mathbb{E}_{\Omega_{\delta}}\left[\sigma_{0}\right]}=\;(-2)^{n}\cdot & \mbox{\ensuremath{\mathrm{Pf}}}(\mathbf{F}_{[0]}^{\mathcal{B}})+(-2)^{n}\cdot\delta\cdot\re\left[- \frac{1}{4}\partial_{z}\log r_{\Omega}\left(z\right)\Big|_{z=0} \cdot D_{\mathbf{E}_{[0]}^{\mathcal{B}}}\mbox{\ensuremath{\mathrm{Pf}}}\left(\mathbf{F}_{[0]}^{B}\right)\right]+o(\delta)\,,
\end{align*}
where $z=x+iy$ and $\partial_{z}=\frac{1}{2}(\partial_{x}-i\partial_{y})$.
\end{theorem}
\begin{remark*}
Theorems \ref{thm:spin-sym-energy} and \ref{thm:spin-sens-energy}
yield that the infinite-volume limits of $\mathbb{E}_{\Omega_{\delta}}\left[\sigma_{0}\epsilon_{[0]}(\delta\mathcal{B})\right]/\mathbb{E}_{\Omega_{\delta}}\left[\sigma_{0}\right]$ and $\mathbb{E}_{\Omega_{\delta}}\left[\epsilon(\delta\mathcal{B})\right]$ exist and are given explicitly by 
\begin{align*}
\lim_{\Omega\to\mathbb{C}}\mathbb{E}_{\Omega_{\delta}}\left[\epsilon(\delta\mathcal{B})\right]=(-2)^{n}\cdot\mbox{Pf}(\mathbf{F}^{\mathcal B})\,,\qquad \mbox{and} \qquad& \lim_{\Omega\to\mathbb{C}}\frac{\mathbb{E}_{\Omega_{\delta}}\left[\sigma_{0}\epsilon_{[0]}(\delta\mathcal{B})\right]}{\mathbb{E}_{\Omega_{\delta}}\left[\sigma_{0}\right]}=(-2)^{n}\cdot\mbox{Pf}(\mathbf{F}_{[0]}^{\mathcal B})\,.
\end{align*}
\end{remark*}
For spin pattern fields, our results translate to the following:
\begin{corollary}[Conformal invariance of pattern probabilities]
\label{thm:spin-pattern}Let $\mathcal{F}$ be a finite connected
collection of faces of $\mathbb{C}_{1}$ including $0$. For any $\rho\in\{\pm1\}^{\mathcal{F}}$
we have:

\begin{align*}
\delta^{-1}\left(\mathbb{P}_{\Omega_{\delta}}\left[\sigma|_{\delta\mathcal{F}}\in\{\pm\rho\}\right]-\mathbb{P}_{\mathbb{C}_{1}}\left[\sigma|_{\mathcal{F}}\in\{\pm\rho\}\right]\right)\xrightarrow[\delta\to0]{} & \langle\langle\mathcal{F}, \{\pm \rho\}\rangle\rangle_{\Omega}\,,\\
\delta^{-1/8}\left(\mathbb{P}_{\Omega_{\delta}}[\sigma|_{\delta\mathcal{F}}=\rho]-\mathbb{P}_{\mathbb{C}_{1}}[\sigma|_{\mathcal{F}}=\rho]\right)\xrightarrow[\delta\to0]{} & \langle\langle\mathcal{F},\rho \rangle\rangle_{\Omega}'\,,
\end{align*}
where the functions $\langle \langle \cdot \rangle \rangle_{\Omega}$ and $\langle \langle \cdot \rangle \rangle' _\Omega$ depend only on $\Omega$, and where:

\begin{itemize}
\item infinite-volume limits $\mathbb{P}_{\mathbb{C}_{1}}\left[\sigma|_{\mathcal{F}}\in\{\pm\rho\}\right]={\displaystyle \lim_{\Omega\to\mathbb{C}}}\mathbb{P}_{\Omega_{1}}[\sigma|_{\mathcal{F}}\in\{\pm\rho\}]$
and $\mathbb{P}_{\mathbb{C}_{1}}[\sigma|_{\mathcal{F}}=\rho]$
are explicit.
\item $\langle\langle\mathcal{F},\{\pm \rho\}\rangle\rangle_{\Omega}$ and 
$\langle\langle\mathcal{F},\rho\rangle\rangle'_{\Omega}$ are
explicit and are such that for the map $\varphi:\mathbb{D}\to\Omega$ as in Theorem~\ref{thm:spin-sym-energy}, 
\begin{align*}
\langle\langle\mathcal{F},\{\pm \rho\}\rangle\rangle_{\Omega} & =r_{\Omega}^{-1}(0)\langle\langle\mathcal{F},\{\pm \rho\}\rangle\rangle_{\mathbb{D}}\,, \qquad \mbox{and} \qquad\langle\langle\mathcal{F},\rho\rangle\rangle_{\Omega}' =r_{\Omega}^{-\frac{1}{8}}(0)\langle\langle\mathcal{F},\rho\rangle\rangle_{\mathbb{D}}'\,.
\end{align*}
\end{itemize}
\end{corollary}

\noindent As a result of Corollary~\ref{thm:spin-pattern}, our results include explicit expressions for all the finite-dimensional distributions of $\mathbb P_{\mathbb Z^2}$ as finite linear combinations of a certain Fourier integral given in Theorem~\ref{thm:slit-plane-harmonic-measure}.

\subsection{Applications}

In this subsection, we briefly detail three applications of our results: the lattice local field conjecture of the introduction, relations between Markov chain dynamics flip rates and the geometry
of the domain where the dynamics live, and explicit computations of pattern probabilities under the Gibbs measure.

\subsubsection{\label{subsec:Lattice-local-fields}Lattice Local Fields and CFT}

Returning to the conjectured Ising-CFT correspondence in terms of
lattice fields, we observe that any lattice local field $\Phi_{\delta}^{F}\left(x\right)$
is such that $F$ can be expressed as a linear combination of indicator
functions of spin-pattern events in a microscopic neighborhood of
$x$; a spin-symmetric lattice local field can in particular be written
in terms of indicators of spin-symmetric pattern events.

Then Theorems \ref{thm:spin-sym-energy}\textendash \ref{thm:spin-sens-energy}
give the infinite-volume limits, and first-order CFT corrections of
the one point function of any lattice local field $\Phi_{\delta}^{F}(x)$
in terms of those of spin-symmetric and spin-antisymmetric pattern
fields, whose one-point functions can be obtained explicitly. We believe extending this to multi-field correlations of fields with scaling dimension $D\leq 1$ should carry over from~\cite{hon2010,chelkak-hongler}. 

In the other direction, though
our main statements only go up to first-order corrections ($\delta^{1/8}$
or $\delta$), the methods of this paper can, in principle, be employed
to reduce the computation of higher order CFT corrections to correlations of any local lattice field to questions of discrete complex analysis. Of course, then obtaining the necessary sharper discrete complex analytic expansions is itself a major obstacle to the extension of such results. All the same, using the present framework along with, hypothetically, improved discrete complex analysis asymptotics, should yield that all Ising lattice
local fields are, as conjectured in the Introduction, either zero
in correlations, or have scaling dimensions $\Delta\in\mathbb{N}\cup(\mathbb{N}+\frac{1}{8})$,
as predicted from the Ising CFT. 

\subsubsection{\label{subsec:Local-Markov-chain}Local Markov Chain Dynamics}

There are a number of Markov chain dynamics for which the Ising measure
is the stationary measure; as a result, an efficient way to sample
the Ising model is to run such a Markov chain for long times. 

Of particular importance are local dynamics, such as the Glauber dynamics,
where one picks a spin at random, and flips it with a probability
given by the state of spins in a microscopic neighborhood of it (the
simplest one using only the four neighbors). 
At critical temperature interesting dynamical behavior arises (see
\cite{lubetzky-sly}). In particular, as our results explain, the
geometry of the domain $\Omega$ has a measurable (i.e. inverse polynomial-sized)
effect on the local dynamics of the Markov chain. 

For such a dynamics, our results allow one to describe, once we are
at equilibrium, the relevant observables to compute the average flip
rates of the system: those are indeed given in terms of the occurence
probabilities of various spin patterns (typically spin-symmetric patterns). 

In particular Corollary \ref{thm:spin-pattern} gives us the following:
at criticality, for any Glauber dynamics (see e.g., \cite[Section 2.1]{lubetzky-sly}), we can derive exact information about spin-symmetric
pattern probabilities, how they behave at constant order, and how
the first-order correction depends on a geometric
quantity. 

Knowing the long-term history of a Glauber dynamics in a microscopic
neighborhood of a point enables the computation of various spin pattern
probabilities and hence lattice local field one-point functions. Higher order corrections of these terms 
in turn give geometric information beyond the conformal radius of the domain. A particularly interesting question, for which our results provide relevant tools, is the following one, due to Benjamini (private communication to the
second author):\emph{ does the complete (i.e., unbounded in time)
knowledge of the flip history of a single spin allow one to recover
the shape of the domain $\Omega$, up to isometry? }

\subsubsection{Explicit Computations}
Explicit calculation of infinite-volume limits and finite-size corrections
of pattern probabilities in the critical Ising model is of general
interest, and may be particularly useful for the program of Application~1.3.2. Such computation requires the explicit matrices of Theorems
\ref{thm:spin-sym-energy}\textendash \ref{thm:spin-sens-energy},
which are expressed in terms of the full-plane fermion and spin-fermion
observables: some values of the former are given in e.g., \cite{ken2000}; we characterize the latter as a Fourier integral 
(see Theorem~\ref{thm:slit-plane-harmonic-measure}) and give some of its values in Figure~\ref{fig:explicit-values}. In particular, the entries of the matrices in Theorems~\ref{thm:spin-sym-energy}--\ref{thm:spin-sens-energy} are given as finite linear combinations of a slit-plane harmonic measure, whose values are explicitly computable as a Fourier integral.

We present an example computation of the infinite-volume limit and
first-order conformal correction of $\mathbb{E}_{\Omega_{\delta}}[\sigma_{0}\sigma_{2\delta}]$
in Corollary \ref{cor:diag-cor} of Appendix \ref{sec:explicit-pattern-probabilities},
where, since the spins live on the rotated square lattice, this is
a pair of diagonally ``adjacent'' spins. The first and second order corrections to this term, and their representation in terms of discrete complex analysis will be used in the Ising stress tensor
on the lattice level (see \cite{benoist-hongler-stress} and, for an alternative approach to the stress tensor, \cite{cgs-stress}). 

As a computation of spin-antisymmetric fields, Corollary~\ref{cor:l-shaped} gives the values of the infinite-volume limit and conformal correction to the spin weighted ``L''-shaped correlation $\mathbb E_{\Omega_\delta}[\sigma_0 \sigma_{(1+i)\delta} \sigma_{2\delta}]/\mathbb E_{\Omega_\delta}[\sigma_0]$.  

\subsection{Proof Outline}\label{subsec:proof-outline}

In this subsection we outline the strategy for proving our main results:
Theorems~\ref{thm:spin-sym-energy}\textendash \ref{thm:spin-sens-energy}. The proof combines ideas from~\cite{hon2010,chelkak-hongler}, and we try to focus the outline on the places where substantial new ingredients are needed. 
The steps in proving the main theorems broadly consist of the following.

Section \ref{sec:discrete-complex-analysis} defines standard concepts
in discrete complex analysis as well as the discrete Riemann boundary
value problems solved by certain discrete observables.
Section \ref{sec:bounded-domain-observables} begins by defining the
two-point discrete \emph{fermion} and \emph{spin-fermion} (given by $(\alpha,\zeta) \mapsto F_{\Omega_{\delta}}^{\alpha,\zeta}$ and $F_{[\Omega_{\delta},0]}^{\alpha,\zeta}$
respectively) via the low-temperature expansion of the Ising
model and disorder lines, as well as their full-plane analogues. 
\begin{itemize}
\item In \S\ref{subsec:two-point-observable}, we define the
bounded domain observables on $\Omega_{\delta}$, as previously defined in~\cite{hosm2013,chelkak-hongler}. 
\item In \S\ref{sec:full-plane-observables}, we introduce their
full-plane analogues: the full plane fermion $H^\alpha_{\mathbb C_1}(z)$ is given explicitly
by a formula due to Kenyon. For the special value of $\alpha_0= \frac 12$, the full plane spin-fermion $H^{\alpha_0}_{[\mathbb C_1,0]}(z)$ was given by~\cite{chelkak-hongler}. Here, we prove existence of the infinite-volume
limit of the spin-fermion $H^{\alpha}_{[\mathbb C_1,0]}$ for every $\alpha$, and express it as a finite linear combination of discrete harmonic
measures on $\mathbb{C}_{1}\backslash\mathbb{R}_{>0}$. Moreover, we
give an explicit representation formula using Fourier techniques for this discrete harmonic measure (see
Theorem~\ref{thm:slit-plane-harmonic-measure}), allowing computation of $H^{\alpha}_{[\mathbb C_1,0]}(z)$ for arbitrary $\alpha$.
\end{itemize}

\noindent Section \ref{sec:multipoint-observables} defines and analyzes $n$-point
analogues of the two-point fermion and spin-fermion. This section is notationally heavy, but many of its proofs are straightforward adaptations of proofs in~\cite{hon2010}.
\begin{itemize}
\item We first recall the multipoint fermion\emph{
}defined in \cite{hon2010} and slightly, but crucially, generalize its properties to the setting
where its arguments are permitted to be adjacent edges. 
\item Motivated by this definition, we consider a multipoint version of
the spin-fermion observable in Definition \ref{def:multipoint-observable-2}
and prove that the same properties hold after minor modifications. 
\item In Proposition \ref{prop:correlation-fused-observable} we relate
specific values taken by the multipoint fermion\emph{ }and spin-fermion to the $n$-point\emph{ }spin-symmetric and spin-antisymmetric
correlations of Theorems~\ref{thm:spin-sym-energy}\textendash \ref{thm:spin-sens-energy}.
\item These results allow us to arrive at the spin-antisymmetric analogues
of the Pfaffian formulae of \cite{hon2010}, connecting spin-antisymmetric
$n$-point Ising correlations to the Pfaffian of a matrix with entries consisting
only of the two-point spin-fermion (see Proposition \ref{prop:spin-energy-fermion-correlation-pfaffian-observable}). 
\end{itemize}

Section \ref{sec:continuous-observables} defines two-point continuous observables,
and proves that they are the renormalized scaling limits (as $\delta\to0$)
of the discrete two-point observables.
\begin{itemize}
\item In \S\ref{subsec:Continuous-Observables} we introduce the
continuous analogues of the discrete Riemann boundary value problem
defined in Section \ref{sec:discrete-complex-analysis} and give their
full-plane solutions and bounded domain solutions $h_{\Omega}^{\alpha}$
and $h_{[\Omega,0]}^{\alpha}$. Again this was already done for the fermion in~\cite{hon2010,hosm2013} and for the spin-fermion in the particular case of $h^{\alpha=\frac 12}_{[\Omega,0]}$~\cite{chelkak-hongler}; extra care is needed in constructing the continuous bounded domain and full-plane spin-fermions for arbitrary source point $\alpha$. 
\item The heart of Section~\ref{sec:continuous-observables}, \S\ref{subsec:observable-convergence}
proves convergence of a rescaled, renormalized discrete spin-fermion to a conformally covariant quantity obtained by Taylor
expanding $h_{[\Omega,0]}^{\alpha}$, for arbitrary $\alpha$. In adapting the proof of convergence in~\cite{chelkak-hongler} to the case where $\alpha\neq \frac 12$, we require refined analysis of the observables near their branch points and singularities. Here we encounter some discrete complex analytic peculiarities regarding discretizations of the function $i\sqrt z$ which are independently interesting.
\end{itemize}

Section~\ref{subsec:main-theorem-proofs} combines the Pfaffian formulae
of \S\ref{sec:observables-and-ising-correlations} expressing $n$-point correlations
in terms of two-point discrete observables, with the convergence
results of \S\ref{sec:continuous-observables}, to prove Theorems \ref{thm:spin-sym-energy}\textendash \ref{thm:spin-sens-energy}.

In Appendix \ref{app:full-plane-observables-construct}, we prove
the validity of our explicit construction of the discrete harmonic
measure on $\mathbb{C}_{1}\backslash\mathbb{R}_{>0}$ and provide
a recursive formula to obtain its value at any lattice point as a
finite linear combination of Fourier integrals.
In Appendix \ref{app:contour-weights}, Proposition \ref{prop:weight-welldefined},
we provide a combinatorial proof of the well-definedness of the discrete
multipoint spin-fermion we introduce in Definition~\ref{def:multipoint-observable-2}. 
As mentioned earlier, Appendix~\ref{sec:explicit-pattern-probabilities}
consists of explicit computations of the infinite-volume limit and
first-order correction of the correlations of two diagonally
adjacent spins and three spins in an ``L'' shape.

\subsection{\label{subsec:Extra-Notation}Extra Notation and Glossary}

We now introduce extended notation that will be used globally throughout the
paper. This notation mimics very closely that of \cite{chelkak-hongler}, and we try to make note of places where our conventions differ.

\subsubsection{Relevant Constants}
The following constants will recur throughout the paper.

\vspace{.3cm}
\begin{tabular}{lccr}
\toprule%
$\beta_c = \frac 12 \log (1+\sqrt 2)$ & \qquad\qquad $\mu= \frac{\sqrt 2}2$ &\qquad\qquad  $\lambda = e^{i \pi/4}$ & \qquad\qquad 
\\
\midrule
\bottomrule
\end{tabular}

\subsubsection{Graph Notation}

We list below the additional graph notation that will be used throughout
the paper. 

\begin{itemize}
\item For two adjacent vertices $a,b\in\mathcal{V}_{\Omega_{\delta}}$ the
edge $e=\{a,b\}$ is identified with the line segment in $\Omega$
connecting $a$ and $b$; we define the set of \emph{medial vertices}
$\mathcal{V}_{\Omega_{\delta}}^{m}$ as the set of edge midpoints;
given an edge $e\in\mathcal{E}_{\Omega_{\delta}}$, we denote its
midpoint by $m(e)\in\mathcal{V}_{\Omega_{\delta}}^{m}$, and conversely
for $m\in\mathcal{V}_{\Omega_{\delta}}^{m}$ the corresponding edge
$e(m)\in\mathcal{E}_{\Omega_{\delta}}$.
\item We call \emph{corners} the points that are at distance $\delta/2$
from the vertices in one of the four $\pm1,\pm i$ directions. Following~\cite{chelkak-hongler}, we set 
\begin{align*}
\mathcal{V}_{\Omega_{\delta}}^{1}:=\mathcal{V}_{\Omega_{\delta}}+\frac{\delta}{2}, \quad\mathcal{V}_{\Omega_{\delta}}^{i}:=\mathcal{V}_{\Omega_{\delta}}-\frac{\delta}{2}, \quad \mathcal{V}_{\Omega_{\delta}}^{\lambda}:=\mathcal{V}_{\Omega_{\delta}}-\frac{i\delta}{2}, \quad \mbox{ and } \mathcal{V}_{\Omega_{\delta}}^{\bar{\lambda}}:=\mathcal{V}_{\Omega_{\delta}}+\frac{i\delta}{2}\,.
\end{align*}
The set of corners $\mathcal{V}_{\Omega_{\delta}}^{c}$ is the union
$\mathcal{V}_{\Omega_{\delta}}^{1}\cup\mathcal{V}_{\Omega_{\delta}}^{i}\cup\mathcal{V}_{\Omega_{\delta}}^{\lambda}\cup\mathcal{V}_{\Omega_{\delta}}^{\bar{\lambda}}$.
\item The domain of definitions for most discrete functions in the following
sections is the set of both corners and medial vertices, or $\mathcal{V}_{\Omega_{\delta}}^{cm}:=\mathcal{V}_{\Omega_{\delta}}^{c}\cup\mathcal{V}_{\Omega_{\delta}}^{m}$.
We declare a medial vertex and a corner \emph{adjacent} if they are
$\frac{\delta}{2}$ apart from each other.
\item The \emph{boundary faces $\partial\mathcal{F}_{\Omega_{\delta}}$,
boundary medial vertices} $\partial\mathcal{V}_{\Omega_{\delta}}^{m}$,
\emph{boundary edges} in $\partial\mathcal{E}_{\Omega_{\delta}}$
are those faces, medial vertices, and edges in $\mathbb{C}_{\delta}$
that are incident to but not contained in $\mathcal{F}_{\Omega_{\delta}}$,
$\mathcal{V}_{\Omega_{\delta}}^{m}$, and $\mathcal{E}_{\Omega_{\delta}}$. 
\item Given a boundary edge $z$ (resp., boundary medial vertex), we define the \emph{unit
normal outward vector} $\nu_{z}$ as the unit vector in the direction
of the vertex in $\mathbb{C}\setminus\Omega$ viewed from the vertex
inside $\Omega$.
\end{itemize}

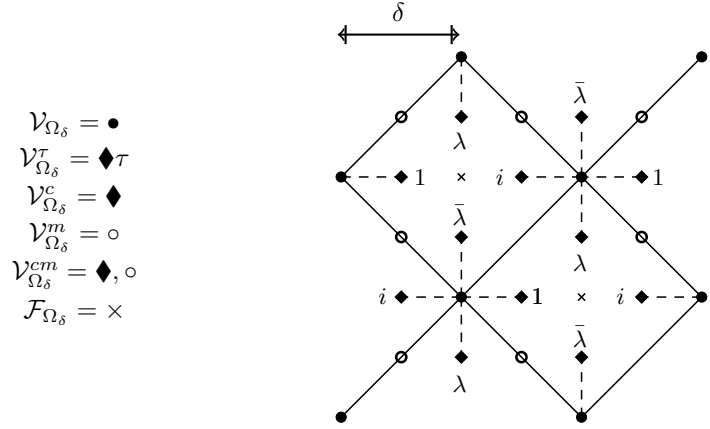
\begin{figure}
\centering
  \begin{tikzpicture}
  \begin{scope} at (0,0) 
  	\node[color=black] at (0,1.5cm) {$\mathcal V_{\Omega_\delta}= \bullet$};
  	\node[color=black] at (0,1) {$\mathcal V_{\Omega_\delta}^\tau = \blacklozenge \tau$};
	\node[color=black] at (0,.5cm) {$\mathcal V_{\Omega_\delta}^c = \blacklozenge$};
	\node[color=black] at (0,0cm) {$\mathcal V_{\Omega_\delta}^m = \circ$};
	\node[color=black] at (0,-.5cm) {$\mathcal V_{\Omega_\delta}^{cm}= \blacklozenge,\circ$};
	\node[color=black] at (0,-1cm) {$\mathcal F_{\Omega_\delta}= \times$};
  \end{scope}
  \node (fig2) at (6cm,0) { 
	\includegraphics{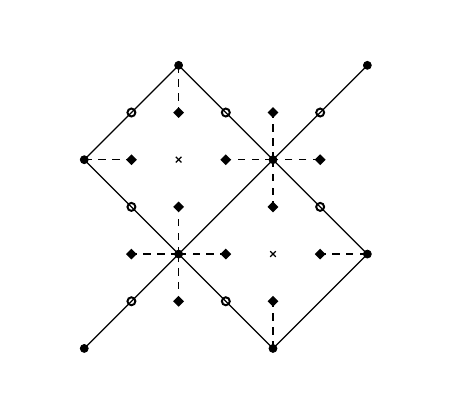}};
\begin{scope}
	\node[font=\small] at (5.12cm,.31cm) {$\bar\lambda$};
	\node[font=\small] at (5.12cm,1.29cm) {$\lambda$};
	\node[font=\small] at (4.6cm,.8cm) {$1$};
	\node[font=\small] at (5.66cm,.8cm) {$i$};
	\node[font=\small] at (6.16cm,-.78cm) {$1$};
	\node[font=\small] at (4.1cm,-.78cm) {$i$};
	\node[font=\small] at (5.12cm,-1.95cm) {$\lambda$};
	\node[font=\small] at (6.72cm,-1.31cm) {$\bar\lambda$};
	\node[font=\small] at (6.72cm,-.33cm) {$\lambda$};
	\node[font=\small] at (6.16cm,-.78cm) {$1$};
	\node[font=\small] at (6.72cm,1.93cm) {$\bar\lambda$};
	\node[font=\small] at (7.26cm,-.78cm) {$i$};
	\node[font=\small] at (7.76cm,.8cm) {$1$};
	\draw[<|-|>, thick] (3.5cm,2.7cm)-- (5.12cm,2.7cm);
	\node at (4.31cm,3cm) {$\delta$};
\end{scope}
\end{tikzpicture}
\vspace{-1cm}
\caption{The graph notation on discretizations of $\Omega$ where $\tau\in \{1,i,\lambda,\bar\lambda\}$; the notation on $[\Omega_\delta,0]$ is analogously defined.}\label{fig:graph-notation}
\end{figure}

\subsubsection{Graph lifts to the double cover}
For the discrete functions with monodromy which will be introduced
in Section 3, we work with graphs lifted to the double cover $\left[\Omega,0\right]$
of $\Omega\setminus\{0\}$.
\begin{itemize}
\item We denote by $\left[\mathbb{C},0\right]$ the double cover of the
plane $\mathbb{C}$ ramified at $0$, i.e. the surface on which the
function $z\mapsto\sqrt{z}\in\mathbb{C}\setminus\left\{ 0\right\} $
is naturally defined; above each point of $\mathbb{C}\setminus\left\{ 0\right\} $
lie exactly two points of $\left[\mathbb{C},0\right]$. We will sometimes
use $z\in\left[\mathbb{C},0\right]$ to refer to its projection on
$\mathbb{C}$ in unambiguous cases.
\item If $z_{1},z_{2}\in\left[\mathbb{C},0\right]$ are two points above
$w_{1},w_{2}\in\mathbb{C}\setminus\left\{ 0\right\} $ such that $\operatorname{Re}\left(\frac{w_{1}}{w_{2}}\right)>0$,
we say that they are \emph{on the same sheet} if $\operatorname{Re}\left(\frac{\sqrt{z_{1}}}{\sqrt{z_{2}}}\right)>0$
and that they are \emph{on opposite sheets} if $\operatorname{Re}\left(\frac{\sqrt{z_{1}}}{\sqrt{z_{2}}}\right)<0$. 
\item If $z\in\left[\mathbb{C},0\right]$ lies above $w\in\mathbb{C}\setminus\left\{ 0\right\} $,
we define $z+x\in\left[\mathbb{C},0\right]$, for $x\in\mathbb{C}$ small enough, as the point above $w+x$ that is on the same sheet
as $z$. 
\item We define complex conjugation on the double cover by conjugating the
square root. In other words, the complex conjugate $\bar{z}$ of $z\in\left[\mathbb{C},0\right]$
is defined by the condition that $\sqrt{\bar{z}}=\overline{\sqrt{z}}$. 
\item We call functions with monodromy $-1$ around $0$ \emph{spinors};
these are naturally defined on $\left[\mathbb{C},0\right]$.
\item We denote by $[\Omega_{\delta},0]$ the double cover of $\Omega_{\delta}$
ramified at $0$; in other words, the vertices, medial vertices, and
corners get lifted from $\Omega_{\delta}$ to yield the lifted vertex,
edge and corner sets. We use similar notations for the lifted vertex,
edge, and corner sets as above by replacing $\Omega_{\delta}$ with
$\left[\Omega_{\delta},0\right]$. Moreover, $\left[\Omega_{\delta},0\right]$
can be naturally viewed as a subgraph of $\left[\mathbb{C}_{\delta},0\right]$
via the natural inclusion $\left[\Omega,0\right]\subset\left[\mathbb{C},0\right]$. 
\item Identify the branches of the double cover $\left[\mathbb{C},0\right]$
using the function $\sqrt{z}$ as follows:

\begin{tabular}{ll}
\toprule
$\mathbb X := \mathbb C \setminus \mathbb R_{<0}$ & \qquad $\mbox{ with } \mathbb X^+ = \{z\in [\mathbb C,0]: \operatorname{Re}(\sqrt z)> 0\} \mbox{ and } \mathbb X^- = \{z\in [\mathbb C,0]: \operatorname{Re}(\sqrt z)< 0\}$ \\
\midrule 
$\mathbb Y := \mathbb C \setminus \mathbb R_{>0}$ & \qquad $\mbox{ with } \mathbb Y^+ = \{z\in [\mathbb C,0]: \operatorname{Im}(\sqrt z)> 0\} \mbox{ and } \mathbb Y^- = \{z\in [\mathbb C,0]: \operatorname{Im}(\sqrt z)< 0\}$\\
\midrule
\bottomrule
\end{tabular}

On the discrete
level, define the lift of $\mathcal{V}_{\Omega_{\delta}}^{1}$ to
$\mathbb{X}^{\pm}$ as $\mathbb{X}_{\delta}^{\pm}$, and the lift
of $\mathcal{V}_{\Omega_{\delta}}^{i}$ to $\mathbb{Y}^{\pm}$ as
$\mathbb{Y}_{\delta}^{\pm}$.
\end{itemize}

\subsubsection{Orientations}
We define orientations and s-orientations for corners and medial vertices.
\begin{itemize}
\item Given an edge $e=\{a,b\}\in\mathcal{E}_{\Omega_{\delta}}$, we denote
the two orientations of $e$ by the complex numbers $\left(a-b\right)/\left|a-b\right|$
and $\left(b-a\right)/\left|b-a\right|$. We can subdivide $e$ into
two \emph{half edges} $\{a,m(e)\}$ and $\{m(e),b\}$; their union
is identified with the whole edge $e$. An orientation $o=o(e)$
is \emph{compatible} with a half edge $\{a,m(e)\}$ if $a-m(e)$ points
to the same direction: i.e., $o=(a-m(e))/|a-m(e)|$. 
\item We call an \emph{oriented medial vertex} and denote by $m^{o}$ an
edge midpoint $m\left(e\right)$ together with an orientation of the
edge $e$. We denote by $\mathcal{V}_{\Omega_{\delta}}^{o}$ the set
of oriented medial vertices. 
\item For a corner $c$, we define its orientation $o=o\left(c\right)$ as
the complex number $\left(v-c\right)/\left|v-c\right|$, where $v$
is the nearest vertex to $c$. 
\item To an orientation
$o$ we further associate two \emph{s-orientations} corresponding to the two choices of square root for $o$; we often denote an s-orientation by $(\sqrt{o})^{2}$, indicating this choice.
\end{itemize}

\subsubsection{Glossary}
For the reader's convenience, we compile some of the most important terminology and quantities used across the paper (see also Fig.~\ref{fig:graph-notation} for the graph notation).
We first recall the various graphs we work with: if $\Omega$ is a simply-connected smooth domain containing the origin and $\overline \Omega$ is its complex conjugate,

\vspace{.3cm}
\begin{tabular}{lllll}
\toprule
$\mathbb C_1 = (1+i)\mathbb Z^2+1$ \qquad & \qquad $\mathbb C_\delta = \delta \mathbb C_1$ \qquad & \qquad
$\Omega_\delta = \Omega \cap \mathbb C_\delta$ \qquad & \qquad $\Lambda_\delta = \Omega_\delta \cap (\overline \Omega)_\delta$ \qquad & \qquad $a_0 = \frac 12$\\
\midrule
\bottomrule
\end{tabular}
\vspace{.3cm}

\noindent and the quantities $[\mathbb C_1,0]$, $[\Omega_\delta,0]$, etc. are analogously defined on the ramified plane $[\mathbb C,0]$. When proofs are independent of the choice, we let $D_\delta$ denote \emph{either of $\Omega_\delta$ or $[\Omega_\delta,0]$}. 

The domain-dependent quantities of interest match with~\cite{chelkak-hongler} and read as follows for fixed $z\in \Omega$.

\vspace{.3cm}
\begin{tabular}{lll}
\toprule
Geometric quantities  &  \\
\midrule 
$\varphi(\omega)$ & conformal map $\varphi:\mathbb D\to\Omega$ with $\varphi(0)=z$ & \\
\midrule
$r_\Omega(z)$ & $r_\Omega(z)=|\varphi'(0)|$ & conformal radius of $z$ \\
\midrule
$\mathcal A_\Omega$ & $\mathcal A_\Omega = -\tfrac 14 \partial_z \log r_\Omega(z) \big|_{z=0}$ & Remark~\ref{rem:A-omega}\\
\midrule
\bottomrule
\end{tabular}
\vspace{.3cm}

In what follows, we present the notation for the fundamental observables we deal with; in order to reduce multipoint observables to such two-point functions via Pfaffian relations we need much heavier notation, that is restricted to \S\ref{sec:multipoint-observables}. In the sequel, $a,z$ will be medial vertices or corners and $\alpha$ and $\zeta$ will be their s-oriented counterparts, e.g., $\alpha=a^{(\sqrt o)^2}$. 

\vspace{.3cm}
\begin{tabular}{lll}
\toprule
Fermion observables & & \\
\midrule
$F_{\Omega_\delta}^{\alpha,\zeta}$ & discrete real fermion  & Definition~\ref{def:fermion-fermion-observable} \\
\midrule
$H_{\Omega_\delta}^{\alpha}(z)$ & discrete complexified fermion  & Definition~\ref{def:complex-fermion-fermion-observable}\\
\midrule
$F_{\mathbb C_1}^{\alpha,\zeta}$ & discrete full-plane fermion & Theorem~\ref{thm:fermion-fermion-infinite-vol-lim} \\
\midrule
$H_{\mathbb C_1}^{\alpha}(z)$ & discrete full-plane complexified fermion  & Proposition~\ref{thm:fermion-fermion-full-plane-explicit-formula}\\
\midrule
$f_\Omega,h_\Omega,f_{\mathbb C},h_{\mathbb C}$ & continuous counterparts of the above & Definitions~\ref{def:full-plane-fermion-observable},~\ref{def:continuous-fermion-fermion-observable}\\
\midrule
$F^\dagger_{\Omega_\delta}, H^{\dagger}_{\Omega_\delta}, f^\dagger_{\Omega},h^{\dagger}_\Omega$ & e.g., $F^{\dagger,\alpha,\zeta}_{\Omega_\delta} = F^{\alpha,\zeta}_{\Omega_\delta} - F^{\alpha,\zeta}_{\mathbb C_\delta}$ &\\
\midrule
\bottomrule
\end{tabular}
\vspace{.3cm}

\noindent The notation for the spin-fermions is analogous to the above, but on the respective double covers. 

\vspace{.3cm}
\begin{tabular}{lll}
\toprule
Spinor observables & &\\
\midrule
$F_{[\Omega_\delta,0]}^{\alpha,\zeta}$ & discrete real spin-fermion  & Definition~\ref{def:spin-fermion-observable}\\
\midrule
$H_{[\Omega_\delta,0]}^{\alpha}(z)$ & discrete complexified spin-fermion  & Definition~\ref{def:complex-spin-fermion-observable}\\
\midrule
$F_{[\mathbb C_1,0]}^{\alpha,\zeta}$ & discrete full-plane spin-fermion & Theorem~\ref{thm:spin-fermion-infinite-vol-lim}, item (E) \\
\midrule
$H_{[\mathbb C_1,0]}^{\alpha}(z)$ & discrete full-plane complexified spin-fermion  & Proposition~\ref{prop:spin-fermion-chi-paper-formula}\\
\midrule
$S_{[\Omega_\delta,0]}^\alpha(z), A_{[\Omega_\delta,0]}^\alpha(z)$ & symmetrized and anti-symmetrized observables & Definition~\ref{def:sym-antisym-observables}\\
\midrule
$G_{[\mathbb C_1,0]}(z), \tilde G_{[\mathbb C_1,0]}^\pm(z)$ & auxiliary functions & Definitions~\ref{def:aux-functions} \\
\midrule
$C_\alpha=C_{a^o}$ & $C_\alpha = -\re \left[ i \sqrt o (\tilde G^{+}_{[\mathbb C_1,0]}-\tilde G^-_{[\mathbb C_1,0]})(a)\right]$ & Corollary~\ref{cor:displacement-scaling-explicit}\\
\midrule
\bottomrule
\end{tabular}
\vspace{.3cm}

\noindent The lower-case versions of the spin-fermions above again are their continuous counterparts, and when there is a $\dagger$ superscript, that denotes the difference of the bounded-domain and full-plane spin-fermions.

\section{\label{sec:discrete-complex-analysis}Discrete Complex Analysis}

In this section, we review basic notions of discrete complex analysis
that will be useful in this paper. We use discrete complex analysis
for the following:
\begin{itemize}
\item To relate the Ising correlations to Pfaffians of fermion and
spin-fermion observables.
\item To obtain explicit formulae for the full-plane observables.
\item To establish the convergence of the two-point observables and study
their local behavior.
\end{itemize}

\subsection{S-holomorphicity}
\begin{definition}
\label{def:s-holomorphicity}Associate to each corner $c\in\mathcal{V}_{\Omega_{\delta}}^{\tau}$
with $\tau\in\{1,i,\lambda,\bar{\lambda}\}$, the line $l\left(c\right):=\tau\mathbb{R}$ in the complex plane.
A function $H_{\delta}$ defined on corners and medial vertices of
a discrete domain $\Omega_{\delta}$ is said to be \emph{s-holomorphic}
at a corner $c\in\mathcal{V}_{\Omega_{\delta}}^{\tau}$ if for any
adjacent medial vertex $a\in\mathcal{V}_{\Omega_{\delta}}^{m}$ we
have 
\begin{equation}
H_{\delta}\left(c\right)=\mathsf{P}_{l\left(c\right)}\left[H_{\delta}\left(a\right)\right]:=\frac{1}{2}\left(H_{\delta}\left(a\right)+\tau^{2}\bar{H_{\delta}}\left(a\right)\right),\label{eq:s-hol}
\end{equation}
where $\mathsf{P}_{l\left(c\right)}$ denotes orthogonal projection
in the complex plane onto the line $l\left(c\right)$. The function $H_{\delta}$
is said to be s-holomorphic at a medial vertex $a\in\mathcal{V}_{\Omega_{\delta}}^{m}$
if Eq. (\ref{eq:s-hol}) holds for all corners $c$ adjacent to $a$.
A function is said to be s-holomorphic on $\Omega_{\delta}$
if it is s-holomorphic at every $c\in \mathcal V^c_{\Omega_\delta}$.
\end{definition}
\begin{remark}
\label{rem:real-imagin-parts}If a function $H_{\delta}$ is s-holomorphic
on a discrete domain then it is purely real on the corners of type
$1$ and purely imaginary on the corners of type $i$. We call respective
restrictions to those corners the \emph{real part} and the \emph{imaginary
part} of $H_{\delta}$.
\end{remark}

\begin{remark}
\label{rem:s-hol-d-hol}S-holomorphicity implies usual \emph{discrete
holomorphicity} of the real and imaginary parts, defined by a lattice
version of Cauchy-Riemann equations (\cite{smirnov-ii}). If $H_{\delta}$
is s-holomorphic, then the following discrete derivative vanishes:
\[
\bar{\partial}_{\delta}H_{\delta}(x):=H_{\delta}\left(x+\frac{\lambda\delta}{\sqrt{2}}\right)-H_{\delta}\left(x-\frac{\lambda\delta}{\sqrt{2}}\right)+i\left(H_{\delta}\left(x-\frac{\bar{\lambda}\delta}{\sqrt{2}}\right)-H_{\delta}\left(x+\frac{\bar{\lambda}\delta}{\sqrt{2}}\right)\right)=0
\]
for $x\in\mathcal{V}_{\Omega_{\delta}}^{\lambda}\cup\mathcal{V}_{\Omega_{\delta}}^{\bar{\lambda}}$.
One similarly defines $\partial_{\delta}$ by taking a negative sign
in front of $i$. We extend the definition to $x\in\mathcal{V}_{\Omega_{\delta}}\cup\mathcal{F}_{\Omega_{\delta}}$
by setting $\partial_{\delta}H_{\delta}(x):=\partial_{\delta}H_{\delta}(x-\frac{i\delta}{2})+\partial_{\delta}H_{\delta}(x+\frac{i\delta}{2})$. Note the differences in our definitions compared to their continuous counterparts, as the discrete derivatives are taken in rotated directions (thus differing by a phase factor); however, we will not take direct scaling limits of the operator and this poses no problem.

The information defined on corners of type $1,i$ is enough to recover
an s-holomorphic function on $\mathcal{V}_{\Omega_{\delta}}^{cm}$:
one can start from a discrete holomorphic function defined on corners
of type 1 and $i$, reconstruct values on medial vertices based on
their projections onto $\mathbb{R}$ and $i\mathbb{R}$, then project
to corners of type $\lambda$ and $\bar{\lambda}$ (discrete holomorphicity
guarantees well-definedness at those corners); see \cite[Remark 3.1]{chelkak-hongler}.
\end{remark}
\begin{definition}
\label{def:discrete-laplacian}We define the \emph{discrete Laplacian}
$\Delta_{\delta}$ by 
\[
\Delta_{\delta}H_{\delta}\left(x\right)=H_{\delta}\left(x+\delta+i\delta\right)+H_{\delta}\left(x-\delta+i\delta\right)+H_{\delta}\left(x-\delta-i\delta\right)+H_{\delta}\left(x+\delta-i\delta\right)-4H_{\delta}\left(x\right)\,.
\]
This quantity makes sense on any discrete domain of rotated square
type, for example $\mathcal{V}_{\Omega_{\delta}}^{1}$ or $\mathcal{V}_{\Omega_{\delta}}^{i}$.
A function $H_{\delta}$ on such a lattice is said to be \emph{discrete
harmonic} if $\Delta_{\delta}H_{\delta}\left(x\right)=0$ for all
$x$ at which $\Delta_{\delta}H_{\delta}$ is defined. Analogously,
it is \emph{discrete sub-harmonic} if $\Delta_{\delta}H_{\delta}\geq0$,
and \emph{discrete super-harmonic }if $\Delta_{\delta}H_{\delta}\leq0$.
\end{definition}
\begin{remark}
\label{rem:parts-harmonicity}The real and imaginary parts of an s-holomorphic
function on a planar domain are discrete harmonic on their respective
lattices, $\mathcal V^1_{\Omega_\delta}$ and $\mathcal V^i_{\Omega_\delta}$. This is a direct consequence of Remark \ref{rem:s-hol-d-hol}:
discrete holomorphicity converts discrete outward derivatives from
the center point in the Laplacian into discrete derivatives in the
angular direction, and going in a closed loop around the center point
gives zero.
\end{remark}
\begin{remark}
\label{rem:double-cover-complex-analysis}The notions of discrete
complex analysis thus far introduced have been defined on the planar
domain $\Omega_{\delta}$, but they generalize to $\left[\Omega_{\delta},0\right]$
in a straightforward manner since the double cover is locally isomorphic
to a planar domain (cf. Section \ref{subsec:Extra-Notation}).
However, great care is needed in applying Remark~\ref{rem:parts-harmonicity}
because, if the center point of the Laplacian is one of the corners
on the monodromy face labeled by $0$, the loop around the center point must
enclose the monodromy; as a result its lift to $[\Omega_\delta,0]$ is not closed, and thus discrete holomorphicity \emph{does not} imply harmonicity at the real and imaginary corners on the face
$0$. We may still obtain harmonicity of a discrete holomorphic spinor
at one of those two types of corners if we assume in addition that
it vanishes at the other, since the sum of discrete derivatives will
vanish as though the spinor does not branch at $0$.
\end{remark}

\subsection{Discrete Singularities}

Discrete singularities appear as violations of the s-holomorphicity projection relations
relations. To study these, we define front and back values at a singularity in order to introduce the notion of 
\emph{discrete residue} of a function $H_{\delta}$ at an oriented medial vertex. 

\begin{definition}
Let $H_\delta$ be a function defined on an oriented medial vertex $\alpha= a^o$ and its adjacent four corners. Let $c_1,c_2$ be the two corners adjacent to $a$ in the direction of $o$ (i.e., $c_1 = a+ \frac {\sqrt 2 \mathrm{Re}(o)\delta}2$ and $c_2 = a+\frac {\sqrt 2\mathrm {Im}(o)\delta}2$). We define the \emph{front} value $H_\delta (\alpha_+)$ as the unique value such that 
\begin{align*}
H_\delta \left(c_1\right) = \mathsf P_{l\left(c_1\right)}\left[H_\delta(\alpha_+)\right]\,, \qquad \mbox{and}\qquad H_\delta(c_2)=\mathsf P_{l\left(c_2\right)}\left[H_\delta(\alpha_+)\right]\,.
\end{align*} 
Likewise if $c_3,c_4$ are the two corners adjacent to $a$ in the direction of $-o$ (i.e., $c_3 = a- \frac {\sqrt 2 \mathrm{Re}(o)\delta}2$ and $c_4 = a-\frac {\sqrt 2\mathrm {Im}(o)\delta}2$), we set the \emph{back} value $H_\delta(\alpha_-)$ as the unique value such that 
\begin{align*}
H_\delta \left(c_3\right) = \mathsf P_{l\left(c_3\right)}\left[H_\delta(\alpha_-)\right]\,, \qquad \mbox{and}\qquad H_\delta(c_4)=\mathsf P_{l\left(c_4\right)}\left[H_\delta(\alpha_-)\right]\,.
\end{align*}
\end{definition}

\begin{definition}
\label{def:discrete-singularity}The \emph{discrete residue} of $H_\delta$
at $\alpha$ is the difference $\operatorname{Res}_{\alpha}(H_\delta):=H_{\delta}\left(\alpha_{+}\right)-H_{\delta}\left(\alpha_{-}\right)$.
\end{definition}
By definition $H_{\delta}$ has an s-holomorphic extension to $a$
if and only if the discrete residue is zero at $\alpha$. It is an analog of the residue in the continuous setting in that doing a closed contour sum around $a$ along the edges of the lattice (i.e. summing $H_\delta(e)\vec{e}$ where $\vec{e}$ is the vector pointing from the start to the end of the edge $e$ on the closed counterclockwise path) will yield  $\sqrt{2}oi\delta\operatorname{Res}_{\alpha}(H_\delta)$, for any choice of $o$.

\subsection{\label{subsec:discrete-riemann-boundary}Discrete Riemann Boundary
Value Problems}

The key tool for our analysis is the study of discrete Riemann boundary
value problems. To prove the convergence of the Ising model observables
as the mesh size goes to zero, we will formulate them as the unique
solutions to such problems.

Recall that we denote by $\partial\mathcal{V}_{\Omega_{\delta}}^{m}$
the set of boundary medial vertices, by $\nu_{z}$ the outer normal
at any $z\in\partial\mathcal{V}_{\Omega_{\delta}}^{m},$ i.e. the
orientation at $z$ which points outward from $\Omega_{\delta}$,
and $\partial_{\nu_{z}}$ the outer normal difference, i.e. the value
on the outer adjacent vertex minus the value on the inner adjacent
vertex.

\begin{definition}We say that a function $H_{\delta}:\mathcal{V}_{\Omega_{\delta}}^{cm}\to\mathbb{C}$
defined on corners and medial vertices of a discrete domain $\Omega_{\delta}$
is the solution to the \emph{discrete Riemann boundary value problem}
on $\Omega_{\delta}$ with boundary data $f:\partial\mathcal{V}_{\Omega_{\delta}}^{m}\to\mathbb{C}$
if it is s-holomorphic and $H_{\delta}(z)-f_{\delta}(z)\in\nu_{z}^{-\frac{1}{2}}\mathbb{R}$
for any boundary medial vertex $z\in\partial\mathcal{V}_{\Omega_{\delta}}^{m}$;
note that the definition is independent of the branch of the square
root $\nu_{z}^{-\frac{1}{2}}$. This notion is straightforwardly generalized
to a function on the double cover $H'_{\delta}:\mathcal{V}_{\left[\Omega_{\delta},0\right]}^{cm}\to\mathbb{C}$
and the boundary data $q:\partial\mathcal{V}_{\left[\Omega_{\delta},0\right]}^{m}\to\mathbb{C}$
by adding the assumption that both have monodromy $-1$ around the
origin.
\end{definition}

Before proving a useful uniqueness result for the discrete Riemann
boundary value problems, we introduce the crucial notion of \emph{integration
of the square }of an s-holomorphic function, defined on the vertices
and faces (see also \cite{smirnov-ii,hon2010,chelkak-hongler}). Although the
square of an s-holomorphic function is not s-holomorphic, we can ``line-integrate''
the square of its magnitude to obtain a single-valued function without
monodromy. Its restrictions to the two rotated square lattices respectively
of faces and vertices are not harmonic, but they are respectively
super-harmonic and sub-harmonic, which will allow us to derive estimates
crucial for proofs of the convergence.

$\mathbb{I}_{\delta}\left(H_{\delta}\right)$
is a discrete analogue of the line integral $\re\int\left[H_{\delta}\right]^{2}dz$,
defined as follows.
\begin{proposition}[{\cite[Lemma 3.8]{smirnov-ii}}]
Let $H_{\delta}$ be an s-holomorphic function on $\Omega_{\delta}$. There
exists a function $\mathbb{I}_{\delta}\left[H_{\delta}\right]:\mathcal{F}_{\Omega_{\delta}}\cup\mathcal{V}_{\Omega_{\delta}}\to\mathbb{R}$ uniquely constructed (up to an additive constant) with the rule
\begin{alignat*}{1}
\mathbb{I}_{\delta}\left[H_{\delta}\right](w)-\mathbb{I}_{\delta}\left[H_{\delta}\right](v) & =2\delta\left|H_{\delta}\left(\frac{1}{2}(w+v)\right)\right|^{2}\,,
\end{alignat*}
where $w$ is a face, $v$ is a vertex incident to the face, so that
$\frac{1}{2}(w+v)$ is the corner between them.

It has $\Delta_{\delta}\mathbb{I}_{\delta}\left[H_{\delta}\right]=2\delta\left|\partial_{\delta}H_{\delta}\right|^{2}$
on $\mathcal{F}_{\Omega_{\delta}}$, $\Delta_{\delta}\mathbb{I}_{\delta}\left[H_{\delta}\right]=-2\delta\left|\partial_{\delta}H_{\delta}\right|^{2}$
on $\mathcal{V}_{\Omega_{\delta}}$.
\end{proposition}
The following uniqueness statement for both types of the discrete
Riemann boundary value problems then allows one to characterize s-holomorphic
functions in terms of their boundary values.
\begin{lemma}
\label{lem:uniqueness-discrete-rbvp}If $H_{\delta}$ is a solution
of the discrete Riemann boundary value problem on $\Omega_{\delta}$
with boundary data $0$, it is identically zero. Similarly, if $H'_{\delta}$
is a solution of the discrete Riemann boundary value problem on $\left[\Omega_{\delta},0\right]$
with boundary data $0$, it is identically zero.
\end{lemma}
\begin{proof}
The case of $H_{\delta}$ is treated in \cite[Proposition 28]{hon2010},
but we summarize it here. Given s-holomorphicity and $\mathsf{P}_{\nu_{z}^{-\frac{1}{2}}}H_{\delta}=0$,
we can calculate $\partial_{\nu_{z}}\mathbb{I}_{\delta}(H_{\delta})(z)=\sqrt{2}\delta\left|\mathsf{P}_{i\nu_{z}^{-\frac{1}{2}}}H_{\delta}(z)\right|^{2}$.
Then by using the discrete divergence formula (\cite[Lemma 6]{hon2010})
and the Laplacian, we can bound from above the orthogonal component
of $H_{\delta}$ on the boundary:
\begin{alignat*}{1}
0\leq\sum_{z\in\partial\mathcal{V}_{\Omega_{\delta}}^{m}}\partial_{\nu_{z}}\mathbb{I}_{\delta}(H_{\delta})(z)=\sum_{v\in\mathcal{V}_{\Omega_{\delta}}}\Delta_{\delta}\mathbb{I}_{\delta}(H_{\delta})(v) & \leq0\,,
\end{alignat*}
which implies that $H_{\delta}\equiv0$ on $\partial\mathcal{V}_{\Omega_{\delta}}^{m}$
and that $\Delta_{\delta}\mathbb{I}_{\delta}\left[H_{\delta}\right]=-2\delta\left|\partial_{\delta}H_{\delta}\right|^{2}\equiv0$
in $\mathcal{V}_{\Omega_{\delta}}$, so $H_{\delta}\equiv0$ in $\mathcal{V}_{\Omega_{\delta}}$.

For $H_{\delta}'$, note that (see \cite[Proposition 4.1]{chelkak-izyurov})
we can similarly define the single-valued square integral $\mathbb{I}_{\delta}(H_{\delta}')$
with single valued increments $\mathbb{I}_{\delta}\left[H'_{\delta}\right](w)-\mathbb{I}_{\delta}\left[H'_{\delta}\right](v)=2\delta\left|H'_{\delta}\left(\frac{1}{2}(w+v)\right)\right|^{2}$.
While its restriction to faces fails to be sub-harmonic at the monodromy
face in general, $\mathbb{I}_{\delta}(H'_{\delta})$ on vertices is nonetheless
super-harmonic everywhere with positive outer difference, and we can
apply the same argument as in the $H_{\delta}$ case.
\end{proof}

\section{\label{sec:bounded-domain-observables}Discrete Two-point Observables}

In this section, we introduce discrete observables, which connect
Ising model correlations to discrete complex analysis. Bounded domain
observables are defined by summing Boltzmann weights over the set
of contours made of the edges in the lattice, alluding to a path integral
formulation.

In this section, we define the two-point functions, in terms of which
the correlations will be formulated in Propositions \ref{prop:energy-correlations-pfaffian-observable}
and \ref{prop:spin-energy-fermion-correlation-pfaffian-observable}.
In \S\ref{sec:multipoint-observables}, multipoint versions of these
observables are introduced to prove these statements.

\subsubsection{\label{subsec:low-temperature-expansion}Low-temperature Expansion}

In this paper, we will use the \emph{low-temperature expansion} of
the Ising model: we represent a spin configuration by the set of edges
separating faces of opposite spins. Through this representation, the
probability of a set of edges $\omega$ is proportional to $e^{-2\beta\left|\omega\right|}$.
When considering $+$ boundary conditions, the relevant set of edges
form a collection of loops (sets of distinct edges $\{e_1,...,e_k\}$ such that $e_i$ is incident to $e_{i+1}$ for every $1\leq i\leq k$ with $e_{k+1} = e_1$). In identifying an edge configuration $\omega$ in which every vertex is incident to an even number of edges, with a \emph{collection of loops}, there is a possible ambiguity at vertices incident to four edges. For concreteness, we fix the convention that at such ambiguous vertices, loops proceed by joining northwest edges to northeast edges and southwest edges to southeast edges. We further prove that all quantities we consider are independent of choice of convention at ambiguous vertices. Denote by $\mathcal C_{\Omega_{\delta}}$ the set of all such $\omega$ (subsets of edges of $\mathcal E_{\Omega_\delta}$ with every vertex incident an even number of edges), corresponding to collections of closed loops in $\Omega_\delta$.  

As a result, for the critical Ising model
with $+$ boundary conditions, the low-temperature expansion of the
partition function is thus obtained by summing over the set $\mathcal{C}_{\Omega_{\delta}}$:
\[
\mathcal{Z}_{\Omega_{\delta}}:=\sum_{\omega\in\mathcal{C}_{\Omega_{\delta}}}e^{-2\beta_{c}\left|\omega\right|}\,.
\]

We also note that in this representation, the value of a spin is determined
by the parity of the number of loops around it (independently of the choice of convention above), and it is easy to
see that
\[
\mathbb{E}_{\Omega_{\delta}}\left[\sigma_{0}\right]=\frac{\sum_{\omega\in\mathcal{C}_{\Omega_{\delta}}}e^{-2\beta_{c}|\omega|}(-1)^{\ell(\omega)}}{\mathcal{Z}_{\Omega_{\delta}}}\,,
\]
where $\ell(\omega)$ counts the number of loops in $\omega$ that
surround 0.

\subsubsection{\label{subsec:disorder-lines}Disorder Lines}

The main tool in the study of the 2D Ising model is its fermionic
formulation. In this paper, we use the low-temperature representation
of the Ising fermion. The relevant sets of contours are deformed versions
of $\mathcal{C}_{\Omega_{\delta}}$ above: in addition to the collection of closed loops in $\Omega_\delta$,
there are paths linking a pair of marked points of the lattice. In the language
of Kadanoff and Ceva (see \cite{kadanoff-ceva}), these correspond
to the results of the insertion of disorder operators next to the
spin.

A medial vertex divides the corresponding edge into two \emph{half-edges}.
A \emph{walk} between two medial vertices $a,z$ is a sequence that
consists of a half-edge of $a$, then continues on successively
adjacent, all distinct edges, before reaching a half-edge of $z$. If $\gamma^{a,z}$ is any such a walk, $\mathcal{C}_{\Omega_{\delta}}^{a,z}:=\{\omega\oplus\gamma^{a,z}:\omega\in\mathcal{C}_{\Omega_{\delta}}\}$,
where $\oplus$ denotes the symmetric difference operation (where the symmetric difference of a half-edge and its edge is defined in the natural way), clearly
does not depend on the choice of $\gamma^{a,z}$. Each $\gamma\in\mathcal{C}_{\Omega_{\delta}}^{a,z}$
corresponds to a walk from $a$ and to $z$ and possibly some collection of loops. 

For an element $\gamma\in \mathcal C_{\Omega_\delta}^{a,z}$, we say a walk $\pi(\gamma)\subset \gamma$ from $a$ to $z$ is
an \emph{admissible choice of walk} if whenever it arrives at an ambiguous vertex, i.e. incident to
four edges in $\gamma$, it chooses to connect northeast with northwest edges and southeast with southwest edges (in accordance with the aforementioned convention for loops). Again, we will prove well-definedness of relevant quantities so that the choice of convention here is irrelevant. 
When one or both of $a,z$ are instead corners, the above is defined analogously, where ``half-edge'' is understood to mean the segment joining the corner to its nearest vertex. 

Recall that any choice of orientation on a medial vertex is in the direction of
exactly one of its two incident half-edges (see Section \ref{subsec:Extra-Notation}).
If $\alpha=a^{o}$ and $\zeta=z^{p}$ are oriented medial vertices,
set $\mathcal{C}_{\Omega_{\delta}}^{\alpha,\zeta}$ to be the subset
of $\gamma\in\mathcal{C}_{\Omega_{\delta}}^{a,z}$ including
the particular half-edges given by the respective orientations at
$a$ and $z$.

\subsection{\label{subsec:two-point-observable}Bounded Domain Observables}

In this subsection, we define the fermion and the spin-fermion
observables. The former is a function defined on the discrete domain
$\Omega_{\delta}$, whereas the latter is defined on the double cover,
$\left[\Omega_{\delta},0\right]$.
Using the above definitions of loops, walks, $\mathcal C_{\Omega_\delta}^{\alpha,\zeta}$, and admissible choices of walks, we define some quantities central to the presentation of the (two-point) fermion and spin-fermion observables.

\begin{definition}\label{def:observable-quantities}
If $\alpha=a^o$ and $\zeta= z^p$ are s-oriented medial vertices or corners, define the constants, 
\begin{align*} 
\mathrm{c}_{\upsilon}:=\begin{cases}
1 & \mbox{if $\upsilon\in \mathcal V_{\Omega_\delta}^m$}\\
\cos{\frac{\pi}{8}} & \mbox{if $\upsilon \in \mathcal V^c_{\Omega_\delta}$}
\end{cases}
\qquad \mbox{and} \qquad \lambda_{\alpha,\zeta}:=\frac{\sqrt{p}}{\sqrt{o}}\mathrm{c}_{a}\mathrm{c}_{z}\,.
\end{align*}
For a walk and loops $\gamma\in \mathcal C_{\Omega_\delta}^{\alpha,\zeta}$, define its length $|\gamma|$ as the number of full edges (where the two half-edges at the ends together count as one) in $\gamma$ and for an admissible choice of walk $\pi(\gamma)$ in $\gamma$ from $a$ to $z$, denote its winding (more accurately, turning) angle by $\mathbf{W}(\pi(\gamma))$ as the total change in argument of the velocity vector of the walk $\pi(\gamma)$ from $a$ to $z$ (see \S5.2.1 of~\cite{hon2010}). The choice of counting two half-edges together as one full-edge, is different from the counting in~\cite{chelkak-hongler}; this leads to an appearance of a normalizing factor of ${(\cos\frac{\pi}{8})^2}e^{-2\beta_c}=\frac{1}{2\sqrt 2}$, whenever considering observables with both arguments in corners. 

The following is a real-valued weight on $\gamma\in \mathcal C_{\Omega_\delta}^{\alpha,\zeta}$:
\begin{alignat*}{1}
\phi_{\alpha,\zeta}\left(\gamma\right) & :=i\lambda_{\alpha,\zeta}\:e^{-2\beta_{c}\left|\gamma\right|}\;e^{-\frac{i}{2}\mathbf{W}\left(\pi(\gamma)\right)}\,.
\end{alignat*}
We also recall for a collection of loops $\omega\in \mathcal C_{\Omega_\delta}$ the definition of $\ell(\omega)$ as the number of loops of $\omega$ around $0$ (whose parity is independent of our convention for loops). See \cite[Proposition 67]{hon2010} for the well-definedness (i.e., independence
of the choice of convention for the admissible path $\pi\left(\gamma\right)$) of $\phi_{\alpha,\zeta}$ .

When $\alpha,\zeta$ are s-oriented medial vertices or corners on $[\Omega_\delta,0]$, we define the spin-fermion weight as
\begin{alignat*}{1}
\mathrm{\phi}_{\alpha,\zeta}^{\Sigma}\left(\gamma\right) & :=\phi_{\alpha,\zeta}\left(\gamma\right)\left(-1\right)^{\ell\left(\gamma\setminus\pi(\gamma)\right)}\mathrm{s_{\alpha,\zeta}}\left(\pi(\gamma)\right)\,,
\end{alignat*}
where $\pi(\gamma)$ is any admissible choice of walk, and $\mathrm{s}_{\alpha,\zeta}$ is the \emph{sheet number} defined by 
\[
\mathrm{s_{\alpha,\zeta}}\left(\pi(\gamma)\right)=\begin{cases}
+1 & \mbox{if }\pi(\gamma)\mbox{ lifted to }\left[\Omega_{\delta},0\right]\mbox{ connects }\alpha\mbox{ to }\zeta\\
-1 & \mbox{if $\pi(\gamma)$ lifted to $[\Omega_\delta,0]$ connects }\alpha\mbox{ to }\zeta^*
\end{cases}\,,
\]
where $\zeta^*$ is the point on $[\Omega_\delta,0]$ that is distinct from $\zeta$ but shares its projection with $\zeta$. Here, the real-valued weight $\phi_{\alpha,\zeta}(\gamma)$ is still computed by identifying $\alpha,\zeta$ with their projections to $\Omega_\delta$. 

See Remark 2.2(ii) of \cite{chelkak-hongler} for the well-definedness
of the spin-fermion weight $\phi_{\alpha,\zeta}^{\Sigma}$.

\end{definition}
 
We are now in position to define the real fermion $F_{\Omega_\delta}^{\alpha,\zeta}$ and the real spin-fermion $F_{[\Omega_\delta,0]}^{\alpha,\zeta}$.  
\begin{definition}
\label{def:fermion-fermion-observable}The (real) \emph{fermion observable}
$F_{\Omega_{\delta}}$ is a function of two variables $\left(\alpha,\zeta\right)\mapsto F_{\Omega_{\delta}}^{\alpha,\zeta}$,
where $\alpha:=a^{o}$ and $\zeta:=z^{p}$ are s-oriented corners
or medial vertices of $\Omega_{\delta}$ given by
\begin{alignat*}{1}
F_{\Omega_{\delta}}^{\alpha,\zeta} & :=\frac{1}{\mathcal{Z}_{\Omega_{\delta}}}\sum_{\gamma\in\mathcal{C}_{\Omega_{\delta}}^{\alpha,\zeta}}\mathrm{\phi}_{\alpha,\zeta}\left(\gamma\right)\,.
\end{alignat*}
\end{definition}

\begin{figure}
\begin{center}
\begin{tikzpicture}
\clip (-5cm,-4cm) rectangle (5cm,4cm);
\node{
\includegraphics[scale=1.8]{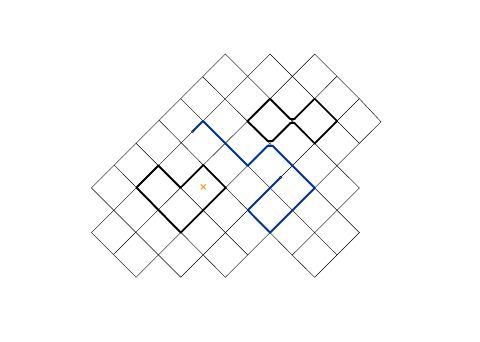}};
\node at (1.45,-.35) {$\alpha$};
\node at (-1.2,1.1) {$\zeta$};
\end{tikzpicture}
\end{center}
\caption{An configuration $\gamma \in \mathcal C^{\alpha,\zeta}_{[\Omega_\delta,0]}$, with an admissible choice of walk $\pi(\gamma)$ in blue, from $\alpha$
to $\zeta$. The winding of the walk has $\mathbf{W}(\pi(\gamma)) =2\pi$.
The loop number $\ell(\gamma)$ is $1$ since
there is precisely one loop with $0$ in its interior and $s_{\alpha,\zeta}(\pi(\gamma))=1$.}
\end{figure}
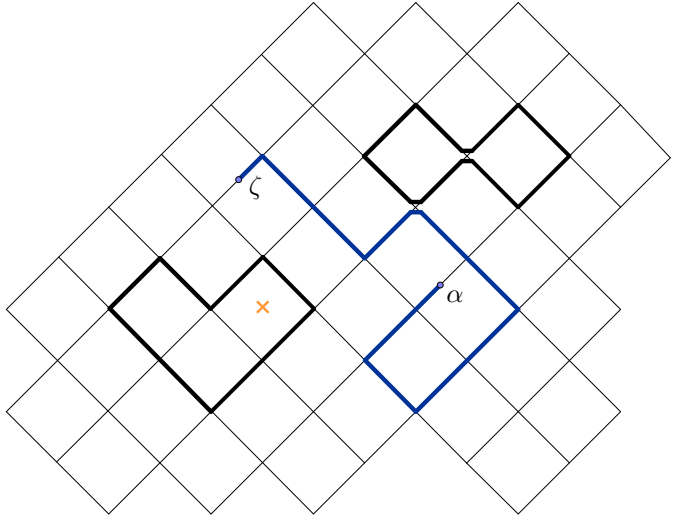

\begin{definition}
\label{def:spin-fermion-observable}The (real) \emph{spin-fermion observable}
$F_{\left[\Omega_{\delta},0\right]}$ is a function of two variables
$(\alpha,\zeta)\mapsto F_{\left[\Omega_{\delta},0\right]}^{\alpha,\zeta}$,
where $\alpha:=a^{o}$ and $\zeta:=z^{p}$ live on the double cover
$\left[\Omega_{\delta},0\right]$ of domain $\Omega_{\delta}$ ramified
at $0$.
Define $F_{\left[\Omega_{\delta},0\right]}^{\alpha,\zeta}$
by
\begin{alignat*}{1}
F_{\left[\Omega_{\delta},0\right]}^{\alpha,\zeta} & :=\frac{1}{\mathcal{Z}_{\Omega_{\delta}}\mathbb{E}_{\Omega_{\delta}}\left[\sigma_{0}\right]}\sum_{\gamma\in\mathcal{C}_{\Omega_{\delta}}^{\alpha,\zeta}}\mathrm{\phi}_{\alpha,\zeta}^{\Sigma}\left(\gamma\right)\,.
\end{alignat*}
Again, when computing $\mathcal{Z}_{\Omega_{\delta}}$, and $\mathcal{C}_{\Omega_{\delta}}^{\alpha,\zeta}$
and $\phi_{\alpha,\zeta}$, identify $\alpha,\zeta$ with their projections to $\Omega_\delta$.
\end{definition}

The well-definedness of $F^{\alpha,\zeta}_{\Omega_\delta}$ and $F^{\alpha,\zeta}_{[\Omega_\delta,0]}$ are implied by well-definedness of $\phi_{\alpha,\zeta}$ and $\phi_{\alpha,\zeta}^{\Sigma}$ respectively.

\begin{remark}
Informally, one can think of $F_{\left[\Omega_{\delta},0\right]}$
as the natural modification to $F_{\Omega_{\delta}}$ when one tries
to reweight it by the value of the spin at $0$. Since the spin and
the disorders are not mutually local (but quasi-local instead), this
gives rise to a multivalued function (with monodromy $-1$ around
$0$).
\end{remark}

\subsubsection{\label{subsec:antisymmetry-observables}Antisymmetry of the Observables}

An important elementary feature of the observables $F_{\Omega_{\delta}},F_{[\Omega_{\delta},0]}$
is their antisymmetry properties, immediate from their definitions.
The fermion observable satisfies the following:
\begin{lemma}
\label{lem:fermion-fermion-antisymmetries}If $\alpha:=a^{o},\zeta:=z^{p}$
are s-oriented corners or medial vertices of $\Omega_{\delta}$
and $\alpha':=a^{o'},\zeta':=z^{p'}$ where $o':=e^{2\pi i}o$ and
$p':=e^{2\pi i}p$, then we have the following antisymmetry properties:
\[
F_{\Omega_{\delta}}^{\alpha,\zeta}=-F_{\Omega_{\delta}}^{\zeta,\alpha}=-F_{\Omega_{\delta}}^{\alpha',\zeta}=-F_{\Omega_{\delta}}^{\alpha,\zeta'}\,.
\]
\end{lemma}
Similarly, the spin-fermion observable satisfies the following antisymmetry
properties.
\begin{lemma}
\label{lem:spin-fermion-antisymmetries}If $\alpha:=a^{o},\zeta:=z^{p}$
are s-oriented corners or medial vertices of $\left[\Omega_{\delta},0\right]$,
and $\alpha',\zeta'$ are as in the previous lemma and $\alpha^{*}:=\left(a^{*}\right)^{o},\zeta^{*}:=\left(z^{*}\right)^{p}$
where $a,a^{*}$ and $z,z^{*}$ are respectively distinct lifts of
the same points in $\Omega\setminus\{0\}$, we have
\[
F_{\left[\Omega_{\delta},0\right]}^{\alpha,\zeta}=-F_{\left[\Omega_{\delta},0\right]}^{\zeta,\alpha}=-F_{\left[\Omega_{\delta},0\right]}^{\alpha',\zeta}=-F_{\left[\Omega_{\delta},0\right]}^{\alpha,\zeta'}=-F_{\left[\Omega_{\delta},0\right]}^{\alpha^{*},\zeta}=-F_{\left[\Omega_{\delta},0\right]}^{\alpha,\zeta^{*}}\,.
\]
\end{lemma}
Recall that we define complex conjugation on the double cover by letting
the square root be conjugated. We similarly conjugate the s-orientations,
and define $\bar{\alpha}:=\bar{a}^{\bar{o}}$ if $\alpha=a^{o}$.
The fact that the contour set of $\Omega_{\delta}$ and its mirror image $\overline{\Omega_{\delta}}$
have a natural bijection arising from the complex conjugation immediately
yields the following.
\begin{lemma}
\label{lem:spin-fermion-contour-antisymmetry}If $\alpha,\zeta$ are
s-oriented corners or medial vertices in $\left[\Omega_{\delta},0\right]$,
we have $F_{\left[\Omega_{\delta},0\right]}^{\alpha,\zeta}=-F_{\left[\overline{\Omega_{\delta}},0\right]}^{\bar{\alpha},\bar{\zeta}}$.
\end{lemma}

\subsubsection{\label{subsec:complexified-observables}Complexified Observables}

As explained in the previous subsections, the observables introduced
in the previous subsections are real quantities antisymmetric in their
two variables; exploiting those properties we can define the following
complexified versions, which can be analyzed with discrete complex analysis.
\begin{definition}
\label{def:complex-fermion-fermion-observable}Let $\alpha$ be an
s-oriented corner or medial vertex of $\Omega_{\delta}$. For an
(unoriented) medial vertex $z\in \mathcal V^{m}_{\Omega_{\delta}}$, we define the
\emph{complex fermion-fermion observable} $H_{\Omega_{\delta}}^{\alpha}$
by 
\[
H_{\Omega_{\delta}}^{\alpha}\left(z\right):=\frac{1}{i\sqrt{p_1}}F_{\Omega_{\delta}}^{\alpha,\zeta^1}+\frac{1}{i\sqrt{p_2}}F_{\Omega_{\delta}}^{\alpha,\zeta^2}\,,
\]
where $\zeta^1:=z^{p_1}$ and $\zeta^2:=z^{p_2}$ are arbitrary s-orientations of $z$ with opposite orientations, i.e. $p_2=e^{\pm\pi i}p_1$. The resulting quantity is easily seen to be well-defined regardless of the choice of s-orientations. Similarly for a corner $c\in \mathcal V^c_{\Omega_{\delta}}$ with s-orientation $\kappa=c^q$, define 
\[
H_{\Omega_{\delta}}^{\alpha}\left(c\right):=\frac{1}{i\sqrt{q}}F_{\Omega_{\delta}}^{\alpha,\kappa}\,.
\]
\end{definition}
Define the complexified spin-fermion observable in the same way:
\begin{definition}
\label{def:complex-spin-fermion-observable}Let $\alpha$ be an s-oriented
corner or medial vertex of $\left[\Omega_{\delta},0\right]$, let
$z$ be a medial vertex of $\left[\Omega_{\delta},0\right]$, and
let $c$ be a corner of $\left[\Omega_{\delta},0\right]$. Using the
same notation as in Definition \ref{def:complex-fermion-fermion-observable},
we define the \emph{complex spin-fermion observable} $H_{\left[\Omega_{\delta},0\right]}$
by
\begin{eqnarray*}
H_{\left[\Omega_{\delta},0\right]}^{\alpha}\left(z\right) & := & \frac{1}{i\sqrt{p_1}}F_{\left[\Omega_{\delta},0\right]}^{\alpha,\zeta^1}+\frac{1}{i\sqrt{p_2}}F_{\left[\Omega_{\delta},0\right]}^{\alpha,\zeta^2}\,,\\
H_{\left[\Omega_{\delta},0\right]}^{\alpha}\left(c\right) & := & \frac{1}{i\sqrt{q}}F_{\left[\Omega_{\delta},0\right]}^{\alpha,\kappa}\,.
\end{eqnarray*}
\end{definition}
\begin{remark}
\label{rem:real-complex-projection} Note that, given a complexified
observable, the real observables can be recovered by $F_{\Omega_{\delta}}^{\alpha,\zeta}=i\sqrt{p}\mathsf{P}_{\frac{1}{i\sqrt{p}}\mathbb{R}}\left[H_{\Omega_{\delta}}^{\alpha}(z)\right]=\re\left[i\sqrt{p}H_{\Omega_{\delta}}^{\alpha}(z)\right]$ for medial vertices $\zeta= z^{p}$, and obviously at corners $\kappa = c^q$, we have $F_{\Omega_\delta}^{\alpha,\kappa}= \re [i\sqrt q H_{\Omega_\delta}^{\alpha}(c)]$.
\end{remark}
\begin{definition}
\label{def:sym-antisym-observables}Let $\Lambda_{\delta}=\Omega_{\delta}\cap\overline{\Omega_{\delta}}$.
Then define the symmetrized and antisymmetrized observables:
\begin{align*}
S_{[\Omega_\delta,0]}^{\alpha}:=\frac{1}{2}\left[H_{\left[\Omega_{\delta},0\right]}^{\alpha}+H_{\overline{\left[\Omega_{\delta},0\right]}}^{\bar{\alpha}}\right],\qquad A_{[\Omega_\delta,0]}^{\alpha}:=\frac{1}{2}\left[H_{\left[\Omega_{\delta},0\right]}^{\alpha}-H_{\overline{\left[\Omega_{\delta},0\right]}}^{\bar{\alpha}}\right]\,.
\end{align*}
\end{definition}
The following lemma then follows immediately from Lemma~\ref{lem:fermion-fermion-antisymmetries} and the definition of $H_{[\Omega_{\delta},0]}$. 
\begin{lemma}
\label{lem:sym-antisym-observables}Let $\alpha$ be an s-oriented
corner or medial vertex in $\left[\Omega_{\delta},0\right]$, and
$z$ be a corner or medial vertex in $\left[\Lambda_{\delta},0\right]$.
Then,
\begin{alignat*}{1}
\mbox{on }\mathcal{V}_{\left[\Omega_{\delta},0\right]}^{1}\cap\mathbb{R}_{>0},\mathcal{V}_{\left[\Omega_{\delta},0\right]}^{i}\cap\mathbb{R}_{<0}, \qquad \mbox{we have} \qquad & S_{[\Omega_\delta,0]}^{\alpha}=H_{\left[\Omega_{\delta},0\right]}^{\alpha} \mbox{ and }A_{[\Omega_\delta,0]}^{\alpha}=0\,;\\
\mbox{on }\mathcal{V}_{\left[\Omega_{\delta},0\right]}^{i}\cap\mathbb{R}_{>0},\mathcal{V}_{\left[\Omega_{\delta},0\right]}^{1}\cap\mathbb{R}_{<0}, \qquad \mbox{we have}\qquad & A_{[\Omega_\delta,0]}^{\alpha}=H_{\left[\Omega_{\delta},0\right]}^{\alpha}\mbox{ and }S_{[\Omega_\delta,0]}^{\alpha}=0\,.
\end{alignat*}
\end{lemma}

\subsection{\label{sec:full-plane-observables}Full Plane Observables}

In this section, we study the infinite-volume limits of the fermion
and spin-fermion observables. By scale invariance, it is enough to
give a characterization on the rotated unit grid $\mathbb{C}_{1}=(1+i)\mathbb{Z}^{2}+1$
placed on increasing domains. On $\mathbb{C}_{\delta}$
we can define $H_{\mathbb{C}_{\delta}}(a\delta):=H_{\mathbb{C}_{1}}(a)$
and $H_{\left[\mathbb{C}_{\delta},0\right]}(a\delta):=H_{\left[\mathbb{C}_{1},0\right]}(a)$.
In the $\delta\to0$ scaling limit, these converge to meromorphic
functions with a singularity at zero.

We first give a unique characterization for the full-plane limits and establish their existence, and then we give an explicit construction.
Using those explicit formulae, we define auxiliary s-holomorphic
functions on the double cover which are discrete forms of $\sqrt{z}$
and $i\sqrt{z}$.

We take the limit $\Omega_{1}\to\mathbb{C}_{1}$ using an increasing
sequence of bounded domains $\Omega_{1}^{1}\subset\Omega_{1}^{2}\subset\cdots\subset\Omega_{1}^{n}\subset\cdots$
such that $\bigcup_{n}\Omega_{1}^{n}=\mathbb{C}_{1}$. The limiting
functions will be seen to be unique, so that they do not depend on
the particular sequence.

\subsubsection{Full Plane Fermion Observable}
The following are straightforward modifications to our setting, of the construction~\cite{hon2010} of the full-plane fermion. 
\begin{theorem}
\label{thm:fermion-fermion-infinite-vol-lim}As $\Omega_{1}\to\mathbb{C}_{1}$,
the complexified fermion observable $H_{\mathbb{C}_{1}}:=\lim_{\Omega\to\mathbb{C}}H_{\Omega_{1}}$
exists and is uniquely characterized by the following properties:

\begin{itemize}
\item if $\alpha=a^{o}$ is an s-oriented medial vertex,
\begin{itemize}
\item $H_{\mathbb{C}_{1}}^{\alpha}$ is s-holomorphic on $\mathbb{C}_{1}\setminus\{a\}$;
\item At $\alpha$ we have the discrete residue $H_{\mathbb{C}_{1}}^{\alpha}\left(\alpha_{+}\right)-H_{\mathbb{C}_{1}}^{\alpha}\left(\alpha_{-}\right)=\frac{1}{\sqrt{o}}$;
\item $H_{\mathbb{C}_{1}}^{\alpha}(l)\to0$ as $|l|\to\infty$.
\end{itemize}
\item For $\zeta=z^{p}$, the full-plane fermion $$F_{\mathbb{C}_{1}}^{\alpha,\zeta}:=i\sqrt{p}\cdot\mathsf{P}_{\frac{1}{i\sqrt{p}}\mathbb{R}}\left[H_{\mathbb{C}_{1}}^{\alpha}(z)\right]$$
satisfies the antisymmetry properties of Lemma \ref{lem:fermion-fermion-antisymmetries}. 
\end{itemize}
\end{theorem}
\begin{proof}
For a given s-oriented medial vertex $\alpha$, such an $H_{\mathbb{C}_{1}}^{\alpha}$
must be uniquely determined: if any two satisfy the above properties,
their difference will have an s-holomorphic extension to $a$, but
any entire s-holomorphic function which decays to $0$ at infinity
must be zero, since its real and imaginary parts will be discrete
harmonic functions.

To use the same reasoning when $\alpha$ is an s-oriented corner
with adjacent medial vertices $z,z'$, it suffices to show that the s-holomorphic singularity at $\alpha$, i.e. concretely the value of nonzero $\mathsf{P}_{l(\alpha)}\left[H_{\mathbb{C}_{1}}^{\alpha}(z')\right]-\mathsf{P}_{l(\alpha)}\left[H_{\mathbb{C}_{1}}^{\alpha}(z)\right]$,
is fixed by the medial vertex case above. But the antisymmetry relation
of Lemma \ref{lem:fermion-fermion-antisymmetries} gives $H_{\mathbb{C}_{1}}^{\alpha}(z)=\frac{1}{i\sqrt{p_1}}F_{\mathbb{C}_{1}}^{\alpha,\zeta^1}+\frac{1}{i\sqrt{p_2}}F_{\mathbb{C}_{1}}^{\alpha,\zeta^2}$
for $\zeta^{1}=z^{p_{1}}$ and $\zeta^2=z^{p_2}$, where $p_1$ and $p_2$ are s-orientations of the two opposite orientations of $z$. Since $F_{\mathbb{C}_{1}}^{\alpha,\zeta^{1}}=-F_{\mathbb{C}_{1}}^{\zeta^{1},a}$ and $F_{\mathbb{C}_{1}}^{\alpha,\zeta^{2}}=-F_{\mathbb{C}_{1}}^{\zeta^{2},a}$,
both terms are determined by their values on medial vertices, and
similarly for $H_{\mathbb{C}_{1}}^{\alpha}(z')$.

An explicit formula, Eq. (\ref{eq:fermion-fermion-explicit-1}), for
this full-plane observable, and thus its existence, is given by Proposition~\ref{thm:fermion-fermion-full-plane-explicit-formula}. Then the fact
that the given explicit function is the infinite-volume limit is immediate
from Theorem~\ref{thm:fermion-fermion-observable-convergence}.
\end{proof}
\begin{proposition}
\label{thm:fermion-fermion-full-plane-explicit-formula}Let $a,z\in\mathcal{V}_{\mathbb{C}_{1}}^{m}$
and for an s-orientation $(\sqrt o)^2$ on $a$, write $\alpha=a^{o}$ for
the s-oriented medial vertex. The function 
\begin{align}
H_{\mathbb{C}_{1}}^{\alpha}(z)= & \,\,
\frac{e^{\pi i/8}}{\sqrt{o}}\cos\frac{\pi}{8}\left(C_{0}\left(\frac{\sqrt{2}a}{o}+1,\frac{\sqrt{2}z}{o}\right)+C_{0}\left(\frac{\sqrt{2}a}{o}-i,\frac{\sqrt{2}z}{o}\right)\right) \label{eq:fermion-fermion-explicit-1} \\
 & \,\, +\frac{e^{-3\pi i/8}}{\sqrt{o}}\sin\frac{\pi}{8}\left(C_{0}\left(\frac{\sqrt{2}a}{o}-1,\frac{\sqrt{2}z}{o}\right)+C_{0}\left(\frac{\sqrt{2}a}{o}+i,\frac{\sqrt{2}z}{o}\right)\right) \nonumber
\end{align}
for $z\neq a$ satisfies the properties of Theorem \ref{thm:fermion-fermion-infinite-vol-lim},
where the translation invariant function $C_{0}$ is the dimer coupling
function defined in \cite{ken2000}:
\[
C_{0}(z_{1},z_{2})=\frac{1}{4\pi^{2}}\int_{0}^{2\pi}\int_{0}^{2\pi}\frac{\exp(i(xs-yt))}{2i\sin s+2\sin t}dsdt\,,\quad \text{if }z_{2}-z_{1}=x+iy\,.
\]
Moreover, at $\alpha$, the front and back values of the s-holomorphic singularity are given by $
H_{\mathbb{C}_{1}}^{\alpha}(\alpha_\pm)=\frac{\mu \pm 1}{2\sqrt{o}}
$.

\end{proposition}
\begin{proof}
These properties were verified in \cite[Proposition 22]{hon2010}
for a version on the non-rotated lattice, which we will call $H_{\mathbb{Z}^{2}}$.
We note that for any s-oriented medial vertex $\alpha=a^{o}$,
we have 
\[
H_{\mathbb{C}_{1}}^{\alpha}(z)=e^{-\frac{\pi i}{8}}H_{\mathbb{Z}^{2}}^{\alpha'}(z')
\]
if $\alpha'=\left(a'\right)^{o'}$ is the rotated medial vertex $a'=\frac{a}{1+i}\in\mathcal{V}_{\mathbb{Z}^{2}}^{m}$
oriented to $\sqrt{o'}=e^{-\frac{\pi i}{8}}\sqrt{o}$, and $z'=\frac{z}{1+i}$.
Given that the projection lines in the s-holomorphicity relation Eq.
(\ref{eq:s-hol}) are also rotated by $e^{-\frac{\pi i}{8}}$ from
the definition in \cite{hon2010}, the results are easily seen to
carry over. The explicit front and back values $H_{\mathbb C_1}^{\alpha}(\alpha_\pm)$ follow from straightforward computation.
\end{proof}

\subsubsection{Full Plane Spin-fermion Observable}

First we recall some notation. Define the left slit plane $\mathbb{X}=\mathbb{C}\setminus\mathbb{R}_{<0}$
and the right slit plane $\mathbb{Y}=\mathbb{C}\setminus\mathbb{R}_{>0}$.
The double cover $\left[\mathbb{C},0\right]$ contains two lifts $\mathbb{X}^{\pm}$
of $\mathbb{X}$ and two lifts $\mathbb{Y}^{\pm}$ of $\mathbb{Y}$;
define $\sqrt{z}$ on the double cover such that the superscripts
of $\mathbb{X}^{\pm}$ denote the sign of the real part of $\sqrt{z}$
and those of $\mathbb{Y}^{\pm}$ denote the sign of the imaginary
part of $\sqrt{z}$. In other words, $\mathbb{X}^{+}\cap\mathbb{Y}^{+},\mathbb{X}^{-}\cap\mathbb{Y}^{-}$
are lifts of the upper half plane, and $\mathbb{X}^{+}\cap\mathbb{Y}^{-},\mathbb{X}^{-}\cap\mathbb{Y}^{+}$
are lifts of the lower half plane. We use the process outlined in
Remark \ref{rem:s-hol-d-hol} to define an s-holomorphic extension
from a discrete holomorphic function defined on type $1$ and $i$
corners, so let us define the slit discrete domains $\mathbb{X}_{1}^{1}:=\mathcal{V}_{\mathbb{C}_{1}}^{1}\cap\mathbb{X}\cong\mathcal{V}_{\left[\mathbb{C}_{1},0\right]}^{1}\cap\mathbb{X}^{\pm}$
and $\mathbb{Y}_{1}^{i}:=\mathcal{V}_{\mathbb{C}_{1}}^{i}\cap\mathbb{Y}\cong\mathcal{V}_{\left[\mathbb{C}_{1},0\right]}^{i}\cap\mathbb{Y}^{\pm}$. 

\begin{theorem}
\label{thm:spin-fermion-infinite-vol-lim}As $\Omega_{1}\to\mathbb{C}_{1}$,
the complexified spin-fermion observable $H_{\left[\mathbb{C}_{1},0\right]}:=\lim_{\Omega\to\mathbb{C}}H_{[\Omega_{1},0]}$
exists and is uniquely characterized by the following properties: for every $\alpha = a^{o}\in \mathcal V^{cm}_{[\mathbb C_1 ,0]}$, 

\begin{description}
\item [{A}] $H_{\left[\mathbb{C}_{1},0\right]}^{\alpha}$ has monodromy
$-1$ around $0$.
\item [{B}] $H_{\left[\mathbb{C}_{1},0\right]}^{\alpha}$ is s-holomorphic
on $\left[\mathbb{C}_{1},0\right]\setminus\{a,a^{*}\}$, where $a^*$ is the point in $[\mathbb C_1,0]$ distinct from $a$ which shares its projection onto $\mathbb C_1$ with the projection of $a$.
\item [{C}] if $\alpha=a^{o}$ is an s-oriented medial vertex, we have discrete residue $H_{\left[\mathbb{C}_{1},0\right]}^{\alpha}\left(\alpha_{+}\right)-H_{\left[\mathbb{C}_{1},0\right]}^{\alpha}\left(\alpha_{-}\right)=\frac{1}{\sqrt{o}}$.
\item [{D}] If $\alpha=a^{o}$ is an s-oriented real or imaginary corner,
$P_{l(a)}H_{\left[\Omega_{1},0\right]}^{\alpha}(a\pm\frac{i}{2})=\mp \nori \sqrt{o}$.
\item [{E}] For $\zeta=z^{p}$, the full-plane spin-fermion, $$F_{\left[\mathbb{C}_{1},0\right]}^{\alpha,\zeta}:=i\sqrt{p}\cdot\mathsf{P}_{\frac{1}{i\sqrt{p}}\mathbb{R}}\left[H_{\left[\mathbb{C}_{1},0\right]}^{\alpha}(z)\right]$$
satisfies the antisymmetry properties laid out in Lemmas~\ref{lem:spin-fermion-antisymmetries}\textendash \ref{lem:spin-fermion-contour-antisymmetry}. 
\item [{F}] $H_{\left[\mathbb{C}_{1},0\right]}^{\alpha}(l)\to0$ as $|l|\to\infty$.
\end{description}
\end{theorem}
We first present the following three lemmas, then conclude the proof using them.
\begin{lemma}
\label{lem:uniform-boundedness}There exists a uniform constant $M>0$
such that $\left|H_{\left[\Omega_{1}^{n},0\right]}^{\alpha}\left(z\right)\right|\leq M$
for all $n\geq0$ and any s-oriented corner $\alpha$ and any corner
$z$. 
\end{lemma}
\begin{proof}
The strategy will be to progressively extend the validity of the result to more
and more points of the domain. Below we will denote by $a$ and $z$
both corners and medial vertices interchangeably:
\begin{enumerate}
\item When $\alpha:=\alpha_{0}$ is the imaginary corner on the monodromy
face 0 (specifically, the lift of $\frac{1}{2}$ to $\mathbb{X}_{\delta}$)
and $z=a+1$, $H_{\left[\Omega_{1}^{n},0\right]}^{\alpha}\left(z\right)$
has a probabilistic interpretation as a ratio of magnetizations $\mathbb{E}_{\Omega_{1}^{n}}\left[\sigma_{2}\right]/\mathbb{E}_{\Omega_{1}^{n}}\left[\sigma_{0}\right]$,
which is bounded from above by the finite-energy property of the model.
\item When $\alpha=\alpha_{0}$ and $z$ is on the boundary, we claim that
$\sum_{z\in\partial\mathcal{V}_{\left[\Omega_{1}^{n},0\right]}^{m}}\left|H_{\left[\Omega_{1}^{n},0\right]}^{\alpha}\left(z\right)\right|^{2}\leq\mathrm{Cst}\cdot H_{\left[\Omega_{1}^{n},0\right]}^{\alpha}\left(a+1\right)$
(the right hand side of which is bounded by step $1$). This inequality
follows by considering the discrete analogue $Q_{1}:=\mathbb{I}_{\delta}(H_{\left[\Omega_{1}^{n},0\right]}^{\alpha})$
of $\re\left(\int\left(H^{n,\alpha_{0}}\right)^{2}\right)$ analyzed
in Section \ref{subsec:square-integral}. By Proposition \ref{prop:square-existence},
the restriction of $Q_{1}$ to the vertices is super-harmonic (except
perhaps at $a+1$), the sum of the Laplacians is hence bounded from
above by $\mathrm{Cst}\cdot H_{\left[\Omega_{1}^{n},0\right]}^{\alpha}\left(a+1\right)$.
At the same time the sum of the Laplacians equals the sum of the outer
normal derivatives $\partial_{\nu_{z}}Q_{1}$ on the boundary of $\left[\Omega_{1}^{n},0\right]$,
and these normal derivatives $\partial_{\nu_{z}}Q_{1}$ equal $\sqrt{2}\left|H_{\left[\Omega_{1}^{n},0\right]}^{\alpha}\left(z\right)\right|^{2}$.
Hence we deduce the inequality. 
\item When $\alpha=\alpha_{0}$ and $z$ (corner or medial vertex) is in
the interior, we extend the bound of step 2 by the maximum principle.
\item When $\alpha$ is on the boundary and $z$ is the imaginary corner adjacent to the monodromy ($z=\frac 12$),
the bound follows from the antisymmetry of $H$ and step 2.
\item When $\alpha$ and $z$ are on the boundary, we have that $\left|H_{\left[\Omega_{1}^{n},0\right]}^{\alpha}\left(z\right)\right|$
acquires a probabilistic interpretation: the winding factors out from
the sum in the definition of $H$, and we sum over contours that represent
the low-temperature expansion of an Ising model with $+/-$ boundary
conditions switching at $a$ and $z$. As a result, it is easy
to show that $\left|H_{\left[\Omega_{1}^{n},0\right]}^{\alpha}\left(z\right)\right|$
is the ratio $\mathbb{E}_{\Omega_{1}^{n}}^{\pm}\left[\sigma_{0}\right]/\mathbb{E}_{\Omega_{1}^{n}}^{+}\left[\sigma_{0}\right]$,
where $\pm$ and $+$ indicate the boundary conditions. By monotonocity
of the Ising magnetization in boundary conditions, this ratio is less
than one, which gives us the desired bound.
\item When $\alpha$ is on the boundary and $z$ in the interior, the result
follows from steps 4 and 5 and the maximum principle.
\item When $\alpha$ is in the interior and $z$ is next to the monodromy
or on the boundary, the result follows from steps 3 and 6 by antisymmetry.
\item When $\alpha$ and $z$ are in the interior, the result follows from
the maximum principle. \qedhere
\end{enumerate}
\end{proof}
\begin{lemma}
\label{lem:observable-decay}Any bounded function $H:=H_{\left[\mathbb{C}_{1},0\right]}$
that satisfies the properties A-E decays at infinity.
\end{lemma}
\begin{proof}
We exploit the antisymmetry properties E, as specified in Lemmas~\ref{lem:spin-fermion-antisymmetries}\textendash \ref{lem:spin-fermion-contour-antisymmetry}. The idea is
to symmetrize-antisymmetrize $H$ as in Definition \ref{def:sym-antisym-observables}
by writing it as $S+A$, where $S^{\alpha}=\frac{1}{2}\left(H^{\alpha}+H^{\bar{\alpha}}\right)$
and $A^{\alpha}=\frac{1}{2}\left(H^{\alpha}-H^{\bar{\alpha}}\right)$.
Let us now show that $S$ and $A$ both decay at infinity. 

We have that the restriction of $S$ to real corners vanishes of the
positive half-line. We can make a branch cut where it vanishes and
study the function on both slit-plane sheets separated by the cut.
Since it is uniformly bounded and harmonic except near 0 and $a,\bar{a}$,
one can use planar random walk arguments (Beurling estimate) to show
that the function vanishes at infinity. 

Similarly, the restriction to imaginary corners and analogous restrictions
of $A$ vanish on either the positive or the negative half-line. By
the same arguments as above, we can conclude the proof. 
\end{proof}
\begin{lemma}
\label{lem:full-plane-uniqueness}There is at most one function satisfying
the properties A-F. 
\end{lemma}
\begin{proof}
To prove the uniqueness, it suffices that if we have two such functions,
their difference is zero. Denote by $D^{\alpha}\left(z\right)$ this
difference, which will be everywhere s-holomorphic and decay at infinity. However, as noted
in Remark \ref{rem:double-cover-complex-analysis}, the absence of
s-holomorphic singularities does not guarantee harmonicity on the
monodromy face, and some care is needed there (note that below, we
abuse notation to refer to points of $\left[\mathbb{C},0\right]$
by their projections on $\mathbb{C}$).

For $\alpha_{0}=(\frac 12)^o$, $D^{\alpha_0}$ extends s-holomorphically
to $a_{0}$ by zero by property D, and we have that the real part
of $D^{\alpha_0}$ is everywhere harmonic by Remark \ref{rem:double-cover-complex-analysis}.
As a result, by the maximum principle and discrete holomorphicity,
the real part of $D^{\alpha_0}$ vanishes and $D^{\alpha_0}\equiv0$. 

For an arbitrary corner $a$, by antisymmetry and the previous step,
we have $D^{\alpha}\left(a_{0}\right)=0$. As a result the real part
of $D^{\alpha}\left(z\right)$ is harmonic. Thus, as in the previous
step, $D^{\alpha}\left(z\right)$ vanishes everywhere. 
\end{proof}
\begin{proof}[\textbf{\emph{Proof of Theorem \ref{thm:spin-fermion-infinite-vol-lim}}}] By Lemma \ref{lem:uniform-boundedness}, we have that for each s-oriented
corner $\alpha$ of $\left[\mathbb{C}_{1},0\right]$, the sequence
of harmonic functions $H_{\left[\Omega_{1}^{n},0\right]}^{\alpha}$
is uniformly bounded and hence by standard arguments, it admits convergent
subsequences as $n\to\infty$ on any finite graph. 
Any limit along
such subsequences satisfies properties A-E and as a result tends to
0 at infinity by Lemma \ref{lem:observable-decay}. By Lemma \ref{lem:full-plane-uniqueness},
it is uniquely determined. This shows the convergence of the sequence
itself to a limit which satisfies the conditions of the theorem, which we call $H_{[\mathbb C_1,0]}$. 
\end{proof}

\subsubsection{\label{subsec:analytical-expressions}Analytical Expressions }

In this subsection, we give characterizations of $H_{\left[\mathbb{C}_{1},0\right]}$
in a few special cases, then outline an inductive process to construct
it explicitly in general.

\begin{figure}
\begin{tikzpicture}
\clip (-8cm,-5.5cm) rectangle (8cm,3cm);
\begin{scope}[xshift=1.8cm]
\node{
\includegraphics[scale=1.9]{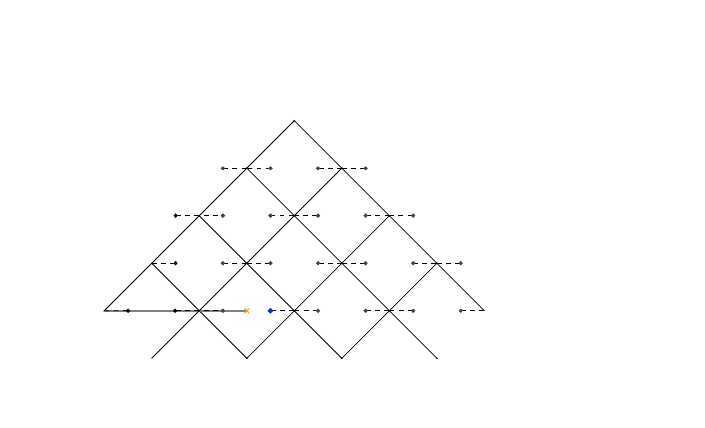}};
\node[font=\small]  at (-2.62,-3.3) {$i$};
\node[font=\small]  at (-2.62,-2.85) {$-i$};

\node[font=\small]  at (-7.22,-2.9) {$0$};
\node[font=\small]  at (-7.22,-3.3) {$0$};

\node[font=\small]  at (-4.14,-2.9) {$0$};
\node[font=\small]  at (-4.14,-3.3) {$0$};

\node[font=\small]  at (-5.75,-3.3) {$i/2$};
\node[font=\small]  at (-5.75,-2.9) {$-i/2$};

\node[font=\small]  at (.48,-3.3) {$0$};

\node[font=\small]  at (3.53,-3.3) {$0$};

\node[font=\small] at (.24,-.22) {$4i-3\sqrt 2 i$};
\node[font=\small] at (-1.2,1.26) {$\frac i{\sqrt 2}-i$};

\node[font=\small] at (-5.65,-1.38) {$- \sqrt 2+\frac 32$};

\node[font=\small] at (-4.05,.15) {$3\sqrt 2-4$};
\node[font=\small] at (-4.4,-1.78) {$\sqrt 2 i-2i$};

\node[font=\small] at (-1.1,-2.9) {$1$};
\node[font=\small] at (2.12,-2.9) {$\frac 12$};

\node[font=\small] at (-2.62,-1.38) {$\sqrt 2-1$};

\node[font=\small] at (.53,-1.38) {$2-\sqrt 2$};

\node[font=\small] at (-2.62,1.66) {$1-\tfrac{1}{\sqrt 2}$};
\node[font=\small] at (-1.05,.15) {$\sqrt 2 -1$};

\node[font=\small] at (-2.75,-.26) {$i-\sqrt 2 i$};
\node[font=\small] at (-1.3,-1.78) {$i-\sqrt 2 i$};

\node[font=\small] at (1.79,-1.82) {$\sqrt 2 i-\frac 32 i $};

\node[font=\small] at (3.53, -1.38)  {$\sqrt 2 -1$};
\node[font=\small] at (1.83, .15) {$-5\sqrt 2+\frac {15}2$};
\node[font=\small] at (.3, 1.66) {$-6+\frac {9}{\sqrt 2}$};
\end{scope}
\end{tikzpicture}
\caption{Some explicit values for the full-plane spin-fermion observable $\norin H_{[\mathbb C_1,0]}^{\alpha_0}(z)$ for $z\in \mathcal V_{[\mathbb C_1,0]}^{1,i}\cap\mathbb{X}^{+}$, where $a_0=\frac 12$. The function has a monodromy about the origin, marked by the orange $\times$, and singularity at $\alpha_0$ marked by the blue $\blacklozenge$.}\label{fig:explicit-values}
\end{figure}
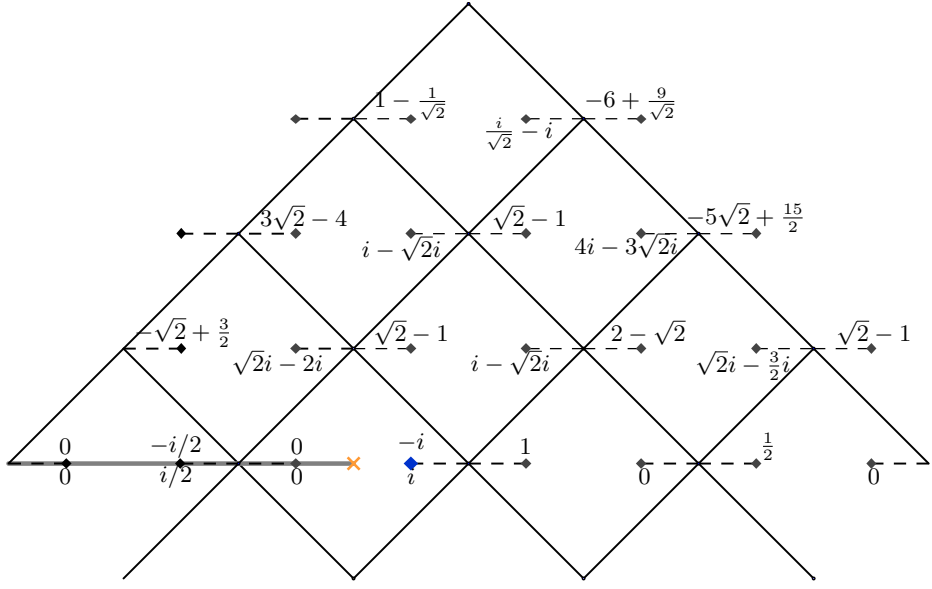

For the observables with monodromy, we have the following characterization
near $0$ from \cite{chelkak-hongler}. Recall $\alpha_{0}=a_{0}^{o}$
where $a_{0}\in\mathbb{X}^{+}$ is the lift of $\frac{1}{2}$ to $\mathbb{X}^{+}$
and $o=(e^{2\pi i})^2$.
\begin{proposition}
\label{prop:spin-fermion-chi-paper-formula}For $z\in\mathcal{V}_{\left[\mathbb{C}_{1},0\right]}^{1}\cup\mathcal{V}_{\left[\mathbb{C}_{1},0\right]}^{i}\setminus\{\frac{1}{2}\}$
we have the characterization 
\[
H_{\left[\mathbb{C}_{1},0\right]}^{\alpha_{0}}(z)=\begin{cases}
\pm\nor\Hm_{3/2}^{\mathbb{X}_{1}^{1}}(z) & \mbox{if }z\in\mathcal{V}_{\left[\mathbb{C}_{1},0\right]}^{1}\cap\mathbb{X}^{\pm}\\
\mp \nori\Hm_{1/2}^{\mathbb{Y}_{1}^{i}}(z) & \mbox{if }z\in\mathcal{V}_{\left[\mathbb{C}_{1},0\right]}^{i}\cap\mathbb{Y}^{\pm}\\
0 & \mbox{otherwise}
\end{cases}\,,
\]
where $\Hm_{a}^{\mathbb{D}_{\delta}}(z)$, for a discrete domain $\mathbb{D}_{\delta}$
and $a\in\mathbb{D}_{\delta}\cup\partial\mathbb{D}_{\delta}$, denotes
the harmonic measure of $a$ as seen from $z$, i.e., the probability
that a simple symmetric random walk on $\mathbb{D}_{\delta}$ started
at $z$ will first hit $a$ when or before exiting $\mathbb{D}_{\delta}$.
\end{proposition}
\begin{proof}
In \cite[Lemma 2.14]{chelkak-hongler} the function defined above
(without the additional normalization factor $
\cos^2{\frac{\pi}{8}}\cdot e^{-2\beta_c}=\nor$) is
proved to be the only function on $\left[\mathbb{C}_{1},0\right]$
which decays at infinity and is s-holomorphic everywhere away from
the singularity at $a_{0}=\frac{1}{2}$ given by $\mathsf{P}_{l(a)}H_{\left[\mathbb{C}_{1},0\right]}^{\alpha}(\frac{1\pm i}{2})=\mp i$.
Thus we can identify it as the unique infinite-volume limit introduced
in Theorem \ref{thm:spin-fermion-infinite-vol-lim} for $\alpha=\alpha_0$.
\end{proof}
\begin{remark}
The zeros in the definition reflect the fact that a slit plane harmonic measure vanishes everywhere on the slit except at the tip, e.g. $\frac{1}{2}$ in case of $\Hm_{1/2}^{\mathbb{Y}_{1}^{i}}(z)$. This function is harmonic on all points of $\mathbb{Y}_{1}^{i}$, but harmonicity fails on the slit (positive real axis), the boundary of the domain.
\end{remark}
The following explicit characterization of the discrete harmonic measure
of the slit plane may be of independent interest. Using this, we provide the values of $\norin H_{[\mathbb C_1,0]}^{\alpha_0}(z)$ near the origin in Figure~\ref{fig:explicit-values}.

\begin{theorem}
\label{thm:slit-plane-harmonic-measure}We have the following expression
for the discrete harmonic measure:
\begin{align*}
H_0(z):=\Hm_{1/2}^{\mathbb{Y}_{1}^{i}}(z=s+ik+\tfrac{1}{2})= & \frac{1}{2\pi}\int_{-\pi}^{\pi}\frac{C^{|k|}(\theta)}{\sqrt{1-e^{-2i\theta}}}e^{-is\theta}d\theta,
\end{align*}
where $C(\theta):=\frac{\cos\theta}{1+\left|\sin\theta\right|}$.
\end{theorem}
\begin{proof}
We defer the proof of this theorem to Appendix \ref{app:full-plane-observables-construct}, Proposition~\ref{prop:harmonic-measure-explicit-appendix}.
\end{proof}
Now we inductively characterize $H^{\alpha}_{[\mathbb C_1,0]}$ in the cases where $\alpha\in \mathcal V^{1,i}_{[\mathbb C_1,0]} \cap \mathbb R_{>0}$.
\begin{proposition}
\label{prop:real-line-full-plane}For an s-oriented corner $\alpha=a^{o}$
with $a\in\mathbb{X}^{+}\cap\mathbb{R}_{\geq0}$, we can recursively
compute $H_{\left[\mathbb{C}_{1},0\right]}^{\alpha}$ starting from
the case $\alpha_{0}=\frac{1}{2}^{o_0},o_0=(e^{2\pi i})^2$, as a finite linear
combination of functions given explicitly in Theorem~\ref{thm:slit-plane-harmonic-measure}. Explicitly, for s-oriented corners $\alpha_0+2n:=(a_0+2n)^{o_0}, \alpha_0+2n+1:=(a_0+2n+1)^{o'}, o'=\left(e^{\frac{\pi i}2}\right)^2$, we have on any of the four half-planes $\mathbb X^\pm \cap \mathbb Y^\pm$, 
\begin{align*}
H_{\left[\mathbb{C}_{1},0\right]}^{\alpha_0+2n}(z)&=
H_{\left[\mathbb{C}_{1},0\right]}^{\alpha_0}(z-2n)-\sum_{m=1}^{n}\frac{H_0(-2m+2)}{2m}H_{\left[\mathbb{C}_{1},0\right]}^{\alpha_0}(z-2(n-m)),\mbox{ and}\\
H_{\left[\mathbb{C}_{1},0\right]}^{\alpha_0+2n+1}(z)&=
iH_{\left[\mathbb{C}_{1},0\right]}^{\alpha_0+2n}(z-1)-i\left(H_0(2n+2)+\frac{H_0(2n)}{2(n+1)} \right)H_{\left[\mathbb{C}_{1},0\right]}^{\alpha_0}(z+1)\,.
\end{align*}

In fact, for any $a\in\mathcal{V}_{[\mathbb{C}_{1},0]}^{1}\cup\mathcal{V}_{[\mathbb{C}_{1},0]}^{i}$
on the real or imaginary axes, we can compute $H_{[\mathbb{C}_{1},0]}^{\alpha}$  using rotational symmetry: if $\alpha=a^{o}$
is any s-oriented corner on the real or imaginary line, and $\alpha'=\left(a'\right)^{o'}$
is the rotated corner $a'=e^{-\frac{\pi i}{2}}a\in\mathcal{V}_{\left[\mathbb{C}_{1},0\right]}^{c}$ oriented
to $o'=(e^{-\frac{\pi i}{4}}\sqrt{o})^2$, and $z'=e^{-\frac{\pi i}{2}}z$, we
have,
\begin{alignat*}{1}
H_{\left[\mathbb{C}_{1},0\right]}^{\alpha}(z) & =e^{-\frac{\pi i}{4}}H_{\left[\mathbb{C}_{1},0\right]}^{\alpha'}(z')\,.
\end{alignat*}
 
\end{proposition}
\begin{proof}
Generalizing from the proof of Proposition \ref{prop:spin-fermion-chi-paper-formula},
we construct a function on $\left[\mathbb{C}_{1},0\right]$ which
decays at infinity and is s-holomorphic everywhere away from the specified
singularity at $a$, and we argue that it is the unique function which
can satisfy the properties specified in Theorem \ref{thm:spin-fermion-infinite-vol-lim}. For convenience, we will take source points $\alpha$ on the sheet $\mathbb X^+$ unless otherwise specified; i.e. the positive real line approached from above on $\mathbb Y^+$.

Assume $a\in\mathcal{V}_{\left[\mathbb{C}_{1},0\right]}^{i}\cap\mathbb{R}_{>0}$
and consider the restriction $H^{\alpha}$ of $H_{\left[\mathbb{C}_{1},0\right]}^{\alpha}$
to the imaginary corners on the slit plane $\mathcal{V}_{\left[\mathbb{C}_{1},0\right]}^{i}\cap\mathbb{Y}^{+}$.
The function $H^{\alpha}$ can be characterized as the unique discrete
harmonic function on the slit plane $\mathcal{V}_{\left[\mathbb{C}_{1},0\right]}^{i}\cap\mathbb{Y}^{+}$
which has the single-valued boundary data $H^{\alpha}(a):=-\nori\sqrt{o}$ and zero elsewhere
on the slit $\mathbb{R}_{>0}$, and decays at infinity. The harmonic
function with these properties can be obtained by translating $H^{\alpha-2}$ for $\alpha-2 := (a-2)^o$ (which takes
its only nonzero boundary value at $a-2$) to the right, and subtracting off a multiple of $H^{\alpha_{0}}$ in order to cancel the nonzero value at $a_{0}=\frac{1}{2}$. Specifically, letting
\[
H^{\alpha}(z)=H^{\alpha-2}(z-2)-H^{\alpha-2}(-\tfrac{3}{2})\cdot\Hm^{\mathbb{Y}^i_1}_{1/2}(z)=H^{\alpha-2}(z-2)-\Hm^{\mathbb{Y}^i_1}_{a-2}(-\tfrac{3}{2})\cdot H^{\alpha_{0}}(z)\,,
\]
then discrete holomorphicity relations imply that the real part is
determined up to a an additive constant; however since the real part must vanish
on $\mathbb{R}_{<0}$, the real part is uniquely determined. In fact,
it is easy to see that, in order to maintain discrete holomorphicity,
the above recursive relation should also hold for the entire $H_{\left[\mathbb{C}_{1},0\right]}^{\alpha}$ as long as we are in the four real-translation invariant half-planes $\mathbb X^\pm \cap \mathbb Y^\pm$. For the explicit identification of the coefficients, we refer to Proposition \ref{prop:recursive-coefficient-proof}.

For $a\in\mathcal{V}_{\left[\mathbb{C}_{1},0\right]}^{1}\cap\mathbb{R}_{>0}$,
we use a similar recursive process but now instead first construct the restriction $H'^\alpha$ of $H_{\left[\mathbb{C}_{1},0\right]}^{\alpha}$ to the real corners on $\mathbb{Y}^+$, starting from the case $a=\frac{3}{2}$. Unlike the imaginary case, we need to consider $-\frac{1}{2}$ as well as $\mathcal{V}_{\left[\mathbb{C}_{1},0\right]}^{1}\cap\mathbb{R}_{>0}$ as part of the slit boundary (where harmonicity fails): since $H_{[\mathbb{C}_{1},0]}^{\alpha}(\frac{1}{2})\neq0$ in general, we cannot assume that the real part is harmonic at $-\frac{1}{2}$ (see Remark \ref{rem:double-cover-complex-analysis}).

In other words, $H'^\alpha$ is the function harmonic on the slit plane $\mathcal{V}_{\left[\mathbb{C}_{1},0\right]}^{1}\cap\mathbb{Y}^+\setminus\{-\frac{1}{2}\}$ which takes nonzero boundary values only at $a$ and $-\frac{1}{2}=-a_0$. As above $H'^\alpha(a)=-\nori\sqrt{o}$. 

For the value at $-a_0:=e^{\pi i}a_0$ with oriented version $-\alpha_0:=(-a_0)^{o'}$, where $o'=(e^{\frac{\pi i}2})^2$ we use antisymmetry in the two inputs to write $H_{\left[\mathbb{C}_{1},0\right]}^{\alpha}(-a_0)=-i\sqrt{o}H_{\left[\mathbb{C}_{1},0\right]}^{-\alpha_0}(a)$. Since by rotation $H_{\left[\mathbb{C}_{1},0\right]}^{-\alpha_0}(a)=-iH_{\left[\mathbb{C}_{1},0\right]}^{\alpha_0}(e^{-\pi i}a)$, we conclude $H_{\left[\mathbb{C}_{1},0\right]}^{\alpha}(-a_0)=-\sqrt{o}H_{\left[\mathbb{C}_{1},0\right]}^{\alpha_0}(e^{-\pi i}a)=-\nori\sqrt{o}\cdot\Hm_{1/2}^{\mathbb{Y}_{1}^{i}}(-a)$. We can match these boundary values with recursion as above.

The general rotation identity can be verified independently with the same
strategy, i.e., identifying the restriction of the left-hand side
to a specific type of corner as the unique harmonic function with
suitable boundary values, which the right-hand side solves.
\end{proof}
\begin{figure}
\begin{center}
\begin{tikzpicture}
\clip (-5cm,-4cm) rectangle (5cm,4cm);
\begin{scope}[xshift=1cm]
\node{
\includegraphics[scale=1.9]{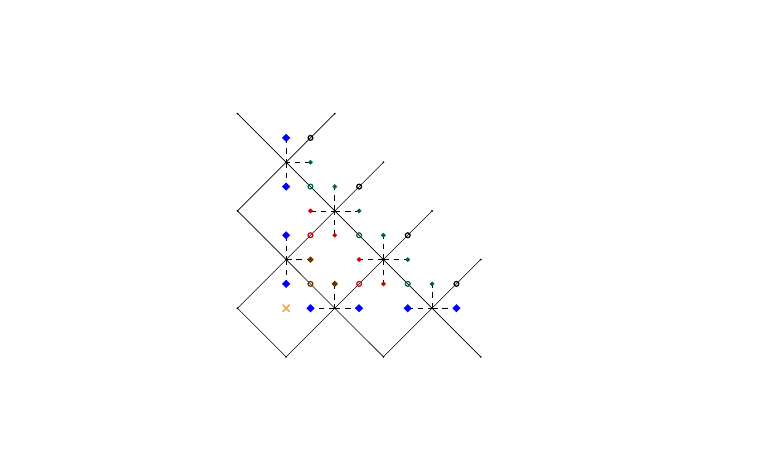}};
\end{scope}
\end{tikzpicture}
\end{center}

\caption{The full plane spinor $H_{[\mathbb{C}_{1},0]}^{\alpha}(z)$ is first defined
at all blue corners $\blacklozenge$ by its antisymmetry relations and Proposition~\ref{prop:real-line-full-plane}. By s-holomorphicity, we can then deduce its value at the brown medial vertex $\circ$, and then project from there onto the two brown corners $\blacklozenge$. We continue this process, next moving to the red $\circ$, then the red $\blacklozenge$, and then green etc.}\label{fig:s-holomorphic-propagation}
\end{figure}

\begin{remark}
\label{rem:recursive-coefficients-explicit}The coefficients
in Proposition \ref{prop:real-line-full-plane} of various translated and rotated versions of $H_{[\mathbb{C}_{1},0]}^{\alpha_0}$ (with the same scaling limit) become important when identifying the scaling limit of the observables in Lemma \ref{lem:chi-base-convergence}. In particular, by Proposition \ref{prop:recursive-coefficient-proof}, the coefficients in the recursive expansion sum to zero in the case of $a\in\mathcal{V}_{\left[\mathbb{C}_{1},0\right]}^{1}\cap\mathbb{R}_{>0}$, which will yield that $\tilde{C}_\alpha=0=C_\alpha$ in Corollary \ref{cor:displacement-scaling-explicit}.
\end{remark}
\begin{corollary}
\label{cor:full-plane-observable-construction}For any $a\in\mathcal{V}_{\left[\mathbb{C}_{1},0\right]}^{cm}$,
we can recursively compute $H_{\left[\mathbb{C}_{1},0\right]}^{\alpha}$
as a finite linear combination of functions given by Theorem~\ref{thm:slit-plane-harmonic-measure}.
\end{corollary}
\begin{proof}
By Proposition~\ref{prop:real-line-full-plane}, $H_{[\mathbb C_1,0]}^\alpha(z)$ is given for all $z\in \mathcal V^{cm}_{[\mathbb C_1,0]}$ whenever $\alpha \in \mathcal V^{1,i}_{[\mathbb C_1,0]} \cap (\mathbb R_{>0}\cup i \mathbb R_{>0})$. The antisymmetry relations it satisfies thus give $H_{[\mathbb C_1,0]}^{\alpha}(z)$ for every $\alpha$, whenever $z\in \mathbb R_{>0}\cup i\mathbb R_{>0}$ (as a finite linear combination of explicit harmonic measures). 

Now observe that s-holomorphicity implies that the value, say, of $H^{\alpha}_{[\mathbb C_1,0]}(\frac {1+i}{2})$ can be recovered from
its values at the $\frac{1}{2},\frac{i}{2}$. From there, one can project the values of $H^{\alpha}_{[\mathbb C_1,0]}(z)$ when $z= 1+\frac i2$ and $\frac 12+i$ (see Figure~\ref{fig:s-holomorphic-propagation}). Continuing this process allows for a recursive construction of
$H_{\left[\mathbb{C}_{1},0\right]}^{\alpha}(z)$ for any $z\in \mathcal V^{cm}_{[\mathbb C_1,0]}$ as a finite linear combination of the explicit functions of Theorem~\ref{thm:slit-plane-harmonic-measure}. 
\end{proof}

\subsubsection{Auxiliary Functions}
\label{subsec:auxiliary-functions}

We introduce here full-plane auxiliary functions $G$ and $\tilde{G}^{\pm}$,
which are everywhere s-holomorphic functions on $[\mathbb{C}_{1},0]$
which do not decay at infinity. The real part of $G$ was defined
in \cite{chelkak-hongler} as a discrete version of the holomorphic
function $\sqrt{z}$ on the double cover; we extend the result in
order to give full s-holomorphic discrete representations of $\nor\sqrt{z}$
and $\nori\sqrt{z}$. Convergence results for these functions will be
proved in Section \ref{subsec:observable-convergence}.

As in previous subsections, we define the functions on the unit grid
$\left[\mathbb{C}_{1},0\right]$, and then scale them by $G_{\left[\mathbb{C}_{\delta},0\right]}(z\delta):=\delta G_{\left[\mathbb{C}_{1},0\right]}(z)$
and $\tilde{G}_{\left[\mathbb{C}_{\delta},0\right]}^{\pm}(z\delta):=\delta\tilde{G}_{\left[\mathbb{C}_{1},0\right]}^{\pm}(z)$.
We define them first on real and imaginary corners by ``integrating''
the harmonic measures, then extend to other points by s-holomorphicity.
The fact that there are two discrete versions of $\nori\sqrt{z}$ is a
peculiarity that will be important in the proof of the main convergence
result in Section \ref{subsec:bounded-domain-convergence} (see
also \cite{dub2015}).
\begin{definition}
\label{def:aux-functions}Define for $z\in\mathcal{V}_{\left[\mathbb{C}_{1},0\right]}^{1}\cup\mathcal{V}_{\left[\mathbb{C}_{1},0\right]}^{i}$,
the auxiliary functions
\[
G_{\left[\mathbb{C}_{1},0\right]}(z):=\begin{cases}
\sum_{n=0}^{\infty}\pm\nor\Hm_{3/2}^{\mathbb{X}_{1}^{1}}(z-2n) & \mbox{if }z\in\mathcal{V}_{\left[\mathbb{C}_{1},0\right]}^{1}\cap\mathbb{X}^{\pm}\\
\sum_{n=0}^{\infty}\pm\nori\Hm_{-3/2}^{\mathbb{Y}_{1}^{i}}(z+2n) & \mbox{if }z\in\mathcal{V}_{\left[\mathbb{C}_{1},0\right]}^{i}\cap\mathbb{Y}^{\pm}\\
0 & \mbox{otherwise}
\end{cases}\,,
\]
and
\[
\tilde{G}_{\left[\mathbb{C}_{1},0\right]}^{\pm}(z):=iG_{\left[\mathbb{C}_{1},0\right]}(z\pm1)\,,
\]
where translation by $1$ is well-defined at any point other than
$\pm\frac{1}{2}$; $G_{\left[\mathbb{C}_{1},0\right]}(\pm\frac{1}{2})=0$
on both sheets so there is no ambiguity in defining $\tilde{G}_{\left[\mathbb{C}_{1},0\right]}^{\pm}(\mp\frac{1}{2})$.
\end{definition}
\begin{remark}
In \cite[Lemma 2.17]{chelkak-hongler} well-definedness and harmonicity
of the real part of $G_{\left[\mathbb{C}_{1},0\right]}$ were proven.
From symmetry, we see the same holds for the imaginary part. Discrete
holomorphicity of $G_{\left[\mathbb{C}_{1},0\right]}$, and thus of
$\tilde{G}_{\left[\mathbb{C}_{1},0\right]}^{\pm}$, is proved in Appendix
\ref{app:full-plane-observables-construct} using the explicit formula
of Theorem~\ref{thm:slit-plane-harmonic-measure}. Using that, we then extend these
to s-holomorphic functions on the corners and medial vertices, again
using the process of Remark~\ref{rem:s-hol-d-hol}.
\end{remark}

\begin{figure}
\hspace{.5in}
\begin{tikzpicture}
\clip (-8cm,-5.5cm) rectangle (8cm,3cm);
\begin{scope}[xshift=1.8cm]
\node{
\includegraphics[scale=1.9]{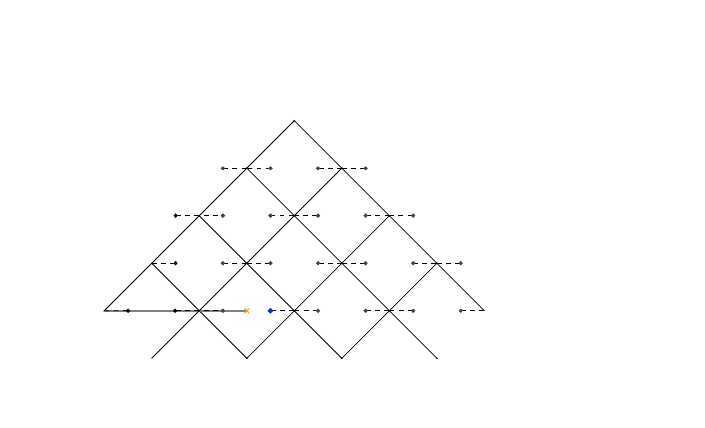}};
\node[font=\small]  at (-2.62,-3.3) {$0$};

\node[font=\small]  at (-4.14,-2.9) {$0$};
\node[font=\small]  at (-4.14,-3.3) {$0$};

\node[font=\small]  at (-7.22,-2.9) {$0$};
\node[font=\small]  at (-7.22,-3.3) {$0$};

\node[font=\small]  at (-5.75,-3.3) {$-i$};
\node[font=\small]  at (-5.75,-2.9) {$i$};

\node[font=\small]  at (.48,-3.3) {$0$};

\node[font=\small]  at (3.53,-3.3) {$0$};

\node[font=\small] at (.24,-.25) {$\frac{i}{2}(8-5\sqrt 2)$};
\node[font=\small] at (-1.4,1.26) {$\frac{i}{4}(5\sqrt 2 -4)$};

\node[font=\small] at (-5.65,-1.38) {$1-\frac {\sqrt 2}{2}$};

\node[font=\small] at (-3.85,.15) {$\frac{\sqrt 2}{2}$};
\node[font=\small] at (-4.5,-1.78) {$\frac{\sqrt 2}{2}i$};

\node[font=\small] at (-1.1,-2.9) {$1$};
\node[font=\small] at (2.12,-2.9) {$\frac 32$};

\node[font=\small] at (-2.42,-1.38) {$\frac{\sqrt 2}{2}$};

\node[font=\small] at (.53,-1.38) {$2-\frac{\sqrt 2}{2}$};

\node[font=\small] at (-2.42,1.66) {$\frac{3\sqrt 2}{4}$};
\node[font=\small] at (-1.05,.15) {$-1+\frac{3\sqrt 2}{2}$};

\node[font=\small] at (-2.9,-.26) {$\frac{\sqrt 2}{2}i$};
\node[font=\small] at (-1.3,-1.78) {$\frac{i}{2}(2-\sqrt 2)$};

\node[font=\small] at (1.79,-1.82) {$\frac{i}{2}(\sqrt 2 -1)$};

\node[font=\small] at (3.53, -1.38)  {$1+\frac{\sqrt 2}{2}$};
\node[font=\small] at (1.93, .15) {$\frac{13}{2}-\frac{7\sqrt 2}{2}$};
\node[font=\small] at (.35, 1.66) {$-6+\frac {21\sqrt 2}{4}$};

\end{scope}
\end{tikzpicture}

\caption{Some explicit values for the auxiliary function $\norin G_{[\mathbb C_1,0]}(z)$ for $z\in \mathcal V_{[\mathbb C_1,0]}^{1,i}\cap \mathbb X^{+}$, as defined in Definition~\ref{def:aux-functions}. The function has monodromy about the origin, represented by the orange $\times$.}\label{fig:explicit-values-G}
\end{figure}

\begin{remark}\label{rem:g-explicit}
Since $G_{\left[\mathbb{C}_{1},0\right]}$ (and thus $\tilde G_{\left[\mathbb{C}_{1},0\right]}^\pm$) is defined using infinite sums, one cannot a priori calculate them exactly. However, once its s-holomorphicity is exhibited in Appendix~\ref{app:full-plane-observables-construct}, we can use a propagation procedure similar to one shown in Fig. \ref{fig:s-holomorphic-propagation} and explained in Corollary~\ref{cor:full-plane-observable-construction} to recursively calculate its values from its values on the real and imaginary axes. Indeed, $G_{\left[\mathbb{C}_{1},0\right]}$ is explicitly computable on the real line since the summands eventually become zero; then we use the rotation identity $e^{\pi i/4} G_{\left[\mathbb{C}_{1},0\right]}(e^{\pi i/2}z) = \frac12\left[\tilde G_{\left[\mathbb{C}_{1},0\right]}^++\tilde G_{\left[\mathbb{C}_{1},0\right]}^-\right](z)$, proved in Proposition \ref{prop:G-rotation}, to find the values on the imaginary axis.

Using this procedure, we provide some explicit values of $G_{[\mathbb C_1,0]}(z)$ near the origin in Figure~\ref{fig:explicit-values-G}
\end{remark}
\section{\label{sec:multipoint-observables}Discrete Multipoint Observables}

In this section, we prove Pfaffian formulae expressing $n$-point
energy correlations, with or without a spin weight, in terms of the real
fermion and spin-fermion two-point functions.

For the $n$-point energy correlations, we formulate them in terms of s-holomorphic multipoint fermion observables introduced
in \cite{hon2010} and follow the strategy there to obtain their Pfaffian formulation with the two-point observables introduced
in Section \ref{sec:bounded-domain-observables}. In~\cite{hon2010}, the arguments of the multipoint observable were required to be distinct, non-adjacent medial vertices; we slightly but crucially generalize this to allow for adjacent medial vertices in Proposition~\ref{prop:observable-s-hol} and Lemma~\ref{lem:projection-xor}, using a combinatorial correspondence between paths sourced at medial vertices and at corners.

When looking at $n$-point energy correlations weighted by a spin,
the process is analogous, and we get fused multipoint spin-fermion
observables, which then reduce to Pfaffians of the two-point spin-fermion
observables of Section~\ref{subsec:two-point-observable}. 

The proofs of these relations connecting the two-point fermion and spin-fermion observables to edge correlations (namely, Propositions~\ref{prop:energy-correlations-pfaffian-observable}--\ref{prop:spin-energy-fermion-correlation-pfaffian-observable}) are quite notationally heavy, but all the extra notation is contained completely to this section. After generalizing the discrete complex analytic properties of the multipoint observables to adjacent medial vertices in Proposition~\ref{prop:observable-s-hol}, the rest of the proof is just a natural extension of the steps outlined in~\cite{hon2010} to prove the desired relations. For an alternative approach using mostly combinatorial arguments, see~\cite{chelkak-kac-ward}.

\subsection{\label{subsec:multi-fermion-observables}Multipoint Observables}

Recall that in Section \ref{sec:bounded-domain-observables} we denoted
by $\mathcal{Z}_{\Omega_{\delta}}$ the low-temperature expansion
of the partition function defined by summing $e^{-2\beta\left|\omega\right|}$
over all $\omega\in\mathcal{C}_{\Omega_{\delta}}$, where $\mathcal{C}_{\Omega_{\delta}}$
is the set of closed loops in $\Omega_{\delta}$. Furthermore, for oriented corners or medial vertices $\alpha,\zeta$, we
defined the contour sets $\mathcal{C}_{\Omega_{\delta}}^{\alpha,\zeta}$
as well as admissible walks.  We also chose the convention that and at ambiguous vertices in the walk, we connect northeast to northwest edges and southeast and southwest edges (see~\S\ref{subsec:two-point-observable}).

We now generalize the two-point observables defined in Section~\ref{sec:bounded-domain-observables}. First, let us define
the generalized contour set $\mathcal{C}_{\Omega_{\delta}}^{\alpha_{1},\ldots,\alpha_{2n}}$
for s-oriented corners or medial vertices $\alpha_{j}=a_{j}^{o_{j}}$ for $j=1,2,\ldots,2n$. For now we
assume that the underlying points $a_{1},\ldots,a_{2n}$ are distinct (we will later generalize to the case where they can take the same value:
see Remark \ref{rem:possible-extension-to-singularity}). In~\cite{hon2010}, these points were all medial vertices, and were required to be non-adjacent (the edges corresponding to them were not allowed to be adjacent).  We do not impose this non-adjacency requirement, and this small extension is important to the proofs. As before, the half-edge of a corner is the line segment connecting it to its nearest vertex.

Each element $\gamma\in\mathcal{C}_{\Omega_{\delta}}^{\alpha_{1},\ldots,\alpha_{2n}}$
is a set containing
\begin{itemize}
\item $2n$ half-edges of $a_{1},\ldots,a_{2n}$ selected by their respective
orientations $o_{1},\ldots,o_{2n}$, and
\item a (possibly empty) collection of edges distinct from the above-mentioned half-edges of $\left\{ a_{j}\right\} _{j=1}^{2n}$,
\end{itemize}
such that any vertex of $\Omega_{\delta}$ is incident to an even
number of the edges and half-edges. Such a $\gamma$ will be an edge-disjoint
union of $n$ walks connecting $\alpha_{1},\ldots\alpha_{2n}$ pairwise
as well as a (possibly empty) collection of edge-disjoint loops. In particular, the definition of the $2$-point
set $\mathcal{C}_{\Omega_{\delta}}^{\alpha,\zeta}$ coincides with
the one given in Section \ref{sec:bounded-domain-observables}.

Generalizing from the $2$-point case in Section \ref{subsec:disorder-lines}, given $\gamma$ we can pick
$n$ admissible walks which connect $\left\{ \alpha_{i}\right\} $
pairwise. We denote by $\Gamma(\gamma)\subset\gamma$, a set of those
$n$ edge-disjoint admissible walks $\{\gamma^{\alpha_{j},\alpha_{k}}\}$ chosen from
the half-edges and edges constituting $\gamma$. We label them so
that $j<k$ for each $\gamma^{\alpha_{j},\alpha_{k}}$. Define the
\emph{crossing parity} $\mathbf{c}(\Gamma(\gamma))$ as the number
of crossings, modulo $2$ when linking $1,\ldots,2n\in\mathbb{R}$
pairwise with generic simple curves in the upper half plane (i.e.
connect $j,k$ if there is a walk in $\Gamma(\gamma)$ connecting
$\alpha_{j},\alpha_{k}$).
Moreover, recall the definitions of $\lambda_{\alpha_i,\alpha_j}$ and $\mathbf{W}(\gamma)$ from Definition~\ref{def:observable-quantities}.

\begin{figure}
\begin{center}
\begin{tikzpicture} 
\clip (-5cm,-4cm) rectangle (5cm,4cm);
\node{
\includegraphics[scale=1.8]{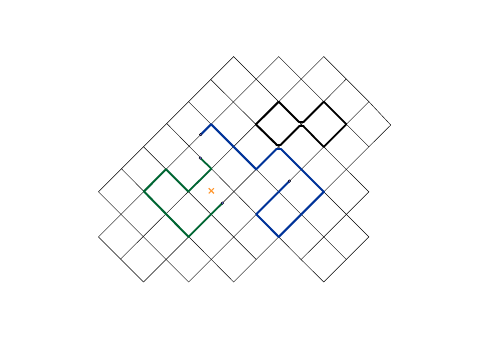}};
\node at (1.45,-.4) {$\alpha_1$};
\node at (-1.2,1.1) {$\alpha_3$};
\node at (-1.6,.4) {$\alpha_4$};
\node at (-.6,-1.1) {$\alpha_2$};
\end{tikzpicture}
\end{center}
\caption{An example of a choice of two admissible walks between $\alpha_{1}$ and $\alpha_{3}$ and $\alpha_{2}$
and $\alpha_{4}$ and a loop. The winding of $\gamma^{\alpha_1,\alpha_3}$ is $2\pi$
while the winding of $\gamma^{\alpha_2,\alpha_4}$ is $-\frac{\pi}{2}$. The crossing parity has $\mathbf{c}(\Gamma(\gamma))= 1$.}
\end{figure}
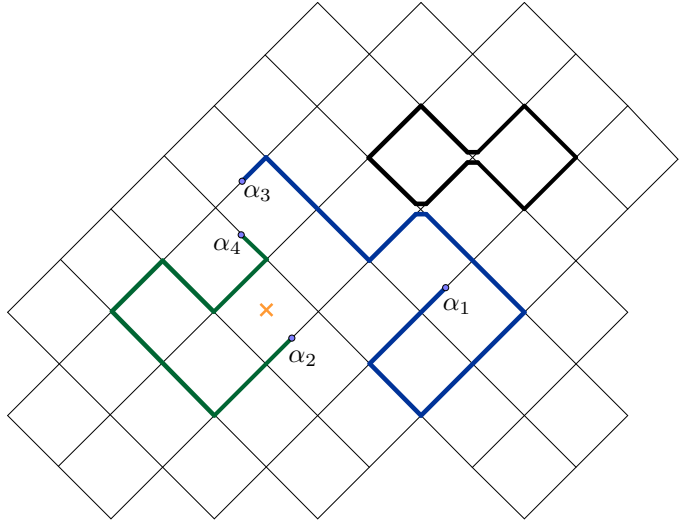

\begin{definition}
\label{def:multipoint-observable-1}For a collection of s-oriented corners or medial vertices $\{\alpha_j\}_{j=1}^{2n}$ in $\Omega_\delta$, define the \emph{multipoint
fermion observable} as
\begin{alignat*}{1}
F_{\Omega_{\delta}}^{\alpha_{1},\ldots,\alpha_{2n}} & :=\frac{1}{\mathcal{Z}_{\Omega_{\delta}}}\sum_{\gamma\in\mathcal{C}_{\Omega_{\delta}}^{\alpha_{1},\ldots,\alpha_{2n}}}\phi_{\left\{ \alpha_{j}\right\} }(\gamma)\,,\\
\phi_{\left\{ \alpha_{j}\right\} }(\gamma) & :=e^{-2\beta_{c}\left|\gamma\right|}(-1)^{\mathbf{c}(\Gamma(\gamma))}\prod_{\gamma^{\alpha_{j},\alpha_{k}}\in\Gamma\left(\gamma\right)}i\lambda_{\alpha_{i},\alpha_{j}}e^{-\frac{i}{2}\mathbf{W}\left(\gamma^{\alpha_{j},\alpha_{k}}\right)}\,.
\end{alignat*}
In analogy with the two point case, if $\zeta^1=z^{p_{1}}$
and $\zeta^{2}=z^{p_{2}}$ are s-orientations of the two opposite orientations of $z:=a_{2n}$, so that $p_2 = e^{\pm \pi i}p_{1}$,
and $\kappa=c^{q}$ is an s-orientation of a corner $c$ in $\Omega_\delta$, we define
the complexification of the multipoint fermion observable,
\begin{alignat}{1}
H_{\Omega_{\delta}}^{\alpha_{1},\ldots,\alpha_{2n-1}}\left(z\right) & :=\frac{1}{i\sqrt{p_{1}}}F_{\Omega_{\delta}}^{\alpha_{1},\ldots,\alpha_{2n-1},\zeta^{1}}+\frac{1}{i\sqrt{p_{2}}}F_{\Omega_{\delta}}^{\alpha_{1},\ldots,\alpha_{2n-1},\zeta^{2}}\,,\label{eq:complexification}\\
H_{\Omega_{\delta}}^{\alpha_{1},\ldots,\alpha_{2n-1}}\left(c\right) & :=\frac{1}{i\sqrt{q}}F_{\Omega_{\delta}}^{\alpha_{1},\ldots,\alpha_{2n-1},\kappa}\,.\nonumber 
\end{alignat}
\end{definition}

\begin{definition}
\label{def:multipoint-observable-2}For a collection of s-oriented corners or medial vertices $\{\alpha_j\}_{j=1}^{2n}$ in $[\Omega_\delta,0]$, define the  \emph{multipoint
spin-fermion observable,} 
\begin{alignat*}{1}
F_{\left[\Omega_{\delta},0\right]}^{\alpha_{1},\ldots,\alpha_{2n}} & :=\frac{1}{\mathcal{Z}_{\Omega_{\delta}}\mathbb{E}_{\Omega_{\delta}}^{+}\left[\sigma_{0}\right]}\sum_{\gamma\in\mathcal{C}_{\Omega_{\delta}}^{\alpha_{1},\ldots,\alpha_{2n}}}\mathrm{\phi}_{\left\{ \alpha_{j}\right\} }^{\Sigma}\left(\gamma\right)\,,\\
\mathrm{\phi}_{\left\{ \alpha_{j}\right\} }^{\Sigma}\left(\gamma\right) & :=\phi_{\left\{ \alpha_{j}\right\} }\left(\gamma\right)\left(-1\right)^{\ell\left(\gamma\setminus\cup\Gamma(\gamma)\right)}\prod_{\gamma^{\alpha_{j},\alpha_{k}}\in\Gamma\left(\gamma\right)}\mathrm{s_{\alpha_j,\alpha_k}}\left(\gamma^{\alpha_{j},\alpha_{k}}\right)\,,
\end{alignat*}
where $\ell$ and $\mathrm{s_{\alpha_j,\alpha_k}}$ are defined as in Definition
\ref{def:observable-quantities}. As in the two-point spin fermion, the contour collection $\mathcal C_{\Omega_\delta}^{\alpha_1,...,\alpha_{2n}}$ and the weight $\phi_{\{\alpha_j\}}$ are computed with respect to the projections of $\alpha_j$ onto $\Omega_{\delta}$, but the sheet choices come in the terms $\mathrm{s_{\alpha,\zeta}}$.
Define the complexified spin-fermion $H_{[\Omega_{\delta},0]}^{\alpha_{1},\ldots,\alpha_{2n-1}}$ on medial vertices and corners of $[\Omega_\delta,0]$ 
analogously to Eq.~(\ref{eq:complexification}).
\end{definition}
\begin{remark}
\label{rem:weight-welldefined}\cite[Propositions 67, 68]{hon2010}
proves well-definedness of $\phi_{\left\{ \alpha_{j}\right\} }\left(\gamma\right)$.
In Proposition \ref{prop:weight-welldefined} we prove the well-definedness
of $\left(-1\right)^{\ell\left(\gamma\setminus\cup\Gamma(\gamma)\right)}\prod\mathrm{s_{\alpha_j,\alpha_k}}(\gamma^{\alpha_j,\alpha_k})$,
and thus of $\mathrm{\phi}_{\left\{ \alpha_{j}\right\} }^{\Sigma}$, so that it is independent of our convention for admissible walks and the choice of pairings of source points; in particular, it is independent of our choice of admissible $\Gamma(\gamma)$.
\end{remark}
In addition to increasing the number of inputs, we can also define
observables summing over a subset $\mathcal{C}_{\Omega_{\delta}:\{e_{k}^{s_{k}}\}}^{\alpha_{1},\ldots,\alpha_{2n}}\subset\mathcal{C}_{\Omega_{\delta}}^{\alpha_{1},\ldots,\alpha_{2n}}$
formed by specifying the inclusion or exclusion of given edges. Given
a collection of edges $\{e_{k}\}_{k=1}^{m}$ in $\Omega_\delta$ (disjoint to the half-edges given by $\{a_{j}\}_{i=1}^{2n}$) and corresponding \emph{inclusion variables} $s_{k}\in\{\bullet,\circ\}$, let
\begin{align*}
\mathcal{C}_{\Omega_{\delta}:\{e_{1}^{s_{1}},\ldots,e_{m}^{s_{m}}\}}^{\alpha_{1},\ldots,\alpha_{2n}}= \big\{\gamma \in \mathcal C^{\alpha_1,...,\alpha_{2n}}_{\Omega_\delta} \,:\, e_k \in \gamma \mbox{ if } s_k= \bullet, \mbox{ and } e_k \notin \gamma \mbox{ if }s_k = \circ\big\}\,.
\end{align*}

\begin{definition}
We define the \emph{restricted fermion and spin-fermion observables}, denoted
$F_{\Omega_{\delta}:\{e_{1}^{s_{1}},\ldots,e_{m}^{s_{m}}\}}^{\alpha_{1},\ldots,\alpha_{2n}}$ and  $F_{\left[\Omega_{\delta},0\right]:\{e_{1}^{s_{1}},\ldots,e_{m}^{s_{m}}\}}^{\alpha_{1},\ldots,\alpha_{2n}}$,
and their complexifications as in Definitions \ref{def:multipoint-observable-1}\textendash \ref{def:multipoint-observable-2},
replacing the contour set $\mathcal{C}_{\Omega_{\delta}}^{\alpha_{1},\ldots,\alpha_{2n}}$
by the restricted contour set $\mathcal{C}_{\Omega_{\delta}:\{e_{1}^{s_{1}},\ldots,e_{m}^{s_{m}}\}}^{\alpha_{1},\ldots,\alpha_{2n}}$.
\end{definition}

The following propositions will characterize the complexified fermion and spin-fermion observables in terms
of discrete complex analysis, proving the connection to the discrete Riemann
boundary value problem defined in Section \ref{sec:discrete-complex-analysis} for possibly adjacent medial vertices $a_1,...,a_{2n}$.

We first modify the real weight
$\phi_{\left\{ \alpha_{k}\right\} }$ defined on $\mathcal{C}_{\Omega_{\delta}}^{\alpha_{1},\ldots,\alpha_{2n}}$
for $2n$ s-oriented vertices $\alpha_{1},\ldots,\alpha_{2n}$
into a \emph{complex weight} $\chi$ defined on 
\begin{align*}
\mathcal{C}_{\Omega_{\delta}}^{\alpha_{1},\ldots,\alpha_{2n-1},a_{2n}}:=\mathcal{C}_{\Omega_{\delta}}^{\alpha_{1},\ldots,\alpha_{2n-1},\alpha_{2n}^{1}}\sqcup\mathcal{C}_{\Omega_{\delta}}^{\alpha_{1},\ldots,\alpha_{2n-1},\alpha_{2n}^{2}}
\end{align*}
for $2n-1$ s-oriented medial vertices or corners $\alpha_{1},\ldots,\alpha_{2n-1}$
and another medial vertex $a_{2n}$. Choose two s-orientations $\alpha_{2n}^{1}=a_{2n}^{p_{1}}$ and $\alpha_{2n}^2=a_{2n}^{p_2}$ of $a_{2n}$ such that the orientations $p_1, p_2$ are the two opposite permissible orientations on $a_{2n}$. Define for $\gamma\in\mathcal{C}_{\Omega_{\delta}}^{\alpha_{1},\ldots,\alpha_{2n}^{j}}\subset \mathcal{C}_{\Omega_{\delta}}^{\alpha_{1},\ldots,\alpha_{2n-1},a_{2n}}$ the complex weight (where dependence on the choice of $\sqrt{p_{j}}$ eventually cancels out),
\[
\chi(\gamma):=\frac{1}{i\sqrt{p_{j}}}\phi_{\{a_{1},\ldots,\alpha_{2n-1},a_{2n}^{j}\}}(\gamma)\,,
\]
noting that the complexified observables can be defined in terms of
sums of this weight. If $\alpha_{2n}=a_{2n}^{o}$ is an s-oriented
corner, there is only one 
corresponding orientation and $\chi(\gamma):=\frac{1}{i\sqrt{o}}\phi_{\left\{ \alpha_{k}\right\} }(\gamma)$.

\begin{proposition}
\label{prop:observable-s-hol}$H_{\Omega_{\delta}:\{e_{1}^{s_{1}},\ldots,e_{m}^{s_{m}}\}}^{\alpha_{1},\ldots,\alpha_{2n-1}}$
and $H_{\left[\Omega_{\delta},0\right]:\{e_{1}^{s_{1}},\ldots,e_{m}^{s_{m}}\}}^{\alpha_{1},\ldots,\alpha_{2n-1}}$
are s-holomorphic wherever defined.
\end{proposition}
\begin{proof}
This was proven for the complexified fermion in \cite[Lemma 74]{hon2010}
in the setting where the $\alpha_{i}$ are non-adjacent medial vertices. We extend this to all possible $\alpha_i$ via the extension of the complexified fermion to corners in addition to medial vertices. The idea is that there is a natural bijection between the set of paths to a medial vertex $e$ and those to an adjacent corner $c$. Namely, if $e(c)$ is the shortest walk from $a_{2n}$ to $c$ consisting of two half-edges both incident to the common vertex $v$, the map $\gamma \mapsto \gamma\oplus e(c)$ is a bijection. One needs to show that the projection in s-holomorphicity relations (\ref{eq:s-hol}) sends the winding weight $e^{-\frac i2 \mathbf{W}(\gamma)}$ to the winding weight of the image, times a factor of $
\cos(\pi/8)$. 

Following this, the next lemma therefore proves s-holomorphicity in the case of $H_{\Omega_{\delta}:\{e_{1}^{s_{1}},\ldots,e_{m}^{s_{m}}\}}^{\alpha_{1},\ldots,\alpha_{2n-1}}$. The summands in the definition of $H_{\left[\Omega_{\delta},0\right]:\{e_{1}^{s_{1}},\ldots,e_{m}^{s_{m}}\}}^{\alpha_{1},\ldots,\alpha_{2n-1}}$
only have additional real factors invariant under the $\cdot\oplus e(c)$
bijection, so the lemma implies the s-holomorphicity for
$H_{\left[\Omega_{\delta},0\right]:\{e_{1}^{s_{1}},\ldots,e_{m}^{s_{m}}\}}^{\alpha_{1},\ldots,\alpha_{2n-1}}$ as well.
\end{proof}

\begin{lemma}
\label{lem:projection-xor}Let $\left\{ a_{j}\right\} _{j=1}^{2n-1}$,
$\{e_{k}\}_{k=1}^{m}$ be distinct medial vertices of $\Omega_{\delta}$
and denote the s-oriented versions of $\left\{ a_{j}\right\} _{j=1}^{2n-1}$
by $\alpha_{j}=a_{j}^{o_{j}}$, and inclusion variables $s_1,...,s_m\in \{\bullet,\circ\}$. Suppose $c$ is an interior corner and $a_{2n}$ is an adjacent interior medial vertex
distinct from $\left\{ a_{j}\right\} _{j=1}^{2n-1}$.
Let $e(c)$ be the shortest walk from $a_{2n}$ to $c$ consisting
of two half-edges both incident to a common vertex $v$. Then the
bijection, 
\[
\gamma\mapsto\gamma\oplus e(c)\,,
\]
from $\mathcal{C}_{\Omega_{\delta}:\left\{ e_{k}^{s_k}\right\} }^{\alpha_{1},\ldots,\alpha_{2n-1},a_{2n}}$
to $\mathcal{C}_{\Omega_{\delta}:\left\{ e_{k}^{s_k}\right\} }^{\alpha_{1},\ldots,\alpha_{2n-1},c}$
satisfies the projection relation, \textup{
\begin{align*}
\mathsf{P}_{l(c)}\chi(\gamma) & =\chi(\gamma\oplus e(c))\,.
\end{align*}
}
\end{lemma}
\begin{proof}
Suppose $\gamma_{f}\in\Gamma(\gamma)$ is the walk ending at $a_{2n}$,
starting at some $\alpha_{s}$. Suppose $c$ is a corner of type $\tau$ adjacent to $a_{2n}$. Then $\Gamma(\gamma\oplus e(c))$,
chosen using paths in $\Gamma(\gamma)$ with $\gamma_{f}$ replaced
by $\gamma_{f}\oplus e(c)$, is clearly an admissible choice of walks
in $\gamma\oplus e(c)$. Note that if $o_{2n}$ is any s-orientation
of $a_{2n}$ compatible with $\gamma_{f}$, $i\frac{\sqrt{o_{2n}}}{\sqrt{o_{s}}}e^{-\frac{i}{2}\mathbf{W}\left(\gamma_{f}\right)}$
is a real quantity. Thus it suffices to show that
\begin{alignat*}{1}
\mathsf{P}_{l(c)}\frac{e^{-2\beta_{c}|\gamma|}}{\sqrt{o_{s}}}e^{-\frac{i}{2}\mathbf{W}\left(\gamma_{f}\right)} & =\cos\frac{\pi}{8}\cdot\frac{e^{-2\beta_{c}|\gamma\oplus e(c)|}}{\sqrt{o_{s}}}e^{-\frac{i}{2}\mathbf{W}\left(\gamma_{f}\oplus e(c)\right)}\,.
\end{alignat*}

Commuting real quantities with projections, we may rewrite the left
hand side as 
\begin{align*}
i\frac{\sqrt{o_{2n}}}{\sqrt{o_{s}}}e^{-\frac{i}{2}\mathbf{W}\left(\gamma_{f}\right)}e^{-2\beta_{c}|\gamma|}\mathsf{P}_{l(c)}\frac{1}{i\sqrt{o_{2n}}}\,.
\end{align*}
There are two cases to consider: the cases when the half-edge $\left\langle a_{2n},v\right\rangle \in\gamma_{f}$
and when $\left\langle a_{2n},v\right\rangle \notin\gamma_{f}$.

\begin{itemize}
\item In the first case, $\left|\gamma\oplus e(c)\right|=\left|\gamma\right|$.
Subsequently $\tau^{2}o_{2n}=-e^{\pm\frac{\pi}{4}i}$, where the sign depends
on the winding change $\mathbf{W}(\gamma\oplus e(c))=\mathbf{W}(\gamma)\mp\frac{\pi}{4}$.
Now we have 
\begin{align*}
\mathsf{P}_{l(c)}\frac{1}{i\sqrt{o_{2n}}}=\frac{1-\tau^{2}o_{2n}}{2i\sqrt{o_{2n}}}=\frac{e^{\pm\frac{\pi}{8}i}}{i\sqrt{o_{2n}}}\cos\frac{\pi}{8}\,,
\end{align*}
and the result follows.
\item In the second case, $\left|\gamma\oplus e(c)\right|=\left|\gamma\right|+1$.
Then $\tau^{2}o_{2n}=e^{\pm\frac{\pi}{4}i}$, where the sign depends
on the winding change $\mathbf{W}(\gamma\oplus e(c))=\mathbf{W}(\gamma)\pm\frac{3\pi}{4}$.
Then 
\begin{align*}
\mathsf{P}_{l(c)}\frac{1}{i\sqrt{o_{2n}}}=\frac{1-\tau^{-2}o_{2n}}{2i\sqrt{o_{2n}}}=\frac{e^{\mp\frac{3\pi}{8}i}}{i\sqrt{o_{2n}}}\sin\frac{\pi}{8}\,.
\end{align*}
Upon noting that $\tan\frac{\pi}{8}=e^{-2\beta_{c}}$, the result follows. \qedhere
\end{itemize}
\end{proof}

\begin{lemma}
\label{lem:observable-residue} The discrete residues at an s-oriented interior oriented medial vertex $\alpha_j$ are
\begin{alignat*}{1}
H_{\Omega_{\delta}:\{e_{1}^{s_{1}},\ldots,e_{m}^{s_{m}}\}}^{\alpha_{1},\ldots,\alpha_{2n-1}}(\alpha_{j+})-H_{\Omega_{\delta}:\{e_{1}^{s_{1}},\ldots,e_{m}^{s_{m}}\}}^{\alpha_{1},\ldots,\alpha_{2n-1}}(\alpha_{j-}) & =\frac{(-1)^{j+1}}{\sqrt{o_{j}}}F_{\Omega_{\delta}:\{e_{1}^{s_{1}},\ldots,e_{m}^{s_{m}}\}}^{\alpha_{1},\ldots,\alpha_{j-1},\alpha_{j+1},\ldots,\alpha_{2n-1}}\,,\\
H_{\left[\Omega_{\delta},0\right]:\{e_{1}^{s_{1}},\ldots,e_{m}^{s_{m}}\}}^{\alpha_{1},\ldots,\alpha_{2n-1}}(\alpha_{j+})-H_{\left[\Omega_{\delta},0\right]:\{e_{1}^{s_{1}},\ldots,e_{m}^{s_{m}}\}}^{\alpha_{1},\ldots,\alpha_{2n-1}}(\alpha_{j-}) & =\frac{(-1)^{j+1}}{\sqrt{o_{j}}}F_{\left[\Omega_{\delta},0\right]:\{e_{1}^{s_{1}},\ldots,e_{m}^{s_{m}}\}}^{\alpha_{1},\ldots,\alpha_{j-1},\alpha_{j+1},\ldots,\alpha_{2n-1}}\,.
\end{alignat*}
where in the two cases $\alpha_1,\ldots, \alpha_{2n-1}$ are all taken respectively on $\Omega_\delta$ and $\left[ \Omega_\delta, 0\right]$.
\end{lemma}
\begin{proof}
We prove this for the fermion and spin-fermion simultaneously, letting $D_\delta$ represent $\Omega_\delta$ or $[\Omega_\delta,0]$. It suffices to show that 
the front and back values of $H_{D_{\delta}:\{e_{1}^{s_{1}},\ldots,e_{m}^{s_{m}}\}}^{\alpha_{1},\ldots,\alpha_{2n-1}}$
at $\alpha_{j}$ are given by
\begin{alignat}{1}\label{eq:H-front-back-val}
H_{D_{\delta}:\{e_{1}^{s_{1}},\ldots,e_{m}^{s_{m}}\}}^{\alpha_{1},\ldots,\alpha_{2n-1}}(\alpha_{j+}) & =\frac{(-1)^{j+1}}{\sqrt{o_{j}}}F_{D_{\delta}:\left\{ e_{1}^{s_{1}},\ldots,e_{m}^{s_{m}},a_{j}^{\circ}\right\} }^{\alpha_{1},\ldots,\alpha_{j-1},\alpha_{j+1},\ldots\alpha_{2n-1}}\,, \nonumber \\
H_{D_{\delta}:\{e_{1}^{s_{1}},\ldots,e_{m}^{s_{m}}\}}^{\alpha_{1},\ldots,\alpha_{2n-1}}(\alpha_{j-}) & =\frac{(-1)^{j}}{\sqrt{o_{j}}}F_{D_{\delta}:\left\{ e_{1}^{s_{1}},\ldots,e_{m}^{s_{m}},a_{j}^{\bullet}\right\} }^{\alpha_{1},\ldots,\alpha_{j-1},\alpha_{j+1},\ldots\alpha_{2n-1}}\,,
\end{alignat}
where in the case $D_\delta = \left[\Omega_\delta, 0 \right]$ the projection of $a_j$ onto $\Omega_\delta$ is considered in $a_{j}^{\circ, \bullet}$.

Let $c_{1}$ denote one of the two corners adjacent to the end vertex of $a_j$ in the direction $o_{j}$. We verify that $H_{D_{\delta}:\{e_{1}^{s_{1}},\ldots,e_{m}^{s_{m}}\}}^{\alpha_{1},\ldots,\alpha_{2n-1}}(\alpha_{j+})$
projects to $H_{D_{\delta}:\{e_{1}^{s_{1}},\ldots,e_{m}^{s_{m}}\}}^{\alpha_{1},\ldots,\alpha_{2n-1}}(c_{1})$.
Let $c_1$ be a corner of type $\tau$.

We note that any $\gamma\in\mathcal{C}_{D_{\delta}:\left\{ e_{1}^{s_{1}},\ldots,e_{m}^{s_{m}}\right\} }^{\alpha_{1},\ldots,\alpha_{2n-1},c_{1}}$
contains the walk $e(c_{1}):=\left\{ \left\langle a_{j},a_{j}+o_{j}\frac{\delta}{\sqrt{2}}\right\rangle ,\left\langle a_{j}+o_{j}\frac{\delta}{\sqrt{2}},c_{1}\right\rangle \right\} $,
so we can start from $\gamma_{f}$ and complete $\Gamma(\gamma)$.
In addition, $\gamma\oplus e(c_{1})\in\mathcal{C}_{D_{\delta}:\left\{ e_{1}^{s_{1}},\ldots,e_{m}^{s_{m}},a_{j}^{\circ}\right\} }^{\alpha_{1},\ldots,\alpha_{j-1},\alpha_{j+1},\ldots,\alpha_{2n-1}}$
and $\Gamma(\gamma)\setminus\left\{ e(c_{1})\right\} $ is naturally
an admissible choice of walk for $2n-2$ points. Since the additional
real factor in the case of $D_{\delta}=\left[\Omega_{\delta},0\right]$
is easily seen to be invariant under the bijection, it now suffices
to show the projection relation
\[
\mathsf{P}_{l(c_{1})}\frac{(-1)^{j+1}}{\sqrt{o_{j}}}\phi_{\{\alpha_{1},\ldots,\alpha_{j-1},\alpha_{j+1},\ldots,\alpha_{2n-1}\}}(\gamma\oplus e(c_{1}))=\chi(\gamma)\,.
\]
This relation can be rewritten as $\mathsf{P}_{l(c_{1})}\frac{1}{\sqrt{o_{j}}}=e^{-2\beta_{c}}\cos\frac{\pi}{8}\cdot\frac{e^{-\frac{i}{2}\mathbf{W}(e(c_{1}))}}{\sqrt{o_{j}}}=\cos\frac{3\pi}{8}\cdot\frac{e^{-\frac{i}{2}\mathbf{W}(e(c_{1}))}}{\sqrt{o_{j}}}$
using explicit formulae and admissible choices $\Gamma(\gamma)$ and
$\Gamma(\gamma\oplus e(c_{1}))$ (where $(-1)^{j+1}$ is precisely the ratio
between the crossing parity factor of $\Gamma(\gamma\oplus e(c_{1}))$
and $\Gamma(\gamma)$; one sees this by drawing the pairing between $j$ and $2n$ very
close to $[j,2n]\subset\mathbb{R}$, so that it crosses exactly $2n-j-1$
other lines).

Note that $o_{j}e^{i\mathbf{W}(e(c_{1}))}=\tau^{2}$ and $\mathbf{W}(e(c_{1}))=\pm\frac{3\pi}{4}$.
Commuting real values with projection, we have, 
\[
\mathsf{P}_{l(c_{1})}\left[\frac{1}{\sqrt{o_{j}}}\right]=\frac{1}{2}\left[\frac{1}{\sqrt{o_{j}}}+\frac{\sqrt{o_{j}}}{\tau^{2}}\right]=\frac{1}{\sqrt{o_{j}}}\frac{1+e^{-i\mathbf{W}(e(c_{1}))}}{2}=\cos\frac{3\pi}{8}\cdot\frac{e^{-\frac{i}{2}\mathbf{W}(e(c_{1}))}}{\sqrt{o_{j}}}\,. \qedhere
\]
\end{proof}

\begin{remark}\label{rem:mu-a}
The explicit front and back values~\eqref{eq:H-front-back-val} shown in the proof above give us a simple correspondence between the full-plane observables and the full-plane (normal and spin-weighted) nearest-pair correlations $\mu, \mu_{a'}$ for edges $a=\delta a' \in \mathcal E_{\Omega_{\delta}}$ defined in~\eqref{eq:mu-e}.

Notice that, by low-temperature expansion, the correlation of the nearest spin pair separated by an edge $a$ can be written $\frac{1}{\mathcal{Z}_{\Omega_{\delta}}}\left[ \sum_{\omega\in\mathcal{C}_{\Omega_{\delta}:\{a^\circ\}}}e^{-2|\omega|} -\sum_{\omega\in\mathcal{C}_{\Omega_{\delta}:\{a^\bullet\}}}e^{-2|\omega|} \right]= F_{\Omega_{\delta}:\{a^\circ\}} -F_{\Omega_{\delta}:\{a^\bullet\}}$. By the values given in~\eqref{eq:H-front-back-val}, this is precisely $\sqrt{o} \left[ H_{\Omega_\delta}^\alpha(\alpha_+)+H_{\Omega_\delta}^\alpha(\alpha_-) \right]$, for any s-oriented version $\alpha$ of $a$. Now taking the infinite-volume limit by sending $\Omega\uparrow\mathbb{C}$ (whose existence is known by Theorem~\ref{thm:fermion-fermion-infinite-vol-lim}), we have 
\begin{align*}
\mu = \sqrt{o} \left[ H_{\mathbb{C}_\delta}^\alpha(\alpha_+)+H_{\mathbb{C}_\delta}^\alpha(\alpha_-) \right]\,,
\end{align*}
matching the values $H_{\mathbb{C}_\delta}^\alpha(\alpha_\pm)=\frac{\mu\pm 1}{2\sqrt{o}}$ given by Proposition \ref{thm:fermion-fermion-full-plane-explicit-formula}.

By analogous reasoning, for every $a=\delta a' \in \mathcal E_{\Omega_\delta}$ we can calculate $\mu_{a'}$ as defined in~\eqref{eq:mu-e} by 
\begin{align*}
\mu_{a'} = \sqrt{o} \left[ H_{\left[ \mathbb{C}_\delta,0 \right]}^\alpha(\alpha_+)+H_{\left[\mathbb{C}_\delta,0\right]}^\alpha(\alpha_-) \right]\,,
\end{align*}
or equivalently for every $a=\delta a' \in \mathcal E_{\Omega_\delta}$ we have $H_{\left[ \mathbb{C}_\delta,0 \right] }^\alpha(\alpha_\pm)=\frac{\mu_{a'} \pm 1}{2\sqrt{o}}$,
where we know the infinite-volume limit $\Omega \uparrow \mathbb C$ of $\sqrt o[H_{[\Omega_\delta,0]}^\alpha(\alpha_+)+H_{[\Omega_\delta,0]}^\alpha(\alpha_-)]$ exists by Theorem~\ref{thm:spin-fermion-infinite-vol-lim}.
\end{remark}

Recall the definition of $\nu_z$ as the outer normal at a boundary medial vertex $z$.

\begin{lemma}
\label{lem:observable-boundary}If $\{\alpha_{i}\}_{i=1}^{2n-1}$ are interior s-oriented medial vertices, $\{e_1^{s_1},...e_m^{s_m}\}$ are edges with corresponding inclusion variables, and $z=a_{2n}$ is a boundary medial vertex, we have that
\begin{alignat*}{1}
H_{\Omega_{\delta}:\{e_{1}^{s_{1}},\ldots,e_{m}^{s_{m}}\}}^{\alpha_{1},\ldots,\alpha_{2n-1}}\left(z\right)\sqrt{\nu_{z}} & \in\mathbb{R}\,;\\
H_{\left[\Omega_{\delta},0\right]:\{e_{1}^{s_{1}},\ldots,e_{m}^{s_{m}}\}}^{\alpha_{1},\ldots,\alpha_{2n-1}}\left(z\right)\sqrt{\nu_{z}} & \in\mathbb{R}\,.
\end{alignat*}
\end{lemma}
\begin{proof}
\cite[Proposition 79]{hon2010} proves the lemma for $H_{\Omega_{\delta}:\{e_{1}^{s_{1}},\ldots,e_{m}^{s_{m}}\}}^{\alpha_{1},\ldots,\alpha_{2n-1}}$
with non-adjacent $\alpha_{i}$. The idea is that if $z$ is on the boundary, a path can only reach $z$ via the half-edge in the inner direction, which fixes the complex phase in the weight; this goes through unchanged for possibly adjacent $\alpha_{i}$. For the spin-fermion observable, since the additional factors in $H_{\left[\Omega_{\delta},0\right]:\{e_{1}^{s_{1}},\ldots,e_{m}^{s_{m}}\}}^{\alpha_{1},\ldots,\alpha_{2n-1}}$
are real, the result holds.
\end{proof}

\subsection{\label{subsec:ising-correlations}Fused Observables and Ising Correlations}

In this subsection, we formulate the Ising correlations in a bounded
domain in terms of fusions of the observables introduced above. We
again write $\alpha_{1},\ldots,\alpha_{2n}$ for distinct s-oriented medial vertices
in $\Omega_{\delta}$ or $\left[\Omega_{\delta},0\right]$ and $e_1,...,e_m$ distinct edges with inclusion variables $s_1,...,s_m \in \{\bullet,\circ\}$.
\begin{definition}
\label{def:fused-observables}Suppose $b_{1},\ldots,b_{N}$ are distinct
medial vertices in $\Omega_{\delta}$ with $b_{N}=\delta b_N'$. Define the \emph{fused fermion
}and\emph{ fused spin-fermion observables} and their complexifications
inductively by
\begin{alignat}{1}
F_{\Omega_{\delta}:\{e_{1}^{s_{1}},\ldots,e_{m}^{s_{m}}\}}^{\alpha_{1},\ldots,\alpha_{2n}:\left[b_{1},\ldots,b_{N}\right]} 
& :=F_{\Omega_{\delta}:\{e_{1}^{s_{1}},\ldots,e_{m}^{s_{m}},b_{N}^{\circ}\}}^{\alpha_{1},\ldots,\alpha_{2n}:\left[b_{1},\ldots,b_{N-1}\right]}-\frac{\mu+1}{2}F_{\Omega_{\delta}:\{e_{1}^{s_{1}},\ldots,e_{m}^{s_{m}}\}}^{\alpha_{1},\ldots,\alpha_{2n}:\left[b_{1},\ldots,b_{N-1}\right]}\,, \label{eq:fusion-recursion}\\
F_{\left[\Omega_{\delta},0\right]:\{e_{1}^{s_{1}},\ldots,e_{m}^{s_{m}}\}}^{\alpha_{1},\ldots,\alpha_{2n}:\left[b_{1},\ldots,b_{N}\right]} & :=F_{\left[\Omega_{\delta},0\right]:\{e_{1}^{s_{1}},\ldots,e_{m}^{s_{m}},b_{N}^{\circ}\}}^{\alpha_{1},\ldots,\alpha_{2n}:\left[b_{1},\ldots,b_{N-1}\right]}-\frac{\mu_{b_N'}+1}{2}F_{\left[\Omega_{\delta},0\right]:\{e_{1}^{s_{1}},\ldots,e_{m}^{s_{m}}\}}^{\alpha_{1},\ldots,\alpha_{2n}:\left[b_{1},\ldots,b_{N-1}\right]}\,,\nonumber 
\end{alignat}
and the usual complexification scheme on $\alpha_{2n}$ (see e.g.
(\ref{eq:complexification})).
\end{definition}
These fusions arise naturally from the process of removing singularities.
Suppose $\alpha_{j}$ is an s-oriented interior medial vertex. Note that, by
Lemma \ref{lem:observable-residue} and Theorems \ref{thm:fermion-fermion-infinite-vol-lim}\textendash \ref{thm:spin-fermion-infinite-vol-lim},
the functions
\begin{align}
H_{\Omega_{\delta}:\{e_{1}^{s_{1}},\ldots,e_{m}^{s_{m}}\}}^{\alpha_{1},\ldots,\alpha_{2n-1}}+(-1)^{j}F_{\Omega_{\delta}:\{e_{1}^{s_{1}},\ldots,e_{m}^{s_{m}}\}}^{\alpha_{1},\ldots,\alpha_{j-1},\alpha_{j+1},\ldots,\alpha_{2n-1}}H_{\mathbb{C}_{\delta}}^{\alpha_{j}}\,, \label{eq:s-holom-extension-unfused} \\
H_{\left[\Omega_{\delta},0\right]:\{e_{1}^{s_{1}},\ldots,e_{m}^{s_{m}}\}}^{\alpha_{1},\ldots,\alpha_{2n-1}}+(-1)^{j}F_{\left[\Omega_{\delta},0\right]:\{e_{1}^{s_{1}},\ldots,e_{m}^{s_{m}}\}}^{\alpha_{1},\ldots,\alpha_{j-1},\alpha_{j+1},\ldots,\alpha_{2n-1}}H_{\left[\mathbb{C}_{\delta},0\right]}^{\alpha_{j}}\,, \label{eq:s-holom-extension-unfused-2}
\end{align}
have discrete residue $0$ at $\alpha_{j}$; thus they extend to $a_{j}$
s-holomorphically. In fact, we have the following extension result
for them, which also applies to the corresponding fused observables.
\begin{lemma}
\label{lem:s-hol-extension}The following fused versions of the functions defined in~\eqref{eq:s-holom-extension-unfused}--\eqref{eq:s-holom-extension-unfused-2} have s-holomorphic extensions to
$a_{j}$ given by the values 
\begin{alignat*}{1}
 & \left[H_{\Omega_{\delta}}^{\alpha_{1},\ldots,\alpha_{2n-1}:[e_{1},\ldots,e_{m}]}+(-1)^{j}F_{\Omega_{\delta}}^{\alpha_{1},\ldots,\alpha_{j-1},\alpha_{j+1},\ldots,\alpha_{2n-1}:\left[e_{1},\ldots,e_{m}\right]}H_{\mathbb{C}_{\delta}}^{\alpha_{j}}\right](a_{j})\\
 & \qquad \qquad :=\frac{(-1)^{j+1}}{\sqrt{o_{j}}}F_{\Omega_{\delta}}^{\alpha_{1},\ldots,\alpha_{j-1},\alpha_{j+1},\ldots\alpha_{2n-1}:\left[e_{1},\ldots,e_{m},a_{j}\right]}\,,\\
 & \left[H_{\left[\Omega_{\delta},0\right]}^{\alpha_{1},\ldots,\alpha_{2n-1}:[e_{1},\ldots,e_{m}]}+(-1)^{j}F_{\left[\Omega_{\delta},0\right]}^{\alpha_{1},\ldots,\alpha_{j-1},\alpha_{j+1},\ldots,\alpha_{2n-1}:\left[e_{1},\ldots,e_{m}\right]}H_{\left[\mathbb{C}_{\delta},0\right]}^{\alpha_{j}}\right](a_{j})\\
 & \qquad \qquad :=\frac{(-1)^{j+1}}{\sqrt{o_{j}}}F_{\left[\Omega_{\delta},0\right]}^{\alpha_{1},\ldots,\alpha_{j-1},\alpha_{j+1},\ldots\alpha_{2n-1}:\left[e_{1},\ldots,e_{m},a_{j}\right]}\,,
\end{alignat*}
where in the second case, $a_j$ in $[\Omega_\delta , 0]$ is identified with its
projection on $\Omega_{\delta}$ in the expression $\left[e_{1},\ldots,e_{m},a_{j}\right]$.
\end{lemma}
\begin{proof}
By $\mathbb R$-linearity of s-holomorphicity and the definition of fused observables, it suffices to show that the unfused observables given by~\eqref{eq:s-holom-extension-unfused}--\eqref{eq:s-holom-extension-unfused-2} extend s-holomorphically to $a_j$ with above right hand side values (without $e_{1},\ldots,e_{m}$).

Since the function~\eqref{eq:s-holom-extension-unfused} has discrete residue $0$ at $a_j$, it has an s-holomorphic extension to $a_j$ given by 
\begin{alignat*}{1}
 & \left[H_{\Omega_{\delta}:\{e_{1}^{s_{1}},\ldots,e_{m}^{s_{m}}\}}^{\alpha_{1},\ldots,\alpha_{2n-1}}+(-1)^{j}F_{\Omega_{\delta}:\{e_{1}^{s_{1}},\ldots,e_{m}^{s_{m}}\}}^{\alpha_{1},\ldots,\alpha_{j-1},\alpha_{j+1},\ldots,\alpha_{2n-1}}H_{\mathbb{C}_{\delta}}^{\alpha_{j}}\right](a_{j})\\
 & \qquad \qquad =H_{\Omega_{\delta}:\{e_{1}^{s_{1}},\ldots,e_{m}^{s_{m}}\}}^{\alpha_{1},\ldots,\alpha_{2n-1}}(\alpha_{j+})+(-1)^{j}F_{\Omega_{\delta}:\{e_{1}^{s_{1}},\ldots,e_{m}^{s_{m}}\}}^{\alpha_{1},\ldots,\alpha_{j-1},\alpha_{j+1},\ldots,\alpha_{2n-1}}H_{\mathbb{C}_{\delta}}^{\alpha_{j}}(\alpha_{j+})\,.
\end{alignat*}
In turn, by the explicit values in the proof of Lemma~\ref{lem:observable-residue}, this is given by 
\begin{alignat*}{1}
\frac{(-1)^{j+1}}{\sqrt{o_{j}}}F_{\Omega_{\delta}:\left\{ e_{1}^{s_{1}},\ldots,e_{m}^{s_{m}},a_{j}^{\circ}\right\} }^{\alpha_{1},\ldots,\alpha_{j-1},\alpha_{j+1},\ldots\alpha_{2n-1}} & +(-1)^{j}F_{\Omega_{\delta}:\{e_{1}^{s_{1}},\ldots,e_{m}^{s_{m}}\}}^{\alpha_{1},\ldots,\alpha_{j-1},\alpha_{j+1},\ldots,\alpha_{2n-1}}H_{\mathbb{C}_{\delta}}^{\alpha_{j}}(\alpha_{j+})\\
 & \qquad \qquad \qquad=\frac{(-1)^{j+1}}{\sqrt{o_{j}}}F_{\Omega_{\delta}:\{e_{1}^{s_{1}},\ldots,e_{m}^{s_{m}}\}}^{\alpha_{1},\ldots,\alpha_{j-1},\alpha_{j+1},\ldots\alpha_{2n-1}:[a_{j}]}\,,
\end{alignat*}
using that $H_{\mathbb{C}_{\delta}}^{\alpha_{j}}(a_{j+})=\frac{\mu+1}{2\sqrt{o_{j}}}$ (see Remark \ref{rem:mu-a}).
Identical computation gives the $\left[\Omega_{\delta},0\right]$
case, where now we use the fact that for every $a_j=\delta a_j'$, we have $H_{[\mathbb C_\delta,0]}^{\alpha_j}(\alpha_{j+})= \frac{\mu_{a_j'}+1}{2\sqrt{o_j}}$.
\end{proof}
\begin{remark}
\label{rem:possible-extension-to-singularity}So far, we have assumed
that the $2n$ inputs $\alpha_{1},\ldots,\alpha_{2n}$ of the real
observable are s-orientations of distinct $a_{1},\ldots,a_{2n}$.
An important observation to be made is that the combinatorial definition
of the real observable is robust enough for the \emph{pairwise-fused}
case, where a medial vertex $e$ appears twice among the $2n$ inputs (say
$a_{j}=a_{k}=e$), as long as their respective orientations $o_{j},o_{k}$
point to opposite directions.

In the complexified case, a medial vertex can appear twice, again oppositely
oriented, among the first $2n-1$ s-oriented medial vertices; if $a_{j}=a_{k}$
on $\Omega_{\delta}$ with opposite orientations $o_j$ and $o_k$, one can verify that the residue at $\alpha_{j}$
is given by
\[
\frac{(-1)^{j+1}}{\sqrt{o_{j}}}F_{\Omega_{\delta}:\{e_{1}^{s_{1}},\ldots,e_{m}^{s_{m}}\}}^{\alpha_{1},\ldots,\alpha_{j-1},\alpha_{j+1},\ldots,\alpha_{2n-1}}-\frac{(-1)^{k+1}}{\sqrt{o_{k}}}F_{\Omega_{\delta}:\{e_{1}^{s_{1}},\ldots,e_{m}^{s_{m}}\}}^{\alpha_{1},\ldots,\alpha_{k-1},\alpha_{k+1},\ldots,\alpha_{2n-1}}\,.
\]

This is precisely a directed superposition of the two residues derived
in Lemma \ref{lem:observable-residue} for s-oriented medial vertices $\alpha_{j},\alpha_{k}$;
similarly Lemma \ref{lem:s-hol-extension} and the resulting Propositions
\ref{prop:unfused-pfaffian}\textendash \ref{prop:fused-pfaffian},
introduced in the next subsection, easily generalize to the pairwise-fused
case.
\end{remark}
We are now in position to connect the fused multipoint observables to multipoint
Ising correlations.
\begin{proposition}
\label{prop:correlation-fused-observable}Suppose $\left\{ b_{k}\right\} _{k=1}^{N}$
is a set of $N$ distinct interior edges in $\Omega_{\delta}$. We
have 
\begin{alignat*}{1}
\mathbb{E}_{\Omega_{\delta}}\left[\prod_{b\in\left\{ b_{k}\right\} }\epsilon({b})\right] & =(-1)^{N}2^{N}F_{\Omega_{\delta}}^{[b_{1},\ldots,b_{N}]}\,,\\
\frac{\mathbb{E}_{\Omega_{\delta}}\left[\sigma_{0}\prod_{b\in\left\{ b_{k}\right\} }\epsilon_{[0]}(b)\right]}{\mathbb{E}_{\Omega_{\delta}}\left[\sigma_{0}\right]} & =(-1)^{N}2^{N}F_{\left[\Omega_{\delta},0\right]}^{\left[b_{1},\ldots,b_{N}\right]}\,.
\end{alignat*}
\end{proposition}
\begin{proof}
The first identity was proved in \cite[Proposition 72]{hon2010} inductively, where our above extensions of the projection relations
to adjacent edges now allow for the $b_{k}$ to be adjacent. Explicitly, denoting $\left\{ e_j^{s_j}\right\}=\left\{ e_1^{s_1},\ldots,e_m^{s_m} \right\}$ for a set of edges (distinct from $b_k$) with inclusion variables $s_j \in \{\bullet,\circ\}$ and recalling $\mathcal{C}_{\Omega_\delta:\left\{ e_j^{s_j}\right\}}$, it is straightforward to show by induction on $N$ (the base case is trivial),
\[
\mathbb{E}_{\Omega_{\delta}}\left[\mathbf{1}_{\left\{ e_j^{s_j}\right\}}\prod_{b\in\left\{ b_{k}\right\} }\epsilon({b}) \right] =(-1)^{N}2^{N}F_{\Omega_{\delta}:\left\{ e_j^{s_j}\right\}}^{[b_{1},\ldots,b_{N}]}\,,
\]
(where for an edge $e$ separating faces $x,y$ with inclusion variable $s\in \{\bullet,\circ\}$, the indicator $\boldsymbol 1_{\{e^s\}}$ denotes an indicator on the event that $\sigma_x = \sigma_y$ if $s=\circ$ and $\sigma_x \neq \sigma_y$ if $s = \bullet$) using the expansion
\begin{align*}
\mathbb{E}_{\Omega_{\delta}}\left[\mathbf{1}_{\left\{ e_j^{s_j}\right\}}\epsilon({b_{1}})\cdots \epsilon({b_{N+1}})\right]  = & \,\, (\mu-1)\mathbb{E}_{\Omega_{\delta}}\left[\mathbf{1}_{\left\{ e_j^{s_j},b_{N+1}^\circ \right\}}\epsilon({b_{1}})\cdots \epsilon({b_{N}})\right] \\
& \,\, + (\mu+1)\mathbb{E}_{\Omega_{\delta}}\left[\mathbf{1}_{\left\{ e_j^{s_j},b_{N+1}^\bullet \right\}}\epsilon({b_{1}})\cdots \epsilon({b_{N}})\right]\,,
\end{align*}
and that by definition $F_{\Omega_{\delta}:\left\{ e_j^{s_j} \right\}}^{[b_{1},\ldots,b_{N}]} = F_{\Omega_{\delta}:\left\{ e_j^{s_j},b_{N+1}^\circ \right\}}^{[b_{1},\ldots,b_{N}]}+F_{\Omega_{\delta}:\left\{ e_j^{s_j},b_{N+1}^\bullet \right\}}^{[b_{1},\ldots,b_{N}]}$.

The second identity follows from an analogous process, where we note that for a collection of loops and walks $\omega$, we have $\sigma_0(\omega)=(-1)^{\ell(\omega)}$ due to the plus boundary condition.
\end{proof}

\subsection{\label{subsec:fusion-relations-pfaffian}Pfaffian Formulae}

Having related the Ising correlations to the fused observables, in
this subsection we will elucidate how the recursive relation (\ref{eq:fusion-recursion})
gives rise to the Pfaffian relation in Section \ref{sec:observables-and-ising-correlations}.
The argument is identical to the one presented in Chapter 6 of \cite{hon2010},
albeit with the stronger lemmas introduced in Section \ref{subsec:multi-fermion-observables}
allowing the $\alpha_{i}$'s to be adjacent.

We first prove that a $2n$-point observable is in fact a Pfaffian
of a matrix of two-point observables. Recall that for a $2n\times2n$
antisymmetric matrix $A=(A_{jk})_{j,k=1,...,2n}$, 
\begin{equation}
\pf A:=\frac{1}{2^{n}n!}\sum_{\sigma\in S^{2n}}\operatorname{sgn}(\sigma)A_{\sigma(1)\sigma(2)}A_{\sigma(3)\sigma(4)}\cdots A_{\sigma(2n-1)\sigma(2n)}\,,\label{eq:pfaffian}
\end{equation}
 and we have the recursive expansion formula,
\begin{equation}
\pf A=\sum_{j=1}^{2n-1}(-1)^{j}A_{j,2n}\pf A_{\widehat{j;2n}}\,,\label{eq:pfaffian-recursion}
\end{equation}
where $A_{\widehat{j;2n}}$ is the matrix where the $j$ and $2n$-th
rows and columns are removed.
\begin{proposition}
\label{prop:unfused-pfaffian}Suppose $\alpha_{j}=a_{j}^{o_{j}},j=1,\ldots,2n$
are distinct (possibly pairwise-fused) s-oriented interior medial
vertices of $D_{\delta}=\Omega_{\delta}\mbox{ or }\left[\Omega_{\delta},0\right]$.
Define the $2n\times2n$ antisymmetric matrix
\[
\mathbf{F}_{D_{\delta}}^{\{\alpha_{j}\}}=\begin{pmatrix}0 & \mbox{\ensuremath{F_{D_{\delta}}^{\alpha_{1},\alpha_{2}}}} & \cdots & \ensuremath{F_{D_{\delta}}^{\alpha_{1},\alpha_{2n-1}}} & \ensuremath{F_{D_{\delta}}^{\alpha_{1},\alpha_{2n}}}\\
 & 0 & \cdots & \ensuremath{F_{D_{\delta}}^{\alpha_{2},\alpha_{2n-1}}} & \ensuremath{F_{D_{\delta}}^{\alpha_{2},\alpha_{2n}}}\\
 &  & 0 & \vdots & \vdots\\
 &  &  & \ddots & \ensuremath{F_{D_{\delta}}^{\alpha_{2n-1},\alpha_{2n}}}\\
 &  &  &  & 0
\end{pmatrix}\,.
\]
Then
\[
F_{D_{\delta}}^{\alpha_{1},\ldots,\alpha_{2n}}= \pf \mathbf{F}_{D_{\delta}}^{\{\alpha_{j}\}}\,.
\]
\end{proposition}
\begin{proof}
The $2\times2$ case is trivial. Given the recursive formula (\ref{eq:pfaffian-recursion})
for the Pfaffian, inductively it suffices to show
\[
F_{D_{\delta}}^{\alpha_{1},\ldots,\alpha_{2n}}=\sum_{j=1}^{2n-1}(-1)^{j}F_{D_{\delta}}^{\alpha_{1},\ldots,\alpha_{j-1},\alpha_{j+1}\ldots,\alpha_{2n-1}}F_{D_{\delta}}^{\alpha_{j},\alpha_{2n}}\,.
\]

The strategy is to use the boundary value problem uniqueness result
(Lemma \ref{lem:uniqueness-discrete-rbvp}) to show that the function
$H_{D_{\delta}}^{\alpha_{1},\ldots,\alpha_{2n-1}}(a_{2n})-\sum_{j=1}^{2n-1}(-1)^{j}F_{D_{\delta}}^{\alpha_{1},\ldots,\alpha_{j-1},\alpha_{j+1}\ldots,\alpha_{2n-1}}H_{D_{\delta}}^{\alpha_{j}}(a_{2n})$
is identically zero. The boundary condition is obviously satisfied;
the fact that all singularities at $\alpha_{1},\ldots,\alpha_{2n-1}$
are removable, i.e. have residue zero, is immediate from
Lemma \ref{lem:observable-residue}.
\end{proof}
We now extend the Pfaffian representation to the fused observables.
For any edge $e$, write $e^{+}=e^{o^{+}}$, $e^{-}=e^{o^{-}}$ for a pair of s-orientations 
$o^{\pm}$ of $e$, so that $o^{+}=e^{\pi i}o^{-}$. Recall the use of the $\dagger$ to denote $$F_{\Omega_{\delta}}^{\dagger}= F_{\Omega_\delta}-F_{\mathbb C_\delta} \qquad \mbox{and}\qquad F_{[\Omega_{\delta},0]}^{\dagger}= F_{[\Omega_\delta,0]}-F_{[\mathbb C_\delta,0]}\,.$$
For edges $e_1,...,e_m$ define the $2m \times 2m$ antisymmetric matrix with $F^{\dagger e_i^+, e_i^-}_{D_\delta}$ on the anti-diagonal $i+j =2m+1$, $i\leq m$, and more generally, entries, 
\[
\mathbf{F}_{D_{\delta}}^{[\{e_{k}\}]}:=
\begin{pmatrix}0 & \mbox{\ensuremath{F_{D_{\delta}}^{e_{1}^{+},e_{2}^{+}}}} & \cdots & F_{D_{\delta}}^{e_{1}^{+},e_{m}^{+}} & F_{D_{\delta}}^{e_{1}^{+},e_{m}^{-}} & \cdots & F_{D_{\delta}}^{e_{1}^{+},e_{2}^{-}} & F_{D_{\delta}}^{\dagger e_{1}^{+},e_{1}^{-}}\\
 & 0 & \cdots & F_{D_{\delta}}^{e_{2}^{+},e_{m}^{+}} & F_{D_{\delta}}^{e_{2}^{+},e_{m}^{-}} & \cdots & F_{D_{\delta}}^{\dagger e_{2}^{+},e_{2}^{-}} & F_{D_{\delta}}^{e_{2}^{+},e_{1}^{-}}\\
 &  & 0 & \vdots &  & \iddots & \vdots & \vdots\\
 &  &  & \ddots & F_{D_{\delta}}^{\dagger e_{m}^{+},e_{m}^{-}} &  & F_{D_{\delta}}^{e_{m}^{+},e_{2}^{-}} & F_{D_{\delta}}^{e_{m}^{+},e_{1}^{-}}\\
 &  &  &  & 0 & \cdots & F_{D_{\delta}}^{e_{m}^{-},e_{2}^{-}} & F_{D_{\delta}}^{e_{m}^{-},e_{1}^{-}}\\
 &  &  &  &  & \ddots & \vdots & \vdots\\
 &  &  &  &  &  & 0 & F_{D_{\delta}}^{e_{2}^{-},e_{1}^{-}}\\
 &  &  &  &  &  &  & 0
\end{pmatrix}\,.
\]
\begin{proposition}
\label{prop:fused-pfaffian}Suppose $\alpha_{j}=a_{j}^{o_{j}}$ for
\textup{$j=1,\ldots,2n$} and $e_{k}$ for $k=1,\ldots,m$ are distinct
(possibly pairwise-fused) interior medial vertices of $D_{\delta}=\Omega_{\delta}\mbox{ or }\left[\Omega_{\delta},0\right]$.
Define the block antisymmetric $2(m+n)\times2(m+n)$ matrix $F_{D_{\delta}}^{\{a_{j}\}:[\{e_{k}\}]}$
by
\begin{alignat*}{1}
\mathbf{F}_{D_{\delta}}^{\{a_{j}\}:[\{e_{k}\}]} & :=\begin{pmatrix}\mathbf{F}_{D_{\delta}}^{[\{e_{k}\}]} & -\left[\mathbf{F}_{D_{\delta}}^{\{a_{j}\}\times[\{e_{k}\}]}\right]^{T}\\
\mathbf{F}_{D_{\delta}}^{\{a_{j}\}\times[\{e_{k}\}]} & \mathbf{F}_{D_{\delta}}^{\{a_{j}\}}
\end{pmatrix}\text{, where}\\
\mathbf{F}_{D_{\delta}}^{\{a_{j}\}\times[\{e_{k}\}]} & :=\left(\begin{array}{cccccc}
F_{D_{\delta}}^{\alpha_{1},e_{1}^{+}} & \cdots & F_{D_{\delta}}^{\alpha_{1},e_{m}^{+}} & F_{D_{\delta}}^{\alpha_{1},e_{m}^{-}} & \cdots & F_{D_{\delta}}^{\alpha_{1},e_{1}^{-}}\\
\vdots &  & \vdots & \vdots &  & \vdots\\
F_{D_{\delta}}^{\alpha_{2n},e_{1}^{+}} & \cdots & F_{D_{\delta}}^{\alpha_{2n},e_{m}^{+}} & F_{D_{\delta}}^{\alpha_{2n},e_{m}^{-}} & \ldots & F_{D_{\delta}}^{\alpha_{2n},e_{1}^{-}}
\end{array}\right)\,.
\end{alignat*}
Then 
\[
F_{D_{\delta}}^{\alpha_{1},\ldots,\alpha_{2n}:[e_{1},\ldots,e_{m}]}=\pf\mathbf{F}_{D_{\delta}}^{\{a_{j}\}:[\{e_{k}\}]}\,.
\]
In particular, since $F_{D_{\delta}}^{\alpha_{1},\ldots,\alpha_{2n}:[e_{1},\ldots,e_{m}]}$
does not depend on the choice of s-orientations $o_j^+,o_j^-$ on $e_{k}$,
the Pfaffian does not.
\end{proposition}
\begin{proof}
Without loss of generality, we will assume $D_{\delta}=\Omega_{\delta}$;
the case $D_{\delta}=[\Omega_{\delta},0]$ can be treated identically.
We use induction on $m$. The case $m=0$ is given by Proposition
\ref{prop:unfused-pfaffian}. Now we assume the result holds for
$m$, and consider the case $m+1$. By Lemma \ref{lem:s-hol-extension},
if $e_{m+1}^{+}=e_{m+1}^{o}$ is an s-orientation of $e_{m+1}$,
we can extend to the removed singularity
\begin{alignat*}{1}
\left[H_{\Omega_{\delta}}^{\alpha_{1},\ldots,\alpha_{2n},e_{m+1}^{+}:[e_{1},\ldots,e_{m}]}-F_{\Omega_{\delta}}^{\alpha_{1},\ldots,\alpha_{2n}:\left[e_{1},\ldots,e_{m}\right]}H_{\mathbb{C}_{\delta}}^{e_{m+1}^{+}}\right](e_{m+1}) & :=\frac{1}{\sqrt{o}}F_{\Omega_{\delta}}^{\alpha_{1},\ldots\alpha_{2n}:\left[e_{1},\ldots,e_{m+1}\right]}\,.
\end{alignat*}

Then by the projection relations given by s-holomorphicity, we can deduce
\[
F_{\Omega_{\delta}}^{\alpha_{1},\ldots,\alpha_{2n},e_{m+1}^{+},e_{m+1}^{-}:[e_{1},\ldots,e_{m}]}-F_{\Omega_{\delta}}^{\alpha_{1},\ldots,\alpha_{2n}:\left[e_{1},\ldots,e_{m}\right]}F_{\mathbb{C}_{\delta}}^{e_{m+1}^{+},e_{m+1}^{-}}:=F_{\Omega_{\delta}}^{\alpha_{1},\ldots\alpha_{2n}:\left[e_{1},\ldots,e_{m+1}\right]}\,.
\]
 By the inductive hypothesis, we have the desired Pfaffian formulations of $F_{\Omega_{\delta}}^{\alpha_{1},\ldots,\alpha_{2n},e_{m+1}^{+},e_{m+1}^{-}:[e_{1},\ldots,e_{m}]}$
and $F_{\Omega_{\delta}}^{\alpha_{1},\ldots,\alpha_{2n}:\left[e_{1},\ldots,e_{m}\right]}$.
Expanding the Pfaffian along the last column as in (\ref{eq:pfaffian-recursion})
the result easily follows.
\end{proof}

\subsection{\label{sec:observables-and-ising-correlations}Observables and Ising
Correlations}

In this subsection we present the connection between Ising model correlations
and spin-weighted correlations and the two-point observables defined
in Section \ref{subsec:two-point-observable}. These formulae follow
immediately from the results of Sections \ref{subsec:ising-correlations}
and \ref{subsec:fusion-relations-pfaffian}.

\subsubsection{Spin-symmetric correlations}

Recall that we have defined a spin-symmetric correlation as the expectation
of a product of energy densities which scale with the mesh size $\delta$.
Our characterization of the correlation consists of a Pfaffian involving
the real observables introduced in Section \ref{sec:bounded-domain-observables}.
For an edge $e$, we will denote by $e^{+}:=e^{o^{+}},e^{-}:=e^{o^{-}}$
for any pair of s-orientations $o^{+},o^{-}$ such that $o^{+}=e^{\pi i}o^{-}$;
the following characterization shows in particular that the Pfaffian
does not depend on such choice.
\begin{proposition}
\label{prop:energy-correlations-pfaffian-observable}For any collection
of distinct (possibly adjacent) interior edges $e_{1},\ldots, e_{m}$
in $\Omega_{\delta}$, 
\[
\mathbb{E}_{\Omega_{\delta}}\left[\prod_{e\in\left\{ e_{k}\right\} }\epsilon(e)\right]=\left(-1\right)^{m}2^{m}\mathrm{Pf}\left(\mathbf{F}_{\Omega_{\delta}}^{\left[\left\{ e_{k}\right\} \right]}\right)\,,
\]
 where for any admissible s-orientations, $o_1^\pm,...,o_m^\pm$, the antisymmetric matrix $\mathbf{F}_{\Omega_{\delta}}^{\left[\left\{ e_{k}\right\} \right]}$
is given by
\[
\begin{pmatrix}0 & \mbox{\ensuremath{F_{\Omega_{\delta}}^{e_{1}^{+},e_{2}^{+}}}} & \cdots & F_{\Omega_{\delta}}^{e_{1}^{+},e_{m}^{+}} & F_{\Omega_{\delta}}^{e_{1}^{+},e_{m}^{-}} & \cdots & F_{\Omega_{\delta}}^{e_{1}^{+},e_{2}^{-}} & F_{\Omega_{\delta}}^{\dagger e_{1}^{+},e_{1}^{-}}\\
 & 0 & \cdots & F_{\Omega_{\delta}}^{e_{2}^{+},e_{m}^{+}} & F_{\Omega_{\delta}}^{e_{2}^{+},e_{m}^{-}} & \cdots & F_{\Omega_{\delta}}^{\dagger e_{2}^{+},e_{2}^{-}} & F_{\Omega_{\delta}}^{e_{2}^{+},e_{1}^{-}}\\
 &  & 0 & \vdots &  & \iddots & \vdots & \vdots\\
 &  &  & \ddots & F_{\Omega_{\delta}}^{\dagger e_{m}^{+},e_{m}^{-}} &  & F_{\Omega_{\delta}}^{e_{m}^{+},e_{2}^{-}} & F_{\Omega_{\delta}}^{e_{m}^{+},e_{1}^{-}}\\
 &  &  &  & 0 & \cdots & F_{\Omega_{\delta}}^{e_{m}^{-},e_{2}^{-}} & F_{\Omega_{\delta}}^{e_{m}^{-},e_{1}^{-}}\\
 &  &  &  &  & \ddots & \vdots & \vdots\\
 &  &  &  &  &  & 0 & F_{\Omega_{\delta}}^{e_{2}^{-},e_{1}^{-}}\\
 &  &  &  &  &  &  & 0
\end{pmatrix}\,,
\]
where $F_{\Omega_{\delta}}^{\dagger\alpha,\zeta}:=F_{\Omega_{\delta}}^{\alpha,\zeta}-F_{\mathbb{C}_{\delta}}^{\alpha,\zeta}$.
\end{proposition}
\begin{proof}
Starting from the characterization of Proposition \ref{prop:correlation-fused-observable},
we can calculate the fused observable via an application of Proposition
\ref{prop:fused-pfaffian}.
\end{proof}

\subsubsection{Spin-antisymmetric Correlations}

Recall that a spin-weighted correlation was defined as the expectation
of a product of the spin at $0$ and energy densities on adjacent
sites. We characterize it as a Pfaffian analogously to the previous
subsection, but with $\Omega_{\delta}$ replaced by its double cover
and the two-point fermion replaced with the spin-fermion observable.
Accordingly, we again fix orientations $e^{+},e^{-}$ as well as a
choice of lift in $\left[\Omega_{\delta},0\right]$ for an edge $e$
in $\Omega_{\delta}$, on which the value of the Pfaffian does not
depend.
\begin{proposition}
\label{prop:spin-energy-fermion-correlation-pfaffian-observable}For
any collection of distinct (possibly adjacent) interior edges $e_{1},\ldots, e_{m}$
in $\Omega_{\delta}$, 
\[
\frac{1}{\mathbb{E}_{\Omega_{\delta}}\left[\sigma_{0}\right]}\mathbb{E}_{\Omega_{\delta}}\left[\sigma_{0}\prod_{e\in\left\{ e_{k}\right\} }\epsilon_{[0]}(e)\right]=\left(-1\right)^{m}2^{m}\mathrm{Pf}\left(\mathbf{F}_{\left[\Omega_{\delta},0\right]}^{\left[\left\{ e_{k}\right\} \right]}\right)\,,
\]
 where for any admissible s-orientations $o_1^\pm,...,o_m^\pm$, the antisymmetric matrix $\mathbf{F}_{\left[\Omega_{\delta},0\right]}^{\left[\left\{ e_{k}\right\} \right]}$
is given by
\[
\begin{pmatrix}0 & \mbox{\ensuremath{F_{\left[\Omega_{\delta},0\right]}^{e_{1}^{+},e_{2}^{+}}}} & \cdots & F_{\left[\Omega_{\delta},0\right]}^{e_{1}^{+},e_{m}^{+}} & F_{\left[\Omega_{\delta},0\right]}^{e_{1}^{+},e_{m}^{-}} & \cdots & F_{\left[\Omega_{\delta},0\right]}^{e_{1}^{+},e_{2}^{-}} & F_{\left[\Omega_{\delta},0\right]}^{\dagger e_{1}^{+},e_{1}^{-}}\\
 & 0 & \cdots & F_{\left[\Omega_{\delta},0\right]}^{e_{2}^{+},e_{m}^{+}} & F_{\left[\Omega_{\delta},0\right]}^{e_{2}^{+},e_{m}^{-}} & \cdots & F_{\left[\Omega_{\delta},0\right]}^{\dagger e_{2}^{+},e_{2}^{-}} & F_{\left[\Omega_{\delta},0\right]}^{e_{2}^{+},e_{1}^{-}}\\
 &  & 0 & \vdots &  & \iddots & \vdots & \vdots\\
 &  &  & \ddots & F_{\left[\Omega_{\delta},0\right]}^{\dagger e_{m}^{+},e_{m}^{-}} & \cdots & F_{\left[\Omega_{\delta},0\right]}^{e_{m}^{+},e_{2}^{-}} & F_{\left[\Omega_{\delta},0\right]}^{e_{m}^{+},e_{1}^{-}}\\
 &  &  &  & 0 & \cdots & F_{\left[\Omega_{\delta},0\right]}^{e_{m}^{-},e_{2}^{-}} & F_{\left[\Omega_{\delta},0\right]}^{e_{m}^{-},e_{1}^{-}}\\
 &  &  &  &  & \ddots & \vdots & \vdots\\
 &  &  &  &  &  & 0 & F_{\left[\Omega_{\delta},0\right]}^{e_{2}^{-},e_{1}^{-}}\\
 &  &  &  &  &  &  & 0
\end{pmatrix}\,,
\]
where $F_{\left[\Omega_{\delta},0\right]}^{\dagger\alpha,\zeta}:=F_{\left[\Omega_{\delta},0\right]}^{\alpha,\zeta}-F_{\left[\mathbb{C}_{\delta},0\right]}^{\alpha,\zeta}$.
\end{proposition}
\begin{proof}
Again starting from the characterization of Proposition \ref{prop:correlation-fused-observable},
we can calculate the fused observable via Proposition \ref{prop:fused-pfaffian},
using now the equations corresponding to the spin-fermion. 
\end{proof}

\section{\label{sec:continuous-observables}Scaling Limits of Observables}

We have thus far defined the discrete observables which encode probabilistic
information in the form of $n$-point correlations. As a result of
the Pfaffian formulae of Section \ref{sec:observables-and-ising-correlations},
it suffices to consider two-point discrete observables since all multipoint
correlations can now be written in terms of only two-point observables.
In this section, we introduce the continuous observables, which are
precisely defined to have continuous analogues of the properties satisfied
by the discrete observables; see Section \ref{sec:discrete-complex-analysis}.
With the appropriate scaling, the discrete observables will be shown to converge
to these continuous observables. 

As the heart of convergence proofs is the extension of the convergence results of~\cite{chelkak-hongler} for $H_{[\Omega_\delta,0]}^{\alpha}$ to the case where the singularity point $\alpha$ is no longer at $\alpha_0$, familiarity with the proofs of convergence in~\cite{chelkak-hongler} for the case $\alpha=\alpha_0$ is very helpful to understanding the sequel.

\subsection{\label{subsec:square-integral}Integration of the Square}

Write $\nu_{z}$ for the unit outward normal vector at $z\in\partial\Omega$.
We have the following characterization of the complex fermion
observable $H_{\Omega_{\delta}}$ in terms of discrete complex analysis:
\begin{proposition}[see \cite{hon2010}]
\label{prop:analytic-characterization-fermion-fermion-observable}Let
$\alpha=a^{o}$ be an s-oriented medial vertex of $\Omega_{\delta}$
that is not on the boundary. The function $H_{\Omega_{\delta}}^{\alpha}$
is the unique function such that: 

\begin{itemize}
\item $H_{\Omega_{\delta}}^{\alpha}$ is s-holomorphic on $\Omega_{\delta}\setminus\left\{ a\right\} $;
\item $H_{\Omega_{\delta}}^{\alpha}$ has discrete residue $1$ at $\alpha$:
$H_{\Omega_{\delta}}^{\alpha}\left(\alpha_{+}\right)-H_{\Omega_{\delta}}^{\alpha}\left(\alpha_{-}\right)=\frac{1}{\sqrt{o}}$; 
\item $H_{\Omega_{\delta}}^{\alpha}\left(z\right)\sqrt{\nu_{z}}\in\mathbb{R}$
for any boundary medial vertex $z$.
\end{itemize}
\end{proposition}
Similarly, we have the following characterization of the complex spin-fermion
observable $H_{\left[\Omega_{\delta}0\right]}$.
\begin{proposition}
\label{prop:analytic-characterization-spin-fermion-observable}Let
$\alpha=a^{o}$ be an s-oriented medial vertex of $\left[\Omega_{\delta},0\right]$
that is not on the boundary. The function $H_{\left[\Omega_{\delta},0\right]}^{\alpha}$
is the unique function such that:

\begin{itemize}
\item $H_{\left[\Omega_{\delta},0\right]}^{\alpha}$ has monodromy $-1$
around $0$;
\item $H_{\left[\Omega_{\delta},0\right]}^{\alpha}$ is s-holomorphic on
$\left[\Omega_{\delta},0\right]\setminus\left\{ a,a^{*}\right\} $,
where $a,a^{*}$ are on opposite sheets of $\left[\Omega_{\delta},0\right]$;
\item $H_{[\Omega_{\delta},0]}^{\alpha}$ has discrete residue $1$ at $\alpha$:
$H_{\left[\Omega_{\delta},0\right]}^{\alpha}\left(\alpha_{+}\right)-H_{\left[\Omega_{\delta},0\right]}^{\alpha}\left(\alpha_{-}\right)=\frac{1}{\sqrt{o}}$;
\item $H_{\left[\Omega_{\delta},0\right]}^{\alpha}\left(z\right)\sqrt{\nu_{z}}\in\mathbb{R}$
for any boundary medial vertex $z$.
\end{itemize}
\end{proposition}
\begin{proof}[Proof of Propositions \ref{prop:analytic-characterization-fermion-fermion-observable}\textendash \ref{prop:analytic-characterization-spin-fermion-observable}]
Monodromy for the latter is clear from the sheet factor $\mathrm{s_{\alpha,\zeta}}$.
Proposition \ref{prop:observable-s-hol} and Lemmas \ref{lem:observable-residue}
and \ref{lem:observable-boundary} provide the remaining properties.
Uniqueness follows from Lemma \ref{lem:uniqueness-discrete-rbvp}.
\end{proof}
We now consider the square integral $Q_{\delta}^{\alpha}$ of the
observables. These are the same discrete square integral analogues
$\mathbb{I}_{\delta}(H_{\delta})=\re\int\left(H_{\delta}\right)^{2}$
introduced in Section \ref{subsec:discrete-riemann-boundary} in
the case where $H_{\delta}$ is one of $H_{\Omega_{\delta}}^{\alpha},H_{\left[\Omega_{\delta},0\right]}^{\alpha}$,
but we need to analyze their properties near the singularity at $\alpha$
and the monodromy. These square integrals will be our primary means
of estimating both observables, but for conciseness, we will assume
that we are working with $H_{\left[\Omega_{\delta},0\right]}^{\alpha}$.
Neither the construction nor the properties are changed in the $H_{\Omega_{\delta}}^{\alpha}$
case, except for the fact that there are no longer complications arising
from the branching at $0$.
\begin{proposition}
\label{prop:square-existence}Suppose $\alpha=a^{o}$ is an s-oriented
corner in $\left[\Omega_{\delta},0\right]$. Consider the single-valued
integral of the square $Q_{\delta}^{\alpha}:=\mathbb{I}_{\delta}\left[H_{\left[\Omega_{\delta},0\right]}^{\alpha}\right]:\mathcal{F}_{\Omega_{\delta}}\cup\mathcal{V}_{\Omega_{\delta}}\to\mathbb{R}$
constructed with the usual rule
\begin{alignat*}{1}
Q_{\delta}^{\alpha}(w)-Q_{\delta}^{\alpha}(v) & =2\delta\left|H_{\left[\Omega_{\delta},0\right]}^{\alpha}\left(\frac{1}{2}(w+v)\right)\right|^{2}\,,
\end{alignat*}
where $w$ is a face, $v$ is a vertex incident to the face, so that
$\frac{1}{2}(w+v)$ is the corner between them (note that at the singularity $a$, $\left|H_{\left[\Omega_{\delta},0\right]}^{\alpha}\left(a\right)\right|^{2}=\norsquare$),
and the Dirichlet boundary condition
\[
Q_{\delta}^{\alpha}(w)=0\mbox{ for }w\in\partial\mathcal{F}_{\Omega_{\delta}}\,.
\]
It has

\begin{itemize}
\item $\Delta_{\delta}Q_{\delta}^{\alpha}=2\delta\left|\partial_{\delta}H_{\left[\Omega_{\delta},0\right]}^{\alpha}\right|^{2}$
on $\mathcal{F}_{\Omega_{\delta}}\setminus\left\{ 0,a-o\frac{\delta}{2}\right\} $,
$\Delta_{\delta}Q_{\delta}^{\alpha}=-2\delta\left|\partial_{\delta}H_{\left[\Omega_{\delta},0\right]}^{\alpha}\right|^{2}$
on $\mathcal{V}_{\Omega_{\delta}}\setminus\left\{ a+o\frac{\delta}{2}\right\} $;
\item The outer normal derivative $\partial_{out}Q_{\Omega}^{\alpha}=\sqrt{2}\left|H_{\left[\Omega_{\delta},0\right]}^{\alpha}\right|^{2}$
on $\partial\mathcal{V}_{\Omega_{\delta}}^{m}$.
\end{itemize}
\end{proposition}
\begin{proof}
The proof follows from direct computations in Chapter 2 of \cite{hon2010}
and Section 3.3 of \cite{chsm2012}. Note that as in \cite[Proposition 3.6]{chelkak-hongler},
the singularity at $a$ (two projections from neighboring medial vertices
differing only by a sign) and branching at $0$ does not affect well-definedness
of $Q_{\delta}^{\alpha}$, but does affect the Laplacian at $0,a\pm o\frac{\delta}{2}$. 
\end{proof}
\begin{remark}
\label{rem:boundary-modification-trick}In keeping with the Dirichlet
boundary condition, we can define $\tilde{Q}_{\delta}^{\alpha}$ which
simply modifies the value of $Q_{\delta}^{\alpha}$ on $\partial\mathcal{V}_{\Omega_{\delta}}$
to be zero. This affects the Laplacian and normal derivative, but
one can define an alternate Laplacian $\tilde{\Delta}_{\delta}$ modified
at the boundary which gives $\tilde{\Delta}_{\delta}\tilde{Q_{\delta}^{\alpha}}=\Delta_{\delta}Q_{\delta}^{\alpha}$;
see \cite[Preposition 3.6]{chelkak-hongler}.
\end{remark}

\begin{remark}
\label{rem:dagger-integral}Similarly, one can define the integral
$Q_{\delta}^{\dagger\alpha}:=\mathbb{I}_{\delta}\left[H_{\left[\Omega_{\delta},0\right]}^{\alpha}-H_{\left[\mathbb{C}_{\delta},0\right]}^{\alpha}\right]$;
this has the advantage of removing the singularity at $\alpha$, so
that we have sub- and super-harmonicity at all points except for $0$.
\cite[Remark 3.8]{chelkak-hongler} notes that in the special case
$a=\frac{\delta}{2}$ we in fact do have sub-harmonicity at $a$,
owing to the fact that $\left[H_{\left[\Omega_{\delta},0\right]}^{\alpha}-H_{\left[\mathbb{C}_{\delta},0\right]}^{\alpha}\right](a):=0$.
\end{remark}

\subsection{Continuous Observables}\label{subsec:Continuous-Observables}

The continuous observables are the solutions of the continuous Riemann
boundary value problem corresponding to the discrete b.v.p. of Lemma
\ref{lem:uniqueness-discrete-rbvp}. We prove that they are in fact
the scaling limits of their discrete counterparts in Section \ref{subsec:convergence-statement}.

\subsubsection{\label{subsec:Continuous-Full-Plane-Observable}Continuous Full-Plane
Observables}

We begin by defining the continuous full-plane observables. In analogy with the discrete case, we define an s-orientation $o$ of a point $a$ in a continuous domain $\Omega$ as a choice of any unit complex number $o$ with a specified square root.
\begin{definition}
\label{def:full-plane-fermion-observable}Let $\alpha:=a^{o}$ and
$\zeta:=z^{p}$ be two s-oriented points of $\mathbb{C}$ with
$a\neq z$. We define the continuous fermion observable 
$f_{\mathbb{C}}$ by 
\[
f_{\mathbb{C}}^{\alpha}\left(\zeta\right):=i\sqrt{p}\cdot\mathsf{P}_{\frac{1}{i\sqrt{p}}\mathbb{R}}\left[h_{\mathbb{C}}^{\alpha}(z)\right]=\re\left[i\sqrt{p}\cdot h_{\mathbb{C}}^{\alpha}(z)\right]\,,
\]
where the complex observable $h_{\mathbb{C}}$ is defined by 
\[
h_{\mathbb{C}}^{\alpha}\left(z\right):=\frac{1}{\sqrt{2}\pi}\frac{\sqrt{o}}{z-a}\,.
\]
\end{definition}
Analogously to the previous definition, we proceed to define the continuous full-plane
spin-fermion observable. Since the source point $\delta\alpha$ of
the discrete two-point spin-fermion tends to the monodromy point $0$
as $\delta\to0$, we fix it as $\alpha\in\mathcal V_{[\mathbb{C}_{1},0]}^{cm}$, and formally denote the dependence of the continuous spin-fermion on $\alpha$ by $d\alpha$ (not to be confused with $\delta \alpha\in \mathcal V_{[\mathbb C_\delta,0]}^{cm}$, which is the scaling of $\alpha$ by $\delta>0$).
\begin{definition}
\label{def:continuous-full-plane-spin-fermion-observable}Let $\alpha:=a^{o}$
be an s-oriented corner or medial vertex of $\left[\mathbb{C}_{1},0\right]$. Let
$\zeta:=z^{p}$ be an s-oriented point of $\left[\mathbb{C},0\right]$.
We define the continuous spin-fermion observable $f_{\left[\mathbb{C},0\right]}$
by 
\[
f_{\left[\mathbb{C},0\right]}^{d\alpha}\left(\zeta\right):=\re[i\sqrt{p}\cdot h_{\left[\mathbb{C},0\right]}^{d\alpha}(z)]\,,
\]
where the complex observable $h_{\left[\mathbb{C},0\right]}^{d\alpha}$
is defined by
\[
h_{\left[\mathbb{C},0\right]}^{d\alpha}\left(z\right):=\frac{C_{\alpha}}{\sqrt{z}}\,,
\]
and the scaling limit factor $C_{\alpha}\in\mathbb{R}$ is given by 
\begin{align}\label{eq:def-C-alpha}
C_{\alpha}=-\re\left[ i\sqrt{o}\left(\tilde{G}_{[\mathbb{C}_{1},0]}^{+}-\tilde{G}_{[\mathbb{C}_{1},0]}^{-}\right)(a)\right]\,.
\end{align}
Recall, here, that $\tilde{G}^\pm$ are the discrete analogs of $\nori\sqrt{z}$ given by Definition \ref{def:aux-functions}. Note that  when $\alpha$ is an s-oriented 
 corner, we do not need to take the real-part in the definition above as $i\sqrt{o}\tilde G^\pm (a)$ are already real, due to s-holomorphicity. In particular, if $\alpha$ is an s-oriented real or imaginary corner we have that
\[
C_{\alpha}=\begin{cases}
\pm\nor\sqrt{o}\cdot\Hm_{1/2}^{\mathbb{X}_{1}^{i}}(a) & \mbox{if }a\in\mathcal{V}_{\left[\mathbb{C}_{1},0\right]}^{i}\cap\mathbb{X}^{\pm}\\
\mp\nori\sqrt{o}\cdot\Hm_{-1/2}^{\mathbb{Y}_{1}^{1}}(a) & \mbox{if }a\in\mathcal{V}_{\left[\mathbb{C}_{1},0\right]}^{1}\cap\mathbb{Y}^{\pm}
\end{cases}\,.
\]
\end{definition}

\subsubsection{Continuous Bounded Domain Observables}

Let $\Omega\subset\mathbb{C}$ be a bounded simply connected domain
containing $0$. Recall that we denote by $\varphi$ the conformal mapping from
$\Omega$ to the open unit disk $\mathbb{D}$ with $\varphi\left(0\right)=z$ and
$\varphi'$$\left(0\right)>0$. Here we set $z=0$. We transform $\alpha=a^{o}$
 as $\varphi(\alpha)=\varphi(a)^{o'},o'=\left(\sqrt{\varphi'(a)}\sqrt{o}\right)^2/\left| \varphi'(a) \right|$ under $\varphi$ (with a continuous square root branch choice for $\varphi'$ such that $\sqrt{\varphi'(0)}>0$ ).
\begin{definition}
\label{def:continuous-fermion-fermion-observable}Let $\alpha=a^{o},\zeta=z^{p}$
be two s-oriented points of $\Omega$ with $a\neq z$. We define
the fermion observable $f_{\Omega}$ by
\[
f_{\Omega}^{\alpha}\left(\zeta\right):=i\sqrt{p}\mathsf{P}_{\frac{1}{i\sqrt{p}}\mathbb{R}}\left[h_{\Omega}^{\alpha}(z)\right]=\re\left[i\sqrt{p}\cdot h_{\Omega}^{\alpha}(z)\right]\,,
\]
where the complexified fermion observable $h_{\Omega}$ is
defined by 
\[
h_{\Omega}^{\alpha}\left(z\right):=\sqrt{\left|\varphi'(a)\right|}\sqrt{\varphi'(z)}h_{\mathbb{D}}^{\varphi_{}(\alpha)}(\varphi(z))\,,\quad\mbox{where}\quad h_{\mathbb{D}}^{\alpha}(z):=\frac{1}{\sqrt{2}\pi}\left( \frac{1}{\sqrt{o}}\cdot\frac{1}{1-\bar{a}z}+\frac{{\sqrt{o}}}{z-a}\right)\,.
\]
\end{definition}
\begin{definition}
\label{def:continuous-spin-fermion-observable}Let $\alpha=a^{o}$
be an s-oriented corner or medial vertex on $\left[\mathbb{C}_{1},0\right]$,
and $\zeta=z^{p}$ be an s-oriented point of $\Omega$ with $a\neq z$.
We define the spin-fermion observable $f_{\left[\Omega,0\right]}$
by 
\[
f_{\left[\Omega,0\right]}^{d\alpha}\left(\zeta\right):=i\sqrt{p}\mathsf{P}_{\frac{1}{i\sqrt{p}}\mathbb{R}}\left[h_{\left[\Omega,0\right]}^{d\alpha}(z)\right]=\re\left[i\sqrt{p}\cdot h_{\left[\Omega,0\right]}^{d\alpha}(z)\right]\,,
\]
where the complexified spin-fermion observable $h_{\left[\Omega,0\right]}$
is defined by
\[
h_{\left[\Omega,0\right]}^{d\alpha}\left(z\right):=\sqrt{\varphi'(z)}h_{\mathbb{D}}^{d\alpha}(\varphi(z))\,,\quad\mbox{where}\quad h_{[\mathbb{D},0]}^{d\alpha}(z)=\frac{C_{\alpha}}{\sqrt{z}}\,.
\]
\end{definition}
\begin{remark}
\label{rem:continuous-riemann-hilbert}As explained above, these definitions
imply continuous versions of the properties satisfied by the discrete
observables; they share the singularities of the full-plane observables,
and satisfy the boundary condition $h_{\Omega}^{\alpha}(z),h_{[\Omega,0]}^{d\alpha}(z)\in{\nu_{z}}^{-1/2}\mathbb{R}$,
if $\nu_{z}$ is the unit outward normal vector at $z\in\partial\Omega$.
These also uniquely characterize a holomorphic function: see \cite[Proposition 48]{hon2010}
and \cite[Lemma 2.9]{chelkak-hongler}.
\end{remark}

\subsection{\label{subsec:convergence-statement}Observable Convergence: Statements}

We now state the two observable convergence results for the fermion and spin-fermion that are needed for the proof of the main theorem in
Section \ref{subsec:main-theorem-proofs}. The bulk of the remaining work is in the proof of the spin-fermion observable convergence; that proof is deferred until the next subsection.

In what follows, we say an s-holomorphic function $H_{\delta}:\mathcal{V}_{D_{\delta}}^{cm}\to\mathbb{C}$
\emph{converges }to a continuous function $h:D\to\mathbb{C}$ if for
any sequence $a_{\delta}\in\mathcal{V}_{D_{\delta}}^{m}$ such that
$a_{\delta}\to a\in D$ we have
$H_{\delta}(a_{\delta})\to h(a)$. Equivalently, the values of $H_{\delta}$
on type $1$ and $i$ corners respectively converge to the real and
imaginary parts of $h$. %

For notational convenience and concreteness, when we take $z\in D$ to be the argument of an s-holomorphic function $H_\delta$ defined on $\mathcal V_{D_\delta}^{cm}$, we will take a closest medial vertex $z_\delta \in \mathcal V_{D_\delta}^m$ to $z$  and  evaluate it at $z_\delta$. Then %
the convergence is often \emph{uniform} on
a compact set $K\subset D$, i.e. $\left|H_{\delta}(z_{\delta})-h(z)\right|$
is small uniformly in $z\in K$.

By a slight abuse of notation, we will then use the notation $\alpha=a^o$ and $\zeta=z^p$ both for the the s-orientations of $a$ or $z$, and those of $a_\delta$ and $z_\delta$.

\begin{theorem}[{\cite[Theorem 91]{hon2010}}]
\label{thm:fermion-fermion-observable-convergence}As $\delta\to0$,
we have
\[
\frac{1}{\delta}\left(F_{\Omega_{\delta}}^{\alpha}-F_{\mathbb{C}_{\delta}}^{\alpha}\right)\left(\zeta\right)\to\left(f_{\Omega}^{\alpha}-f_{\mathbb{C}}^{\alpha}\right)\left(\zeta\right)\,,
\]
 uniformly for $a,z$ away from $\partial\Omega$.
\end{theorem}
Now let $\alpha:=a_{}^{o}$ and $\zeta:=z^{p}$ be s-oriented corners
on $\left[\mathbb{C}_{1},0\right]$. For $\delta>0$, denote by $\delta\alpha:=\left(\delta a\right)^{o}$
and $\delta\zeta:=\left(\delta z\right)^{p}$ their corresponding
scaled versions on $\left[\mathbb{C}_{\delta},0\right]$ and, for
sufficiently small $\delta$, on $\left[\Omega_{\delta},0\right]$.
Recall that $G,\tilde{G}^{\pm}$ are the discrete analogues of $\nor\sqrt{z},\nori\sqrt{z}$,
respectively, introduced in Definition~\ref{def:aux-functions}.
\begin{theorem}
\label{thm:spin-fermion-observable-convergence}For $\alpha,\zeta$ any s-oriented corners or medial vertices in $[\mathbb C_1,0]$, we have as $\delta\to 0$,
\begin{alignat*}{1}
\frac{1}{\delta}\left(F_{\left[\Omega_{\delta},0\right]}^{\delta\alpha}-F_{\left[\mathbb{C}_{\delta},0\right]}^{\delta\alpha}\right)\left(\delta\zeta\right)\to & \,2\cdot \norin\re\mathcal{A}_{\Omega}\cdot \left(C_{\alpha} \re \left[ i\sqrt{p}G_{[\mathbb{C}_{1},0]}(z)\right] - C_\zeta \re \left[ i\sqrt{o}G_{[\mathbb{C}_{1},0]}(a)\right] \right) \\
 & +2\cdot \norin\im\mathcal{A}_{\Omega}\cdot  \left(C_{\alpha} \re \left[ i\sqrt{p}\tilde{G}^{-}_{[\mathbb{C}_{1},0]}(z)\right] - C_\zeta \re \left[ i\sqrt{o}\tilde{G}^{-}_{[\mathbb{C}_{1},0]}(a)\right] \right)\,,
\end{alignat*}
where, for $\alpha_{0}=\frac{1}{2}^{o}$ as before, $\mathcal{A}_{\Omega}$
is the coefficient in the expansion
\[
h_{\Omega}^{d\alpha_{0}}(z)=\nor\left( \frac{1}{\sqrt{z}}+2\mathcal{A}_{\Omega}\sqrt{z}+O(|z|^{3/2})\right)\,.
\]
\end{theorem}
\begin{proof}
This statement is the consequence of results proved in \S\ref{subsec:observable-convergence}. Theorem \ref{thm:convergence-near-singularity} shows the following in the case where $\alpha,\zeta$ are both s-oriented corners: $\frac1\delta [H_{[\Omega_{\delta},0]}^{\delta\alpha}(\delta z) -H_{[\mathbb C_{\delta},0]}^{\delta\alpha}(\delta z)]$ converges to
\begin{alignat*}{1} \,&2\cdot \norin\cdot \left[ \re\mathcal{A}_{\Omega}\cdot \left(C_{\alpha} G_{[\mathbb{C}_{1},0]}(z) +i\sqrt{o} \left[\tilde{G}_{[\mathbb{C}_{1},0]}^{+}-\tilde{G}_{[\mathbb{C}_{1},0]}^{-}\right]( z) G_{[\mathbb{C}_{1},0]}(a)\right) \right. \\ &\left. + \im\mathcal{A}_{\Omega}\cdot  \left(C_{\alpha}  \tilde{G}^{-}_{[\mathbb{C}_{1},0]}(z) + i\sqrt{o} \left[\tilde{G}_{[\mathbb{C}_{1},0]}^{+}-\tilde{G}_{[\mathbb{C}_{1},0]}^{-}\right]( z)\tilde{G}^{-}_{[\mathbb{C}_{1},0]}(a) \right)\right]\,,
\end{alignat*}
where we used Corollary \ref{cor:displacement-scaling-explicit} in order to identify $\tilde C_\alpha$ with $C_\alpha$ defined by~\eqref{eq:def-C-alpha}. This formula also applies in the case where $z$ is a medial vertex by linearity. Reformulation in the form of the theorem statement follows from the definition of the real observable; we can then swap $\alpha,\zeta$ from the antisymmetry of the real observable and proceed to prove the result for the case where both are medial vertices.
\end{proof}
\begin{remark}\label{rem:A-omega}
It was shown in \cite[Lemma 2.21]{chelkak-hongler} that $\mathcal{A}_{\Omega}=\mathcal{A}_{[\Omega,0]}=-\frac{1}{4}\partial_{z}\log r_{\Omega}(z)\big|_{z=0}$
which is the logarithmic derivative of the conformal radius as viewed
from $0\in\Omega$. This is used in Theorem \ref{thm:spin-energy-multipoint-convergence} to identify the coefficient
in the first-order correction to spin weighted correlations with $-\frac 14 \partial_z \log r_{\Omega}(z)|_{z=0}$.
\end{remark}

\subsection{\label{subsec:observable-convergence}Observable Convergence: Proofs}

We now prove the aforementioned observable convergence theorems, in
particular, the convergence of the full-plane observables and Theorem
\ref{thm:spin-fermion-observable-convergence}.

A tricky point of this subsection is that the explicit coefficient $C_\alpha$ of Definition \ref{def:continuous-full-plane-spin-fermion-observable}, in the claimed limit $h_{\left[\mathbb{C},0\right]}^{d\alpha}(z)=C_\alpha/\sqrt{z}$ of the full-plane spinor $H_{\left[\mathbb{C}_{\delta},0\right]}^{\delta\alpha}$, can only be identified as such once all convergence results are proven; until then we introduce a recursively defined stand-in $\tilde{C}_{\alpha}$ and $\tilde{h}_{\left[\mathbb{C},0\right]}^{d\alpha}(z):=\tilde{C}_{\alpha}/\sqrt{z}$. We then show the equality $\tilde C_\alpha= C_\alpha$ at the end of this subsection in Corollary \ref{cor:displacement-scaling-explicit}.

\subsubsection{\label{subsec:full-plane-convergence}Convergence of the Full-plane
Observables}

We first state direct extensions of some results in \cite{chelkak-hongler}
regarding convergence of the full-plane observables and functions. In \cite{chelkak-hongler},  the renormalization factor $\vartheta(\delta):=\Hm_{0}^{
\mathbb C_\delta \setminus \mathbb R_{<0}}({2\delta}{\lfloor (2\delta)^{-1} \rfloor})$ is used; we are able to calculate the constant explicitly thanks to Proposition~\ref{prop:harmonic-measure-explicit-appendix}. Specifically, writing $N:=\lfloor (2\delta)^{-1} \rfloor$, $\vartheta(\delta)=\frac{1}{2}\cdot\frac{3}{4}\cdots\frac{2N-1}{2N}=\frac{(2N)!}{4^{N}N!^2}$ and Stirling's approximation shows $\vartheta(\delta)\sim\frac{1}{\sqrt{\pi N}}\sim\sqrt{\frac{2\delta}{\pi}}$.
\begin{lemma}
\label{lem:chi-base-convergence}For $\alpha\in\mathcal{V}_{\left[\mathbb{C}_{1},0\right]}^{cm}$, 
we have $\sqrt{\frac{\pi}{{2\delta}}}H_{\left[\mathbb{C}_{\delta},0\right]}^{\delta\alpha}(z)\xrightarrow{\delta\downarrow0}\tilde{h}_{\left[\mathbb{C},0\right]}^{d\alpha}(z):=\frac{\tilde{C}_{\alpha}}{\sqrt{z}}$
uniformly on compact subsets in $\mathbb{C}\setminus\{0\}$ for some $\tilde{C}_\alpha\in\mathbb{R}$, which can be computed recursively.
\end{lemma}
\begin{proof}
\cite[Lemma 2.14]{chelkak-hongler} provides the case where $\alpha=\alpha_0=\frac{1}{2}^{o},o=(e^{2\cdot \pi i})^2$ ($\tilde{C}_{\alpha_0}=\nor$ in our normalization).
The other cases easily follow from the recursive constructions given
in Proposition \ref{prop:real-line-full-plane} and Corollary \ref{cor:full-plane-observable-construction}.
The fact that $\tilde{C}_{\alpha}\in\mathbb{R}$ is inductively apparent from
the fact that for any source point $\alpha$, the function $H_{\left[\mathbb{C}_{\delta},0\right]}^{\delta\alpha}(z)$
vanishes if $z$ is an imaginary corner of the positive real line, or a real
corner of the negative real line.
\end{proof}

\begin{lemma}\label{lem:g-convergence}
We have that $\sqrt{\frac{\pi}{2\delta}}G_{\left[\mathbb{C}_{\delta},0\right]}(z)\xrightarrow{\delta\downarrow0}\nor\sqrt{z}$
and $\sqrt{\frac{\pi}{2\delta}}\tilde{G}_{\left[\mathbb{C}_{\delta},0\right]}^{\pm}(z)\xrightarrow{\delta\downarrow0}\nori\sqrt{z}$
uniformly on compact subsets in $\mathbb{C}\setminus\{0\}$.
\end{lemma}
\begin{proof}
By \cite[Lemma 2.17]{chelkak-hongler}, we have convergence of the
real part of $G_{\left[\mathbb{C}_{\delta},0\right]}$. The lemma
follows by rotation and multiplication by $i$.
\end{proof}

\subsubsection{\label{subsec:bounded-domain-convergence}Convergence of the Bounded
Domain Observables}

We begin by proving the following convergence result, which is a simple
generalization of \cite[Theorem 2.18]{chelkak-hongler}. It provides
a local estimate near the monodromy point $0$ which will be crucial to the proof of the
general global convergence.

As we have introduced $\tilde{h}_{\left[\mathbb{C},0\right]}^{d\alpha}$ as the counterpart of $h_{\left[\mathbb{C},0\right]}^{d\alpha}$ where $C_{\alpha}$ was replaced by $\tilde{C}_{\alpha}$ in Lemma \ref{lem:chi-base-convergence}, we define $\tilde{h}_{\left[\Omega,0\right]}^{d\alpha_{0}}$ using Definition \ref{def:continuous-spin-fermion-observable} where $C_{\alpha}$ in $h_{\left[\Omega,0\right]}^{d\alpha}$ is replaced by $\tilde{C}_{\alpha}$ (i.e. $\tilde{h}_{\left[\Omega,0\right]}^{d\alpha_{0}}(z)=\sqrt{\varphi '(z)}\frac{\tilde{C}_\alpha}{\sqrt{\varphi(z)}}$). Note $C_{\alpha_0}=\tilde{C}_{\alpha_0}=\nor$. Recall also that we defined $\mathcal{A}_{\Omega}$
as the coefficient in the expansion,
\[
h_{\left[\Omega,0\right]}^{d\alpha_{0}}(z)=\tilde{h}_{\left[\Omega,0\right]}^{d\alpha_{0}}(z)=\nor\left(\frac{1}{\sqrt{z}}+2\mathcal{A}_{\Omega}\sqrt{z}+O(\left|z\right|^{3/2})\right)\,.
\]

\begin{lemma}
\label{lem:base-near-case}For $\alpha_{0}=\frac{1}{2}^{o},\sqrt{o}=1$,
and a corner or medial vertex $a$ on $\left[\mathbb{C}_{1},0\right]$,
\[
H_{[\Omega_{\delta},0]}^{\dagger\delta\alpha_{0}}(\delta a)=H_{[\Omega_{\delta},0]}^{\delta\alpha_{0}}(\delta a)-H_{[\mathbb{C}_{\delta},0]}^{\delta\alpha_{0}}(\delta a)=\left(2\re\mathcal{A}_\Omega\cdot G_{[\mathbb{C}_{\delta},0]}+2\im\mathcal{A}_\Omega\cdot\tilde{G}_{[\mathbb{C}_{\delta},0]}^{-}\right)(\delta a)+o(\delta)\,.
\]
\end{lemma}
\begin{proof}
We closely follow the strategy in Section 3.5 of \cite{chelkak-hongler}.
Note that
\[
H_{[\Omega_{\delta},0]}^{\dagger\delta\alpha_{0}}-\left(2\re\mathcal{A}_\Omega\cdot G_{[\mathbb{C}_{\delta},0]}+2\im\mathcal{A}_\Omega\cdot\tilde{G}_{[\mathbb{C}_{\delta},0]}^{-}\right)
\]
 is s-holomorphic, so it suffices to show that it is $o(\delta)$
on real and imaginary corners and propagate.

Recalling the symmetrized and antisymmetrized observables $S_{[\Omega_\delta,0]}^{\alpha}:=\frac{1}{2}\left[H_{\left[\Omega_{\delta},0\right]}^{\alpha}+H_{\overline{\left[\Omega_{\delta},0\right]}}^{\bar{\alpha}}\right], A_{[\Omega_\delta,0]}^{\alpha}:=\frac{1}{2}\left[H_{\left[\Omega_{\delta},0\right]}^{\alpha}-H_{\overline{\left[\Omega_{\delta},0\right]}}^{\bar{\alpha}}\right]
$
and $\Lambda_{\delta}=\Omega_{\text{\ensuremath{\delta}\ }}\cap\overline{\Omega_{\delta}}$
from Definition \ref{def:sym-antisym-observables}, define the following
functions, s-holomorphic everywhere on $\left[\Lambda_{\delta},0\right]$:
\begin{align*}
S_{\delta} & :=S_{[\Omega_{\delta},0]}^{\delta\alpha_{0}}-S_{\left[\mathbb{C}_{\delta},0\right]}^{\delta\alpha_{0}}-2\operatorname{Re}\mathcal{A}_\Omega\cdot G_{\left[\mathbb{C}_{\delta},0\right]}\,,\\
A_{\delta} & :=A_{[\Omega_{\delta},0]}^{\delta\alpha_{0}}-A_{\left[\mathbb{C}_{\delta},0\right]}^{\delta\alpha_{0}}-2\operatorname{Im}\mathcal{A}_\Omega\cdot\tilde{G}_{[\mathbb{C}_{\delta},0]}^{-}\,.
\end{align*}

Given that $S_{\delta}+A_{\delta}=H_{[\Omega_{\delta},0]}^{\dagger\delta\alpha_{0}}-\left(2\re\mathcal{A}_\Omega\cdot G_{[\mathbb{C}_{\delta},0]}+2\im\mathcal{A}_\Omega\cdot\tilde{G}_{[\mathbb{C}_{\delta},0]}\right)$,
it remains to estimate the real and imaginary parts $S_{\delta}^{1}:=S_{\delta}\vert_{\mathcal{V}_{\left[\Lambda_{\delta},0\right]}^{1}\cap\mathbb{X}^{+}},S_{\delta}^{i}:=S_{\delta}\vert_{\mathcal{V}_{\left[\Lambda_{\delta},0\right]}^{i}\cap\mathbb{Y}^{+}}$
and $A_{\delta}^{1}:=A_{\delta}\vert_{\mathcal{V}_{\left[\Lambda_{\delta},0\right]}^{1}\cap\mathbb{Y}^{+}},A_{\delta}^{i}:=A_{\delta}\vert_{\mathcal{V}_{\left[\Lambda_{\delta},0\right]}^{i}\cap\mathbb{X}^{+}}$. Without loss of generality
we show the $o(\delta)$ estimate for $S_{\delta}^{1}(\delta a)$, where $S_{\delta}^{1}$ is harmonic in the slit domain $\mathcal{V}_{\left[\Lambda_{\delta},0\right]}^{1}\cap\mathbb{X}^{+}$ and vanishes on $\mathcal{V}_{\left[\Lambda_{\delta},0\right]}^{1}\cap\mathbb{R}_{<0}$.

Define the discrete circle $w(r):=\{z\in\operatorname{Dom}(S_{\delta}^{1}):r<|z|<r+5\delta\}$
for small $r>0$. The same twist of the discrete Beurling estimate
(\cite[Theorem 1]{lali2004}) as in \cite[Lemma 3.3]{chelkak-hongler}
or our proof of Theorem \ref{thm:spin-fermion-infinite-vol-lim}
gives $\Hm{}_{\{\delta a\}}^{\mathbb{X}_{\delta}^{+}}(z)\leq C\delta^{1/2}|z|^{-1/2}$.
By reversibility of the simple random walk we have $\Hm_{w(r)}^{\mathbb{X}_{\delta}^{+}}(\delta a)\leq C\delta^{1/2}r^{-1/2}$,
which gives an estimate of a harmonic function identically 1 on $w(r)$
and vanishing on the slit. Comparing this with $S_{\delta}^{1}$ on
$w(r)$ and applying the maximum principle in the interior gives 
\begin{alignat*}{1}
\left|S_{\delta}^{1}\left(\delta a\right)\right| & \leq C\delta^{\frac{1}{2}}r^{-\frac{1}{2}}\sup_{w(r)}\left|S_{\delta}^{1}\right|.
\end{alignat*}

Now by convergence of $\sqrt{\frac{\pi}{2\delta}} H_{[\Omega_{\delta},0]}^{\delta\alpha_{0}}$
to $h_{[\Omega,0]}^{d\alpha_0}$, away from the singularity, as $\delta \to 0$ (see Theorem~2.16 of~\cite{chelkak-hongler}), we have 
\begin{alignat*}{1}
\sqrt{\frac{\pi}{2\delta}}S_{\delta}^{1}(z)\to & \re\left[\frac{1}{2}\left(\tilde{h}_{\left[\Omega,0\right]}^{d\alpha_{0}}(z)+\overline{\tilde{h}_{\left[\Omega,0\right]}^{d\alpha_{0}}(\bar{z})}\right)-\tilde{h}_{\left[\mathbb{C},0\right]}^{d\alpha_{0}}(z)-2\mathcal{A}_\Omega\re\sqrt{z}\right]\\
 & =O(|z|^{3/2})\,.
\end{alignat*}
Here we used the fact that $\tilde{h}_{\overline{\left[\Omega,0\right]}}^{\overline{d\alpha_{0}}}(\cdot)=\overline{\tilde{h}_{\left[\Omega,0\right]}^{d\alpha_{0}}(\bar{\cdot})}$
since the right hand side is the unique solution to the boundary value
problem in Remark \ref{rem:continuous-riemann-hilbert}. Thus, we
have $\left|S_{\delta}^{1}(\delta a)\right|\leq C'\delta r$ and since
$r$ is arbitrary we have $S_{\delta}^{1}(\delta a)=o(\delta)$ as
$\delta\to0$.

The estimate
follows analogously for $S_{\delta}^{i},A_{\delta}^{1},A_{\delta}^{i}$, since they share the following properties which were the two properties needed to deduce that $S_\delta^1(\delta a)=o(\delta)$ above:
\begin{enumerate}
\item they are harmonic functions on their respective slit domains and vanish
on the slits; we might extend the slit by a point, specifically the slit for $A_{\delta}^{i}$ includes $\frac{\delta}{2} \in \mathcal{V}_{\left[\Lambda_{\delta},0\right]}^{i}\cap\mathbb{X}^{+}$ given that $A_{\delta}(\frac{\delta}{2})=A^i_{\delta}(\frac{\delta}{2})=0$ (which also implies that $A^1_\delta$ is harmonic at $-\frac{\delta}{2}$). 
\item they are $O(|z|^{3/2})$ on the discrete circle $w(r)$ defined above. \qedhere
\end{enumerate}
\end{proof}
Now we prove global convergence for general source point $\alpha\neq \alpha_0$, away from $0$.
\begin{proposition}
If $\alpha=a^{o}$ is an s-oriented corner or medial vertex on $\left[\mathbb{C}_{1},0\right]$; then as $\delta \to 0$, we have $\sqrt{\frac{\pi}{2\delta}}H_{\left[\Omega_{\delta},0\right]}^{\delta\alpha}(z)\xrightarrow{\delta\downarrow0}\tilde{h}_{\left[\Omega,0\right]}^{d\alpha}(z)$
uniformly on compact subsets of $\Omega\setminus\{0\}$.
\end{proposition}
\begin{proof}
\cite[Theorem 2.16]{chelkak-hongler} proves the case where $\alpha=\alpha_{0}=\frac{1}{2}^{o},\sqrt{o}=1$;
they consider the square integrals $\tilde{Q}_{\delta}^{\delta\alpha_{0}}$
and $Q_{\delta}^{\dagger\delta\alpha_{0}}$ as introduced in Section
\ref{subsec:square-integral} (in their notation, $H_{\delta},H_{\delta}^{\dagger}$),
and show that they converge to the continuous functions $\re\int\left(\tilde{h}_{\left[\Omega,0\right]}^{\delta\alpha_{0}}\right)^{2}$
and $\re\int\left(\tilde{h}_{\left[\Omega,0\right]}^{\delta\alpha_{0}}-\tilde{h}_{\left[\mathbb{C},0\right]}^{\delta\alpha_{0}}\right)^{2}$,
which implies convergence of the integrand (see Section 3.4 of
\cite{chelkak-hongler}). In our notation, they show uniform boundedness, and thus equicontinuity, of $\tilde{Q}_{\delta}^{\delta\alpha_{0}},Q_{\delta}^{\dagger\delta\alpha_{0}}$ in $\delta$ in each subdomain of $\Omega_\delta$, away from the boundary and $0$. Their subsequential limits are then identified with the continuous square integrals above.

We argue that a similar strategy works for all $\alpha$. In fact,
the only difference here is that the sub-harmonicity of $Q_{\delta}^{\dagger\delta \alpha}$
at $0$ fails, since in general $H_{\left[\Omega_{\delta},0\right]}^{\dagger\delta\alpha}(\frac{\delta}{2})\neq0$.
Sub-harmonicity of the square integral is used twice in the proof of \cite[Theorem 2.16]{chelkak-hongler}:
it shows that their $H_{\delta}$ (which corresponds to our $Q_{\delta}^{\dagger\delta\alpha}$) is uniformly bounded near $0$, which is needed to identify the limit, and it is needed to apply the maximum principle near $0$ and obtain \cite[Lemma 3.10]{chelkak-hongler}. We will thus reproduce these two bounds, except that we replace $Q_{\delta}^{\dagger\delta\alpha}$ in their argument by a modified version $Q_{\delta}^{\dagger\dagger\delta\alpha}$, which we now introduce.

By Lemma \ref{lem:base-near-case} and rescaling,
\begin{alignat*}{1}
H_{\left[\Omega_{\delta},0\right]}^{\dagger\delta\alpha_{0}} & (\delta a)=\delta\left(2\re\mathcal{A}_\Omega\cdot G_{[\mathbb{C}_{1},0]}(a)+2\im\mathcal{A}_\Omega\cdot\tilde{G}_{[\mathbb{C}_{1},0]}^{-}(a)+o(1)\right)\,,
\end{alignat*}
and by antisymmetry between the two arguments, it is easy to see that
$H_{[\Omega_{\delta},0]}^{\dagger\delta\alpha}(\delta a_{0})=\nor(A_{\alpha}+o(1))i\delta$
as $\delta \to 0$ for some constant $A_{\alpha}$. Then the modified observable 
\[
H_{[\Omega_{\delta},0]}^{\dagger\dagger\delta\alpha}:=H_{[\Omega_{\delta},0]}^{\dagger\delta\alpha}-\norin\frac{H_{[\Omega_{\delta},0]}^{\dagger\delta\alpha}(\delta a_{0})}{ \delta i}\tilde{G}_{\left[\mathbb{C}_{\delta},0\right]}^{+}=H_{[\Omega_{\delta},0]}^{\dagger\delta\alpha}-(A_{\alpha}+o(1))\tilde{G}_{\left[\mathbb{C}_{\delta},0\right]}^{+}
\]
 is everywhere s-holomorphic and satisfies $H_{[\Omega_{\delta},0]}^{\dagger\dagger\delta\alpha}(\delta a_{0})=0$,
so its integral $Q_{\delta}^{\dagger\dagger\delta\alpha}:=\mathbb{I}_{\delta}\left[H_{[\Omega_{\delta},0]}^{\dagger\dagger\delta\alpha}\right]$
is sub-harmonic on faces and super-harmonic on vertices. It converges
to 
\begin{align*}
\re\int\left(\tilde{h}_{\left[\Omega,0\right]}^{d\alpha}(z)-\tilde{h}_{\left[\mathbb{C},0\right]}^{d\alpha}(z)-\nori A_{\alpha}\sqrt{z}\right)^{2}dz
\end{align*}
uniformly away from $0$, so both the discrete observable and the
continuous function are single-valued and bounded near $0$. This
fact, alternatively to its analogue for $Q_{\delta}^{\dagger\delta\alpha}$,
also implies \cite[(2.8)]{chelkak-hongler}, which identifies the singularity at $0$. The analog of
\cite[Lemma 3.10]{chelkak-hongler} also easily follows by replacing $H_{\delta}^{\dagger}$ in their proof by $Q_{\delta}^{\dagger\dagger\delta\alpha}$ as defined above.
\end{proof}
We now analyze the observable near the singularity at $0$, finally
giving the proof of the main convergence theorem. Balancing the two
discrete analogues $\tilde{G}_{[\mathbb{C}_{\delta},0]}^{\pm}$ of
$\nori \sqrt{z}$ to create a harmonic function which is amenable
to the methods of analysis used thus far is crucial to the proof.
\begin{theorem}[Convergence Content of Theorem \ref{thm:spin-fermion-observable-convergence}]
\textup{\label{thm:convergence-near-singularity}If $z\in\mathcal{V}_{\left[\mathbb{C}_{1},0\right]}^{c}$
and $\alpha=a^{o}$ is an s-oriented corner on $\left[\mathbb{C}_{1},0\right]$,
then as $\delta\to0$, 
\begin{alignat*}{1}
H_{[\Omega_{\delta},0]}^{\dagger\delta\alpha}(\delta z)= & \,\norin\tilde{C}_{\alpha}\left(2\re\mathcal{A}_\Omega\cdot G_{[\mathbb{C}_{\delta},0]}+2\im\mathcal{A}_\Omega\cdot\tilde{G}_{[\mathbb{C}_{\delta},0]}^{-}\right)(\delta z)\\
 & +i\norin\sqrt{o}\left(2\re\mathcal{A}_\Omega\cdot G_{[\mathbb{C}_{1},0]}+2\im\mathcal{A}_\Omega\cdot\tilde{G}_{[\mathbb{C}_{1},0]}^{-}\right)(a)\left[\tilde{G}_{[\mathbb{C}_{\delta},0]}^{+}-\tilde{G}_{[\mathbb{C}_{\delta},0]}^{-}\right](\delta z)\\
 & +o(\delta)\,.
\end{alignat*}
}
\end{theorem}
\begin{proof}
We argue that the same strategy as in the proof of Lemma \ref{lem:base-near-case}
works here. Indeed, after defining
\begin{alignat*}{1}
S_{\delta}:= & \,S_{[\Omega_{\delta},0]}^{\delta\alpha}-S_{\left[\mathbb{C}_{\delta},0\right]}^{\delta\alpha}-2\cdot\norin\tilde{C}_{\alpha}\operatorname{Re}\mathcal{A}_\Omega\cdot G_{\left[\mathbb{C}_{\delta},0\right]}\\
A_{\delta}:= & \,A_{[\Omega_{\delta},0]}^{\delta\alpha}-A_{\left[\mathbb{C}_{\delta},0\right]}^{\delta\alpha}-2\cdot\norin\tilde{C}_{\alpha}\operatorname{Im}\mathcal{A}_{\Omega}\cdot\tilde{G}_{[\mathbb{C}_{\delta},0]}^{-}\\
 & -i\norin \sqrt{o}\left(2\re\mathcal{A}_\Omega\cdot G_{[\mathbb{C}_{1},0]}+2\im\mathcal{A}_\Omega\cdot\tilde{G}_{[\mathbb{C}_{1},0]}^{-}\right)(a)\left[\tilde{G}_{[\mathbb{C}_{\delta},0]}^{+}-\tilde{G}_{[\mathbb{C}_{\delta},0]}^{-}\right]\,,
\end{alignat*}
one sees that the real and imaginary parts of these two functions satisfy properties (1) and (2) at the end of the proof of Lemma~\ref{lem:base-near-case}, sufficient to conclude that $S_\delta^1,S_\delta^i,A_\delta^1,A_\delta^i$ evaluated at $(\delta z)$ are $o(\delta)$. The additional term in $A_\delta$ above is needed
because we require $A_{\delta}(\delta a_{0})=0$;
\begin{alignat*}{1}
\left[A_{[\Omega_{\delta},0]}^{\delta\alpha}-A_{\left[\mathbb{C}_{\delta},0\right]}^{\delta\alpha}\right](\delta a_{0}) & =H_{\left[\Omega_{\delta},0\right]}^{\dagger\delta\alpha}(\delta a_{0})=-\sqrt{o}H_{\left[\Omega_{\delta},0\right]}^{\dagger\delta\alpha_{0}}(\delta a)\\
 & =-\delta\sqrt{o}\left(2\re\mathcal{A}_\Omega\cdot G_{[\mathbb{C}_{1},0]}+2\im\mathcal{A}_\Omega\cdot\tilde{G}_{[\mathbb{C}_{1},0]}^{-}\right)(a)+o(\delta)\,,
\end{alignat*}
so we insert $\tilde{G}^{+}$ to cancel out this nonzero value, then
adjust the coefficient in front of $\tilde{G}^{-}$ to match the global
limit.
\end{proof}

We are now in position to explicitly characterize $\tilde{C}_{\alpha}$. The following is a consequence of Theorem~\ref{thm:convergence-near-singularity}.
\begin{corollary}
\label{cor:displacement-scaling-explicit} For every $\alpha\in \mathcal V_{[\mathbb C_1,0]}^{cm}$, the constant $\tilde{C}_{\alpha}$
defined in Lemma~\ref{lem:chi-base-convergence} is given explicitly by
\[
\tilde{C}_{\alpha}=-\re \left[ i\sqrt{o}\left(\tilde{G}_{[\mathbb{C}_{1},0]}^{+}-\tilde{G}_{[\mathbb{C}_{1},0]}^{-}\right)(a)\right]=:C_\alpha\,,
\]
and therefore $\tilde{h}_{\left[\mathbb{C},0\right]}^{d\alpha}=h_{\left[\mathbb{C},0\right]}^{d\alpha}$.
\end{corollary}
\begin{proof}
First we suppose $a$ is a real or imaginary corner; note that in this case the real part operator in the definition of $C_\alpha$ or real observables is superfluous. In Theorem \ref{thm:convergence-near-singularity}, let $z=\frac{3}{2}$, say on $\mathbb {X}^+$, and let $\zeta = z^p$ with $\sqrt{p}=i$. Since $F_{[\Omega_{\delta},0]}^{\dagger\delta\alpha,\delta \zeta}=\re i\sqrt{p} H_{[\Omega_{\delta},0]}^{\dagger\delta\alpha}(\delta z)=i\sqrt{p} H_{[\Omega_{\delta},0]}^{\dagger\delta\alpha}(\delta z)$,

\begin{alignat*}{1}
(\norin \delta)^{-1} F_{[\Omega_{\delta},0]}^{\dagger\delta\alpha,\delta \zeta}\to& \tilde{C}_{\alpha}\left[i\sqrt{p} \left(2\re\mathcal{A}_\Omega\cdot G_{[\mathbb{C}_{1},0]}+2\im\mathcal{A}_\Omega\cdot\tilde{G}_{[\mathbb{C}_{1},0]}^{-}\right) (z)\right]\\
&- C_\zeta \left[i\sqrt{o}\left(2\re\mathcal{A}_\Omega\cdot G_{[\mathbb{C}_{1},0]}+2\im\mathcal{A}_\Omega\cdot\tilde{G}_{[\mathbb{C}_{1},0]}^{-}\right)(a)\right]\,,\\
(\norin \delta)^{-1}F_{[\Omega_{\delta},0]}^{\dagger\delta\zeta,\delta \alpha}\to& \tilde{C}_{\zeta}\left[i\sqrt{o} \left(2\re\mathcal{A}_\Omega\cdot G_{[\mathbb{C}_{1},0]}+2\im\mathcal{A}_\Omega\cdot\tilde{G}_{[\mathbb{C}_{1},0]}^{-}\right) (a)\right]\\
&- C_\alpha \left[i\sqrt{p}\left(2\re\mathcal{A}_\Omega\cdot G_{[\mathbb{C}_{1},0]}+2\im\mathcal{A}_\Omega\cdot\tilde{G}_{[\mathbb{C}_{1},0]}^{-}\right)(z)\right]\,.
\end{alignat*}
Since the two limits should differ only by sign, the result follows by using that $G_{[\mathbb{C}_{1},0]}(z)\neq 0, \tilde{G}^{\pm}_{[\mathbb{C}_{1},0]}(z)=C_\zeta=0$, and the fact that, by our recursive construction,
$\tilde{C}_{\zeta}=0$ (see Proposition~\ref{prop:real-line-full-plane}, and in particular, Proposition~\ref{prop:recursive-coefficient-proof}).

In the case where $a$ is a medial vertex, note that by s-holomorphicity of $H_{\left[\mathbb{C}_{\delta},0\right]}^{\delta\alpha}$ and antisymmetry of its real counterpart, in their arguments, we can express $\tilde{C}_\alpha$ as a linear combination of $\tilde{C}_{\beta^\pm}$, where $\beta^\pm$ are adjacent real and imaginary (s-oriented) corners. Notice that this linear combination is exactly mirrored in the case of $C_\alpha$, since $\tilde{G}_{[\mathbb{C}_{1},0]}^{+}-\tilde{G}_{[\mathbb{C}_{1},0]}^{-}$ is s-holomorphic; thus the desired equality holds.
\end{proof}

\section{\label{subsec:main-theorem-proofs}Proofs of Theorems}

In this section, we complete the proofs of Theorems \ref{thm:spin-sym-energy}\textendash \ref{thm:spin-sens-energy}
and Corollary \ref{thm:spin-pattern}.

\subsection{Spin-symmetric Fields}

We first prove Theorem (\ref{thm:spin-sym-energy}), establishing
the conformal invariance of spin-symmetric fields.
\begin{definition}
Suppose $a\neq z$ are medial vertices on $\mathbb{C}_{1}$. For s-orientations $o,p$ on $a,z$ respectively, write $\alpha=a^{o},\zeta=z^{p}$.
Define
\begin{align*}
F_{\mathbb{C}_{1}}^{\alpha,\zeta} & =\re\left[i\sqrt{p}H_{\mathbb{C}_{1}}^{\alpha}(z)\right]\,,\\
E_{\mathbb{C}_{1}}^{\alpha,\zeta} & =\re\left[\frac{i\sqrt{p}\overline{\sqrt{o}}}{\sqrt{2}\pi}\right]\,,
\end{align*}
where $H_{\mathbb{C}_{1}}^{\alpha}(z)$ was explicitly defined in (\ref{eq:fermion-fermion-explicit-1}).

Now let $\left\{ e_{k}\right\} =\left\{ e_{1},\ldots,e_{n}\right\} $
be a collection of distinct edges of $\mathbb{C}_{1}$. Set
\[
\left(x_{1},\ldots,x_{n},x_{n+1},\ldots,x_{2n}\right):=\left(e_{1}^{+},\ldots e_{n}^{+},e_{n}^{-},\ldots,e_{1}^{-}\right)\,,
\]
where $e_{j}^{+}:=e_{j}^{o^{+}}$ and $e_{j}^{-}:=e_{j}^{o^{-}}$
denote a choice of opposite s-orientations of $e_{j}$ such that
$o^{+}=e^{\pi i}o^{-}$. 

We write $\mathbf{F}^{\left\{ e_{k}\right\} }$ to denote the $2n\times2n$
antisymmetric matrix with entries $\left(\mathbf{F}^{\left\{e_k\right\}}\right)_{jk}:=F_{\mathbb{C}_{1}}^{x_{j},x_{k}}$ for $j+k\neq2n+1$, and $\left(\mathbf{F}^{\left\{e_k\right\}}\right)_{jk}:=0$ on the anti-diagonal $j+k=2n+1$; we also set $\mathbf{E}^{\left\{ e_{k}\right\} }$ to be the matrix taking values $\left(\mathbf{E}^{\left\{ e_{k}\right\} }\right)_{jk}:=E_{\mathbb{C}_{1}}^{x_{j},x_{k}}$. 
Now define 
\begin{alignat*}{1}
\mathcal{P}^{\left\{ e_{k}\right\} } & :=(-2)^{n}\mathrm{Pf}\left(\mathbf{F}^{\left\{ e_{k}\right\} }\right)\,,\\
\mathcal{Q}^{\left\{ e_{k}\right\} } & :=(-2)^{n}D_{\mathbf{E}^{\left\{ e_{k}\right\} }}\mathrm{Pf}(\mathbf{F}^{\left\{ e_{k}\right\} })\,,
\end{alignat*}
where $D_{\mathbf{E}^{\left\{ e_{k}\right\} }}\mathrm{Pf}$ denotes
the directional derivative of the Pfaffian function in the direction
of $\mathbf{E}^{\left\{ e_{k}\right\} }$.
\end{definition}
\begin{remark}
The values $\mathcal{P}^{\left\{ e_{k}\right\} }$ and $\mathcal{Q}^{\left\{ e_{k}\right\} }$
only depend on the unordered collection $\left\{ e_{k}\right\} $
since they arise as limits of the Pfaffian from Proposition \ref{prop:energy-correlations-pfaffian-observable},
which has an interpretation as a physical quantity only depending
on $\left\{ e_{k}\right\} $.
\end{remark}
\begin{theorem}[Restatement of Theorem \ref{thm:spin-sym-energy}]
\label{thm:energy-multipoint-convergence} Let $\left\{ e_{k}\right\} _{k=1}^{n}$
be a collection of $n$ distinct edges of $\mathbb{C}_{1}$. As $\delta\to0$,
we have
\[
\mathbb{E}_{\Omega_{\delta}}\left[\prod_{e\in\left\{ e_{k}\right\} }\epsilon({\delta e})\right]=\mathcal{P}^{\left\{ e_{k}\right\} }+\delta \cdot r_{\Omega}^{-1} \left(0\right)\cdot\mathcal{Q}^{\left\{ e_{k}\right\} }+o\left(\delta\right)\,,
\]
where $r_{\Omega}\left(z\right)$ is the conformal radius of $\Omega$
at $z\in\Omega$, defined $r_{\Omega}(z):=\left|\varphi'(0)\right|$
where $\varphi:\mathbb{D}\to\Omega$ is a conformal map with $\varphi(0)=z$. 
\end{theorem}
\begin{proof}[\textbf{\emph{Proof of Theorem \ref{thm:energy-multipoint-convergence}}}]
By Proposition \ref{prop:energy-correlations-pfaffian-observable},
we have that 
\[
\mathbb{E}_{\Omega_{\delta}}\left[\prod_{e\in\left\{ e_{k}\right\} }\epsilon({\delta e})\right]=\left(-1\right)^{n}2^{n}\mathrm{Pf}\left(\mathbf{F}_{\Omega_{\delta}}^{\left[\left\{ \delta e_{k}\right\} \right]}\right)\,.
\]
Write $\mathbf{F}_{\Omega_{\delta}}^{\left[\left\{ \delta e_{k}\right\} \right]}=\mathbf{F}_{\mathbb{C}_{\delta}}^{\left[\left\{ \delta e_{k}\right\} \right]}+\left[\mathbf{F}_{\Omega_{\delta}}^{\left[\left\{ \delta e_{k}\right\} \right]}-\mathbf{F}_{\mathbb{C}_{\delta}}^{\left[\left\{ \delta e_{k}\right\} \right]}\right]$.
By scale invariance, $\mathbf{F}_{\mathbb{C}_{\delta}}^{\left[\left\{ \delta e_{k}\right\} \right]}=\mathbf{F}_{\mathbb{C}_{1}}^{\left[\left\{ e_{k}\right\} \right]}$,
which, by definition, satisfies 
$\mathbf{F}_{\mathbb{C}_{1}}^{\left[\left\{ e_{k}\right\} \right]}=\mathbf{F}^{\left\{ e_{k}\right\} }$.

By Theorem \ref{thm:fermion-fermion-observable-convergence}, for
any $\alpha=a^{o},\zeta=z^{p}$, if we set $\delta\alpha:=\left(\delta a\right)^{o}$,
$\delta\zeta:=\left(\delta z\right)^{p}$ we can calculate 
\[
\lim_{\delta\to0}\frac{1}{\delta}\left[F_{\Omega_{\delta}}^{\delta\alpha,\delta\zeta}-F_{\mathbb{C}_{\delta}}^{\delta\alpha,\delta\zeta}\right]=\left[f_{\Omega}^{0^o}-f_{\mathbb{C}}^{0^o}\right](0^{p})=E_{\mathbb{C}_{1}}^{\alpha,\zeta}r_{\Omega}^{-1}\left(0\right)\,,
\]
and the result follows from Taylor expansion of $\mathrm{Pf}\left(\mathbf{F}_{\Omega_{\delta}}^{\left[\left\{ \delta e_{k}\right\} \right]}\right)=\mathrm{Pf}\left(\mathbf{F}^{\left\{ e_{k}\right\} }+\delta\mathbf{E}^{\left\{ e_{k}\right\} }r_{\Omega}^{-1}\left(0\right)+o(\delta)\right)$.
\end{proof}

\subsection{Spin-antisymmetric Fields}

We now generalize the above proof to spin-antisymmetric fields, proving
Theorem \ref{thm:spin-sens-energy}. 

For any s-oriented medial vertex or corner $\zeta=z^{p}$, we
introduce the real quantity
$$G_{[\mathbb C_1,0]}(\zeta) = \re[i\sqrt{p}G_{[\mathbb C_1,0]}(z)]\,,$$
and define the real quantities
$\tilde G^{\pm}_{[\mathbb C_1,0]} (\zeta)$ analogously; 
$G_{\left[\mathbb{C}_{1},0\right]}(z),\tilde{G}_{\left[\mathbb{C}_{1},0\right]}^\pm(z)$ were defined in Section \ref{sec:full-plane-observables}. 
\begin{definition}\label{def:antisymmetric-matrix-entries}
Suppose $a\neq z$ are medial vertices on $\left[\mathbb{C}_{1},0\right]$. For s-orientations $o,p$ respectively on $a,z$, write $\alpha=a^{o},\zeta=z^{p}$
for the s-oriented medial vertices. Set 
\begin{alignat*}{1}
F_{\left[\mathbb{C}_{1},0\right]}^{\alpha,\zeta} & =\re\left[i\sqrt{p}H_{\left[\mathbb{C}_{1},0\right]}^{\alpha}(z)\right]\,,\\
E_{\left[\mathbb{C}_{1},0\right]}^{\alpha,\zeta} & :=2\cdot\norin\bigg([G_{[\mathbb C_1,0]} - i\tilde G^-_{[\mathbb C_1,0]}](\alpha)[\tilde G_{[\mathbb{C}_{1},0]}^+ -\tilde G^-_{[\mathbb C_1,0]}](\zeta) \\
& \,\,\,\,\,\,\qquad\qquad -[G_{[\mathbb C_1,0]} - i\tilde G^-_{[\mathbb C_1,0]}](\zeta)[\tilde G_{[\mathbb{C}_{1},0]}^+ -\tilde G^-_{[\mathbb C_1,0]}](\alpha)\bigg)\,.
\end{alignat*}

Let $\left\{ e_{k}\right\} =\left\{ e_{1},\ldots,e_{n}\right\} $
be a collection of distinct edges of $\mathbb{C}_{1}$. Let $\tilde{e}_{1},\ldots,\tilde{e}_{n}$
be a choice of lifts of $e_{1},\ldots,e_{n}$ to $\left[\mathbb{C}_{1},0\right]$.
Set 
\[
\left(x_{1},\ldots,x_{2n}\right):=\left(\tilde{e}_{1}^{+},\ldots,\tilde{e}_{n}^{+},\tilde{e}_{n}^{-},\ldots,\tilde{e}_{1}^{-}\right)\,,
\]
where $\tilde{e}_{j}^{+}:=\tilde{e}_{j}^{o^{+}}$ and $\tilde{e}_{j}^{-}:=\tilde{e}_{j}^{o^{-}}$
denote a choice of opposite s-orientations of $\tilde{e_{j}}$
such that $o^{+}=e^{\pi i}o^{-}$. 

Define $\mathbf{F}_{\left[0\right]}^{\left\{ e_{k}\right\} }$ as the
$2n\times2n$ antisymmetric matrix with entries $\left(\mathbf{F}_{\left[0\right]}^{\left\{ e_{k}\right\} }\right)_{jk}:=F_{\left[\mathbb{C}_{\delta},0\right]}^{x_{j}x_{k}}$ for $j+k\neq2n+1$, and $\left(\mathbf{F}_{\left[0\right]}^{\left\{ e_{k}\right\} }\right)_{jk}:=0$ on the anti-diagonal, $j+k=2n+1$. Define also $\mathbf{E}_{\left[0\right]}^{\left\{ e_{k}\right\} }$ the $2n\times 2n$ antisymmetric matrix given by
$\left(\mathbf{E}_{\left[0\right]}^{\left\{ e_{k}\right\} }\right)_{jk}:=E_{\left[\mathbb{C}_{\delta},0\right]}^{x_{j}x_{k}}$. Let
 \begin{eqnarray*}
\mathcal{P}_{\left[0\right]}^{\left\{ e_{k}\right\} } & := & (-2)^{n}\mathrm{Pf}\left(\mathbf{F}_{\left[0\right]}^{\left\{ e_{k}\right\} }\right)\,,\\
\mathcal{Q}_{\left[0\right]}^{\left\{ e_{k}\right\} } & := & (-2)^{n}D_{\mathbf{E}_{\left[0\right]}^{\left\{ e_{k}\right\} }}\mathrm{Pf}\left(\mathbf{F}_{\left[0\right]}^{\left\{ e_{k}\right\} }\right)\,.
\end{eqnarray*}
\end{definition}
\begin{remark}
By the same reasoning as in the spin-symmetric case, the values $\mathcal{P}^{\left\{ e_{k}\right\} }$
and $\mathcal{Q}^{\left\{ e_{k}\right\} }$ only depend on the unordered
collection $\left\{ e_{k}\right\} $.
\end{remark}
\begin{theorem}[Restatement of Theorem \ref{thm:spin-sens-energy}]
\label{thm:spin-energy-multipoint-convergence}Let $\left\{ e_{k}\right\} _{k=1}^{n}$
be a set of $n$ edges of $\mathbb{C}_{1}$. For every $1\leq k\leq n$, the quantity $\mu_{e_k}$ defined in~\eqref{eq:mu-e} exists and, independently of s-orientation $o_k$ on $e_k$, is
\begin{align}\label{eq:mu-e-explicit}
\mu_{ e_k} = \sqrt {o_k}\left[H_{[\mathbb C_1,0]}^{e_k^{o_k}}(e_{k+}^{o_k}) + H_{[\mathbb C_1,0]}^{e_k^{o_k}}(e_{k-}^{o_k})\right]\,,
\end{align}
so that $\epsilon_{[0]}(\delta e_k)$ is a well-defined random variable for every $k$, and as $\delta\to0$,
\[
\frac{\mathbb{E}_{\Omega_{\delta}}\left[\sigma_{0}\prod_{e\in\left\{ e_{k}\right\} }\epsilon_{[0]}(\delta e)\right]}{\mathbb{E}_{\Omega_{\delta}}\left[\sigma_{0}\right]}=\mathcal{P}_{\left[0\right]}^{\left\{ e_{k}\right\} }+\delta\cdot\re\left[-\frac{1}{4}\partial_{z}\log r_{\Omega}\left(z\right)\Big|_{z=0} \cdot\mathcal{Q}_{\left[0\right]}^{\left\{ e_{k}\right\} } \right] +o\left(\delta\right),
\]
where $\partial_{z}=\frac{1}{2}(\partial_{x}-i\partial_{y})$ if $z=x+iy$.
In particular, it follows from the results of \cite{chelkak-hongler}
that
\[
\mathbb{E}_{\Omega_{\delta}}\left[\sigma_{0}\prod_{a\in\left\{ e_{k}\right\} }\epsilon_{[0]}({\delta e})\right]=0+\mathcal{C}\cdot\delta^{\frac{1}{8}}\cdot\mathcal{P}_{\left[0\right]}^{\left\{ e_{k}\right\} }\cdot2^{\frac{1}{4}}\cdot r_{\Omega}(0)^{-\frac{1}{8}}+o\left(\delta^{\frac{1}{8}}\right)\,,
\]
where $\mathcal{C}$ is a constant
given explicitly by Eq.\ (1.1) of \cite{chelkak-hongler}.
\end{theorem}
\begin{proof}[\textbf{\emph{Proof of Theorem \ref{thm:spin-energy-multipoint-convergence}}}]
The expression for $\mu_{e}$ for every $e\in \mathcal E_{\mathbb C_{1}}$ was given by Remark~\ref{rem:mu-a}. Now by Proposition \ref{prop:spin-energy-fermion-correlation-pfaffian-observable},
we have that 
\[
\frac{\mathbb{E}_{\Omega_{\delta}}\left[\sigma_{0}\prod_{e\in\left\{ e_{k}\right\} }\epsilon_{[0]}(\delta e)\right]}{\mathbb{E}_{\Omega_{\delta}}[\sigma_{0}]}=\left(-1\right)^{n}2^{n}\mathrm{Pf}\left(\mathbf{F}_{\left[\Omega_{\delta},0\right]}^{\left\{ \delta e_{k}\right\} }\right),
\]
Write $\mathbf{F}_{\left[\Omega_{\delta},0\right]}^{\left\{ \delta e_{k}\right\} }=\mathbf{F}_{\left[\mathbb{C}_{\delta},0\right]}^{\left\{ \delta e_{k}\right\} }+\left[\mathbf{F}_{\left[\Omega_{\delta},0\right]}^{\left\{ \delta e_{k}\right\} }-\mathbf{F}_{\left[\mathbb{C}_{\delta},0\right]}^{\left\{ \delta e_{k}\right\} }\right]$. 

As before, by scale invariance, $\mathbf{F}_{\left[\mathbb{C}_{\delta},0\right]}^{\left\{ \delta e_{k}\right\} }=\mathbf{F}_{\left[\mathbb{C}_{1},0\right]}^{\left\{ e_{k}\right\} }$,
and by definition, $\mathbf{F}_{\left[\mathbb{C}_{1},0\right]}^{\left\{ e_{k}\right\} }=\mathbf{F}_{\left[0\right]}^{\left\{ e_{k}\right\} }$.
By Theorem \ref{thm:spin-fermion-observable-convergence}, for any
$\alpha=a^{o},\zeta=z^{p}$, if we set $\delta\alpha:=\left(\delta a\right)^{o}$,
$\delta\zeta:=\left(\delta z\right)^{p}$ we have 
\[
\lim_{\delta\to0}\frac{1}{\delta}F_{\left[\Omega_{\delta},0\right]}^{\dagger\delta\alpha,\delta\zeta}= - \re\left[\frac{1}{4}\partial_{z}\log r_{\Omega}\left(z\right)\Big|_{z=0} \cdot E_{\left[0\right]}^{\alpha,\zeta}\right]\,,
\]
 and hence we get
\[
\lim_{\delta\to0}\frac{1}{\delta}\left[\mathbf{F}_{\left[\Omega_{\delta},0\right]}^{\left\{ \delta a_{k}\right\} }-\mathbf{F}_{\left[\mathbb{C}_{\delta},0\right]}^{\left\{ \delta a_{k}\right\} }\right]=- \re\left[\frac{1}{4}\partial_{z}\log r_{\Omega}\left(z\right)\Big|_{z=0}  \cdot \mathbf{E}_{\left[0\right]}^{\left\{ a_{k}\right\} }\right]\,.
\]
The first result then follows from Taylor expansion as in the proof
of the previous theorem, and the second follows by multiplying through
by the conformally covariant expansion of $\mathbb{E}_{\Omega_{\delta}}[\sigma_{0}]=\mathcal{C}\cdot2^{\frac{1}{4}}\cdot r_{\Omega}(0)^{-\frac{1}{8}}\cdot\delta^{\frac{1}{8}}+o\left(\delta^{\frac{1}{8}}\right)$
given by \cite{chelkak-hongler}. 
\end{proof}

\subsection{Spin Pattern Probabilities}

Finally, we prove Corollary \ref{thm:spin-pattern} as a consequence
of the above two proofs. 
\begin{proof}[\textbf{\emph{Proof of Corollary \ref{thm:spin-pattern}}}]
We begin by proving the corollary for the spin-symmetric pattern
fields. For a subset $\mathcal{F}\subset\mathcal{F}_{\mathbb{C}_{1}}$,
let $\mathcal{B}$ be the set of all edges separating two adjacent
faces in $\mathcal{F}$. To any spin-symmetric pattern $\pm\rho$
on $\mathcal{F}$, we can associate an edge subset $B\subset\mathcal{B}$
via the usual low-temperature expansion (see Figure \ref{fig:ising-grid}
and Section \ref{subsec:low-temperature-expansion}). Denote the
collection of such edge subsets in $\mathcal{B}$ that are associated
to a spin-symmetric pattern on $\mathcal{F}$ by $P_{\mathcal{F}}(\mathcal{B})\subset P(\mathcal{B})$.
We will index the $2^{\left|\mathcal{F}\right|-1}$-dimensional vector
$P^{\mathcal{F}}=(P_{B})_{B\in P_{\mathcal{F}}(\mathcal{B})}$ of
the probabilities of all spin-symmetric patterns by corresponding
edge subsets. 

Given any such an edge subset $B$, we can calculate a spin-symmetric
correlation $E_{B}=\mathbb{E}_{\Omega_{\delta}}\left[\epsilon(\delta B)\right]=\mathbb{E}_{\Omega_{\delta}}\left[\prod_{e\in B}\epsilon(\delta e)\right]$
and also form another vector $E^{\mathcal{F}}=(E_{B})_{B\in P_{\mathcal{F}}(\mathcal{B})}$
of dimension $2^{\left|\mathcal{F}\right|-1}$.

Clearly, every correlation function $\mathbb{E}_{\Omega_{\delta}}\left[\epsilon(B)\right]$
can be expressed as a linear combination of probabilities of the $2^{|\mathcal{F}|-1}$
spin-symmetric patterns on $\mathcal{B}$. 

Thus we have a $2^{|\mathcal{F}|-1}\times2^{|\mathcal{F}|-1}$
matrix $(EP^{\mathcal{F}})_{BB'}=\prod_{e\in B}(\mu-(2\boldsymbol{1}_{\{e\in B'\}}-1))$,
such that $E^{\mathcal{F}}=(EP^{\mathcal{F}})P^{\mathcal{F}}$. One
can check by hand that the matrix has inverse given by: 
\begin{align*}
(PE^{\mathcal{F}})_{B'B}= & \frac{1}{2^{(2^{|\mathcal{F}|-1})}}(-1)^{\sum_{e\in\mathcal{B}}\boldsymbol{1}_{\{e\in B\}}\oplus\boldsymbol{1}_{\{e\in B'\}}}\prod_{e\in B}(\mu+(2\boldsymbol{1}_{\{e\in B'\}}-1))\,.
\end{align*}
Applying the inverse to $E^{\mathcal{F}}$, consisting of conformally
covariant spin-symmetric correlations from Theorem \ref{thm:spin-sym-energy},
yields the desired result for spin-symmetric patterns. 

For the spin-antisymmetric patterns, an analogous approach but conditioning
on $\sigma_{0}=\pm1$ and replacing $\mu$ by $\mu_e$, combined with the conformally covariant expansion
$\mathbb{E}_{\Omega_{\delta}}[\sigma_{0}]=\mathcal{C}\cdot2^{\frac{1}{4}}\cdot r_{\Omega}(0)^{-\frac{1}{8}}\cdot\delta^{\frac{1}{8}}+o\left(\delta^{1/8}\right)$
from \cite{chelkak-hongler}, gives the desired result.
\end{proof}

\appendix

\section{\label{app:full-plane-observables-construct}The Harmonic Measure on
$\mathbb{C}_{1}\backslash\mathbb{R}_{>0}$}

We start this section by giving an analytic formula for the harmonic
measure of the tip of the slit plane. This uses Fourier series techniques,
which was inspired by \cite{chho2016}. Using the formula, we prove
that the auxiliary functions $G_{\left[\mathbb{C}_{1},0\right]},\tilde{G}_{\left[\mathbb{C}_{1},0\right]}^{\pm}$
are discrete holomorphic.

Since the slit-plane harmonic measures appear as different translations
of the same function, we present the following prototype:
\begin{proposition}\label{prop:harmonic-measure-explicit-appendix}
For $s+ik\in(1+i)\mathbb{Z}^{2}$, the function $H_{0}$ given by,
\begin{align}
H_{0}(z):=\Hm{}_{0}^{(1+i)\mathbb{Z}^{2}\setminus\mathbb{Z}_{\geq0}}(z=s+ik) & =\frac{1}{2\pi}\int_{-\pi}^{\pi}\frac{C^{|k|}(\theta)}{\sqrt{1-e^{-2i\theta}}}e^{-is\theta}d\theta\,,\label{eq:harmonic-fourier}
\end{align}
where $C(\theta):={\displaystyle \frac{\cos\theta}{1+|\sin\theta|}}$
and the square root is evaluated on the principal branch, is the unique discrete
harmonic function on the diagonal slit plane $(1+i)\mathbb{Z}^{2}\setminus\mathbb{Z}_{\geq0}$
with boundary values 1 at the origin and 0 elsewhere on $\mathbb{Z}_{\geq0}$
and as $\to\infty$.
\end{proposition}
\begin{proof}
We first state two Fourier expansions, thanks to the generalized binomial
theorem:
\begin{alignat}{1}
\frac{1}{\sqrt{1-e^{-2i\theta}}} & =\sum_{n=0}^{\infty}(-1)^{n}{-\frac{1}{2} \choose n}e^{-2ni\theta}=1+\frac{1}{2}e^{-2i\theta}+\frac{3}{8}e^{-4i\theta}+\frac{5}{16}e^{-6i\theta}+\ldots\,,\label{eq:harmonic-measure-binomial}\\
\frac{\left|\sin\theta\right|}{\sqrt{1-e^{-2i\theta}}} & =\sqrt{\frac{\sin^{2}\theta}{1-e^{-2i\theta}}}=\frac{1}{2}\sqrt{1-e^{2i\theta}}=\frac{1}{2}\sum_{n=0}^{\infty}(-1)^{n}{\frac{1}{2} \choose n}e^{2ni\theta}\,.\nonumber 
\end{alignat}
The first identity immediately gives the boundary values on $\mathbb{Z}_{\geq0}$.
Discrete harmonicity when $k\neq0$ follows directly from the structure
of the integrand, and when $k=0$ values of the discrete Laplacian
correspond to the Fourier coefficients in the second identity of Eq.
\ref{eq:harmonic-measure-binomial}, so vanishes on $\mathbb{Z}_{<0}$
since there are no negative Fourier modes.

For the decay at infinity, it suffices to show $|H_{0}(s+ik)|\to0$ as $\left|k\right|\to\infty$ uniformly
in $s$ and as $\left|s\right|\to\infty$ for fixed $k$. Note the latter is just the Riemann-Lebesgue Lemma. For the former,
we can use dominated convergence since $\frac{\left|C^{|k|}(\theta)\right|}{\sqrt{\left|1-e^{-2i\theta}\right|}}\downarrow0$
pointwise a.e. as $|k|\to\infty$ and $\frac{\left|C^{|k|}(\theta)\right|}{\sqrt{\left|1-e^{-2i\theta}\right|}}\leq\frac{1}{\sqrt{\left|1-e^{-2i\theta}\right|}}=\frac{1}{\sqrt{2|\sin\theta|}}$,
which is integrable. 
\end{proof}
Now, the above characterization of the harmonic measure of the tip
of the slit plane leads to a recursive construction of harmonic measures
of other points on the slit, as discussed in Proposition \ref{prop:real-line-full-plane}.
Suppose $H_{n}:=\Hm{}_{2n}^{(1+i)\mathbb{Z}^{2}}$ denotes the harmonic
measure of the point $2n$ on the slit plane $(1+i)\mathbb{Z}^{2}\backslash\mathbb{R}_{\geq0}$.
We have the following recursion formula: 
\begin{alignat}{1}
H_{n}(m) & =H_{n-1}(m-2)-H_{n-1}(-2)H_{0}(m)\label{eq:recursive-coefficients}\\
 & =:H_{0}(m-2n)-X_{1}H_{0}(m-2n+2)-X_{2}H_{0}(m-2n+4)-\cdots-X_{n}H_{0}(m)\,.\nonumber 
\end{alignat}

The coefficients $X_{n}:=H_{n-1}(-2)$ can be used to calculate the recursive coefficients in Proposition \ref{prop:real-line-full-plane} and the scaling limit coefficients $C_{\alpha}$ in Section \ref{subsec:Continuous-Full-Plane-Observable}
explicitly. We now give a simple formula for $X_{i}$.
\begin{proposition}
\label{prop:recursive-coefficient-proof}For all $i\geq1$, we have
that $X_{n}=\frac{H_{0}(-2n+2)}{2n}$. Consequently,
\[
1-X_{1}-X_{2}-\cdots-X_{n}=H_{0}(-2n)\,.
\]
\end{proposition}
\begin{proof}
Define the generating functions
\begin{alignat*}{1}
X(z) & :=\sum_{n=1}^{\infty}X_{n}z^{n}=\frac{1}{2}z+\frac{1}{8}z^{2}+\frac{1}{16}z^{3}+\cdots\,,\\
F(z) & :=\sum_{n=0}^{\infty}H_{0}(-2i)z^{n}=1+\frac{1}{2}z+\frac{3}{8}z^{2}+\cdots=\frac{1}{\sqrt{1-z}}\,.
\end{alignat*}

Note that Eq. (\ref{eq:recursive-coefficients}) implies a convolution
identity by taking $m=0$,
\[
H_{0}(-2m)=\sum_{n=0}^{m}X_{n}H_{0}(-2m+2n)\,,
\]
and setting $X_{0}=0$. Thus $XF=F-1$, and $X=1-\sqrt{1-z}=\sum_{n=1}^{\infty}(-1)^{n+1}{1/2 \choose n}z^{n}$.
Given that $H_{0}(-2n)=(-1)^{n}{-1/2 \choose n}$ by Eq. (\ref{eq:harmonic-measure-binomial}),
both results are straightforward.
\end{proof}
\begin{corollary}
The auxiliary functions, $G_{\left[\mathbb{C}_{1},0\right]},\tilde{G}_{\left[\mathbb{C}_{1},0\right]}^{\pm}$,
defined in Definition \ref{def:aux-functions} are discrete holomorphic
on $[\mathbb{C}_{1},0]$.
\end{corollary}
\begin{proof}
Without loss of generality, we will show discrete holomorphicity of
$G_{\left[\mathbb{C}_{1},0\right]}$ at a type-${\lambda}$ corner
$z=s+i\left(k+\frac{1}{2}\right)\in\mathbb{X}^{+}\cap\mathbb{Y}^{+}$
($k\geq0)$. By the fact that 
\[
\sum_{n=-\infty}^{\infty}\left[\Hm_{3/2}^{\mathbb{X}_{1}^{1}}(z+\frac{1+i}{2}+2n)-\Hm_{3/2}^{\mathbb{X}_{1}^{1}}(z-\frac{1+i}{2}+2n)\right]=\lim_{\theta\to0} C^k(\theta)\frac{C(\theta)-1}{\sqrt{1-e^{-2i\theta}}}=0\,,
\]
and discrete holomorphicity of $H^{\alpha_0}_{\left[\mathbb{C}_{1},0\right]}$, we have
\begin{align*}
 & G_{\left[\mathbb{C}_{1},0\right]}(z-\frac{1-i}{2})-G_{\left[\mathbb{C}_{1},0\right]}(z+\frac{1-i}{2})\\
  & \qquad=\nori\sum_{n=1}^{\infty}\left[\Hm_{1/2}^{\mathbb{Y}_{1}^{i}}(z-\frac{1-i}{2}+2n)-\Hm_{1/2}^{\mathbb{Y}_{1}^{i}}(z+\frac{1-i}{2}+2n)\right]\\
 & \qquad=-\nori\sum_{n=1}^{\infty}\left[\Hm_{3/2}^{\mathbb{X}_{1}^{1}}(z+\frac{1+i}{2}+2n)-\Hm_{3/2}^{\mathbb{X}_{1}^{1}}(z-\frac{1+i}{2}+2n)\right]\\
 & \qquad=\nori\sum_{n=0}^{\infty}\left[\Hm_{3/2}^{\mathbb{X}_{1}^{1}}(z+\frac{1+i}{2}-2n)-\Hm_{3/2}^{\mathbb{X}_{1}^{1}}(z-\frac{1+i}{2}-2n)\right]\\
 & \qquad=i\left[G_{\left[\mathbb{C}_{1},0\right]}(z+\frac{1+i}{2})-G_{\left[\mathbb{C}_{1},0\right]}(z-\frac{1+i}{2})\right]\,,
\end{align*}
as desired.
\end{proof}

Now we show that $G$ in fact has some rotation symmetry, which can be exploited to recursively compute its values as outlined in Remark \ref{rem:g-explicit}. The proof relies on the same kind of analysis of discrete harmonic functions as in the proof of Lemma \ref{lem:base-near-case} and therefore we omit some of the details.
\begin{proposition}\label{prop:G-rotation}On $\mathbb C_\delta$, the following holds:
$e^{\pi i/4} \cdot G_{\left[\mathbb{C}_{\delta},0\right]}(e^{\pi i/2}z) = \frac12\left[\tilde G_{\left[\mathbb{C}_{\delta},0\right]}^++\tilde G_{\left[\mathbb{C}_{\delta},0\right]}^-\right](z)$.
\end{proposition}
\begin{proof}
We will write $L_\delta(z)$ for the left hand side, and $R_\delta(z)=\left[ k\tilde G_{\left[\mathbb{C}_{\delta},0\right]}^++(1-k)\tilde G_{\left[\mathbb{C}_{\delta},0\right]}^-\right](z)$ for an as yet undefined real number $k$. Both are s-holomorphic functions, and by Lemma \ref{lem:g-convergence}, both $\sqrt\frac{\pi}{2 \delta} L_\delta(z),\sqrt\frac{\pi}{2 \delta} R_\delta(z)$ converge to $\nori \sqrt z$ on compact subsets away from $0$ as $\delta \to 0$.

First, it is straightforward to check that $G_{\left[\mathbb{C}_{\delta},0\right]}$ vanishes on $\bar\lambda$-corners on the upper half of the imaginary axis, and on $\lambda$-corners on the lower half by the symmetry between its real and imaginary parts (in addition, $G_{\left[\mathbb{C}_{\delta},0\right]}(\bar z) = \overline{G_{\left[\mathbb{C}_{\delta},0\right]}(z)}$). So $L_\delta(z)$ has zero real part on the positive real line, and zero imaginary part on the negative real line. This is also true for $R_\delta(z)$. Now, $L_\delta\left( \pm \frac\delta2 \right)$ is not necessarily zero, so we will choose (taking $\frac\delta2$ on $\mathbb X^+$) $k=-iL_\delta\left( \frac \delta 2\right)$. Then $L_\delta(z)-R_\delta(z)$ is zero at $\frac\delta2$, so we have harmonicity at $-\frac\delta2$ by Remark \ref{rem:double-cover-complex-analysis}.

Without loss of generality, consider the restriction of $\sqrt{\frac{\pi}{2 \delta}}\left[ L_\delta- R_\delta\right]$ to $\mathcal{V}_{\left[\mathbb{C}_{\delta},0\right]}^{i} \cap \mathbb X^+$. It vanishes on the boundary $\mathbb R_{<0}$ and $\frac\delta2$, and as $\delta\to 0$ the values on the boundary $w(1)$ of the discrete ball $B_1(0) \cap \mathcal{V}_{\left[\mathbb{C}_{\delta},0\right]}^{i}$ decays as $o(1)$. By the discrete Beurling estimate (see proof of Lemma \ref{lem:base-near-case}) we can bound $\left| \sqrt{\frac{\pi}{2 \delta}}\left[ L_\delta- R_\delta\right](z\delta)\right| $ for any $z\in\mathcal{V}_{\left[\mathbb{C}_{1},0\right]}^{i} \cap \mathbb X^+$ from above by $C\delta^{1/2}o(1)$. Since by definition $\sqrt{\frac{\pi}{2 \delta}}\left[ L_\delta- R_\delta\right](z\delta)=\sqrt{\frac{\pi \delta}{2}}\left[L_1 - R_1\right] (z)$, $\left[L_1 - R_1\right] (z)=0$, and thus $L_\delta = R_\delta$.

Then we conclude $k=\frac12$ since $(1-k)\tilde G_{\left[\mathbb{C}_{\delta},0\right]}^-\left( e^{\pi i}\frac{\delta}{2} \right)=L_\delta\left( e^{\pi i}\frac{\delta}{2} \right) = iL_\delta\left(\frac{\delta}{2} \right)=ik\tilde G_{\left[\mathbb{C}_{\delta},0\right]}^+\left(\frac{\delta}{2} \right)$ and $-\tilde G_{\left[\mathbb{C}_{\delta},0\right]}^-\left( e^{\pi i}\frac{\delta}{2} \right)=\tilde G_{\left[\mathbb{C}_{\delta},0\right]}^+\left(\frac{\delta}{2} \right)=G_{\left[\mathbb{C}_{\delta},0\right]}\left(\frac{3\delta}{2} \right)$ again by the symmetry between real and imaginary parts of  $G_{\left[\mathbb{C}_{\delta},0\right]}$.
\end{proof}

\section{\label{app:contour-weights}Contour Weights}

Here we prove the well-definedness of the spin-fermionic contour weights introduced
in full generality in Section \ref{subsec:multi-fermion-observables}.

Recall from Sections \ref{subsec:disorder-lines} and \ref{subsec:multi-fermion-observables} the
definition of $\gamma\in\mathcal{C}_{\Omega_{\delta}}^{\alpha_{1},\ldots,\alpha_{2n}}$
and the admissible choices of walks $\{\Gamma(\gamma)\}$ associated
to it. Moreover recall the definition of the multipoint observable
$F_{[\Omega_{\delta},a]}$ from Definition~\ref{def:multipoint-observable-2}. 
\begin{proposition}
\label{prop:weight-welldefined}For any collection of distinct oriented medial vertices
$\alpha_{1},...,\alpha_{2n}$ and any $\gamma\in\mathcal{C}_{\Omega_{\delta}}^{\alpha_{1},\ldots,\alpha_{2n}}$,
for every two admissible choices of walks, $\Gamma(\gamma),\Gamma'(\gamma)$,
we have
\[
\left(-1\right)^{\ell\left(\gamma\setminus\cup\Gamma(\gamma)\right)}\prod_{\gamma^{\alpha_{j},\alpha_{k}}\in\Gamma\left(\gamma\right)}\mathrm{s}_{\alpha_{j},\alpha_{k}}\left(\gamma^{\alpha_{j},\alpha_{k}}\right)=\left(-1\right)^{\ell\left(\gamma\setminus\cup\Gamma'(\gamma)\right)}\prod_{\gamma^{\alpha_{j'},\alpha_{k'}}\in\Gamma\left(\gamma\right)}\mathrm{s}_{\alpha_{j'},\alpha_{k'}}\left(\gamma^{\alpha_{j'},\alpha_{k'}}\right)\,.
\]
As a result, the function $F_{[\Omega_{\delta},a]}$ is well-defined.
\end{proposition}
We will need the following two lemmas for the proof of the above proposition. 
\begin{lemma}
\label{lem:disjoint-loop-xor}If $A$, $B$ are unions of disjoint
loops in $\Omega_{\delta}$, $(-1)^{\ell(A\oplus B)}=(-1)^{\ell(A)}(-1)^{\ell(B)}$.

\begin{proof}
For each of $A$ and $B$, fill in the faces of the lattice with spins,
beginning with the plus boundary conditions, such that there is an
edge between two faces if and only if they differ in sign. For each
loop collection, we have the spin at zero $\sigma_{0}^{A}=(-1)^{\ell(A)}$
and $\sigma_{0}^{B}=(-1)^{\ell(B)}$. 

The result follows after noting that $A\oplus B$ is identified with
the spin configuration constructed by multiplying the spins of configurations
$A$ and $B$ pointwise.
\end{proof}
\begin{lemma}
\label{lem:cycles}Suppose $l$ walks $w_{1},\ldots,w_{l}\in\left\{ \gamma_{1},\ldots,\gamma_{n},\gamma'_{1},\ldots,\gamma'_{n}\right\} $
and $l$ distinct oriented medial vertices $\alpha^{1},\ldots,\alpha^{l}\in\left\{ \alpha_{1},\ldots,\alpha_{2n}\right\} $
form a cycle, i.e. $w_{1}$ connects (projections of) $\alpha^{1}$
with $\alpha^{2}$, $w_{2}$ connects $\alpha^{2}$ with $\alpha^{3}$,
\dots , $w_{l}$ connects $\alpha^{l}$ with $\alpha^{l+1}:=\alpha^{1}$.
Then we have,
\[
\left(-1\right)^{\ell(w_{1}\oplus\cdots\oplus w_{l})}=\mbox{s}_{\alpha^{1},\alpha^{2}}(w_{1})\cdots\mbox{s}_{\alpha^{l},\alpha^{1}}(w_{l})\,.
\]
\end{lemma}
\begin{proof}
Fix a half-line $\Lambda=e^{i\theta}\mathbb{R}_{\geq0}$ such that
it is not parallel to any edges and is disjoint from all $\alpha^{1},\ldots,\alpha^{l}$.
Note that $\Omega\setminus\Lambda$ lifts to $\left[\Omega,0\right]$
as two sheets. We now define two quantities: given a piecewise $C^{1}$
path in $\Omega\setminus\{0\}$, we can count the number of times
$N_{\Lambda}$ that the path crosses $\Lambda$; given two points
$a,z\in\left[\Omega,0\right]\setminus\Lambda$, we define $S_{\Lambda}^{\alpha,\zeta}:=1$
if they belong to the same sheet in $\left[\Omega,0\right]$ of $\Omega\setminus\Lambda$
and $S_{\Lambda}^{\alpha,\zeta}:=-1$ otherwise.

Concatenate the (possibly reversed) walks such that $w:=w_{1}\oplus w_{2}\oplus\cdots\oplus w_{l}$
is a continuous loop on $\Omega\setminus\{0\}$ starting from $\alpha^{1}$.
Then clearly $\ell(w_{1}\oplus\cdots\oplus w_{l})\equiv N_{\Lambda}(w)\mod2$.
Now it suffices to note that for any $j$, $(-1)^{N_{\Lambda}(w_{j})}\mbox{s}_{\alpha^{j},\alpha^{j+1}}(w_{j})=S_{\Lambda}^{\alpha^{j},\alpha^{j+1}}$
so that
\begin{alignat*}{1}
(-1)^{N_{\Lambda}(w)}\prod_{j=1}^{l}\mbox{s}_{\alpha^{j},\alpha^{j+1}}(w_{j}) & =\prod_{j=1}^{l}(-1)^{N_{\Lambda}(w_{j})}\mbox{s}_{\alpha^{j},\alpha^{j+1}}(w_{j})\\
 & =\prod_{j=1}^{l}S_{\Lambda}^{\alpha^{j},\alpha^{j+1}}=S_{\Lambda}^{\alpha^{1},\alpha^{1}}=1\,,
\end{alignat*}
from which the lemma follows.
\end{proof}
\end{lemma}
\begin{proof}[\textbf{\emph{Proof of Proposition \ref{prop:weight-welldefined}}}]
First observe that $\gamma_{1},\ldots,\gamma_{n},\gamma_{1}',\ldots,\gamma'_{n}$
and $a_{1},\ldots,a_{2n}$ are partitioned into disjoint cycles $P_{1},\ldots,P_{l'}$
in the sense of Lemma \ref{lem:cycles} (suppose each $P_{j}$ is
the resulting collection of loops of the form $w_{1}\oplus\cdots\oplus w_{l}$).
Note that $P_{1}\oplus\cdots\oplus P_{l'}=\cup\Gamma(\gamma)\oplus\cup\Gamma'(\gamma)$,
and thus $\left(\gamma\setminus\cup\Gamma(\gamma)\right)\oplus\left(\gamma\setminus\cup\Gamma'(\gamma)\right)\oplus P_{1}\oplus\cdots\oplus P_{l'}=\emptyset$.
By Lemma \ref{lem:disjoint-loop-xor}, 
\begin{alignat*}{1}
\left(-1\right)^{\ell\left(\gamma\setminus\cup\Gamma(\gamma)\right)} & \prod_{\gamma^{\alpha_{j},\alpha_{k}}\in\Gamma\left(\gamma\right)}\mathrm{s}_{\alpha_{j},\alpha_{k}}\left(\gamma^{\alpha_{j},\alpha_{k}}\right)\cdot\left(-1\right)^{\ell\left(\gamma\setminus\cup\Gamma'(\gamma)\right)}\prod_{\gamma^{\alpha_{j'},\alpha_{k'}}\in\Gamma\left(\gamma\right)}\mathrm{s}_{\alpha_{j'},\alpha_{k'}}\left(\gamma^{\alpha_{j'},\alpha_{k'}}\right)\\
= & (-1)^{\ell\left(\gamma\setminus\cup\Gamma(\gamma)\right)}(-1)^{\ell(\gamma\setminus\cup\Gamma'(\gamma))}\prod_{j=1}^{l'}(-1)^{\ell(P_{j})}=(-1)^{\ell(\emptyset)}=1\,.
\end{alignat*}
concluding the proof.
\end{proof}

\section{\label{sec:explicit-pattern-probabilities}Explicit Pattern Probabilities}

In this section we give an example using Theorem \ref{thm:spin-sym-energy} by computing explicitly
the infinite-volume limit of and first-order conformal correction
to a diagonal spin-spin correlation. On rotated lattices, $\Omega_{\delta}\subset\mathbb{C}_{\delta}$
with plus boundary, this corresponds to $\mathbb{E}_{\Omega_{\delta}}[\sigma_{0}\sigma_{2\delta}]$
and is a quantity that appears in the study of the lattice level Ising
stress tensor (\cite{benoist-hongler-stress}). We then explain how the similar computation would be done for an ``L" shaped spin-weighted correlation $\mathbb E_{\Omega_\delta}[\sigma_0 \sigma_{(1+i)\delta} \sigma_{2\delta}]/\mathbb E_{\Omega_\delta} [\sigma_0]$ and give the explicit values one gets from the explicit recursion process outlined to get values of the infinite-volume spin-fermion. 
\begin{corollary}\label{cor:diag-cor}
Consider the Ising model on $\Omega_{\delta}$
with plus boundary conditions and $0\in\mathcal{F}_{\Omega_{\delta}}$; then, as $\delta\to0$,
\begin{align*}
\mathbb{E}_{\Omega_{\delta}}[\sigma_{0}\sigma_{2\delta}]=\frac{2}{\pi}+\delta\cdot\frac{2}{\pi}\cdot r_{\Omega}^{-1}(0)+o(\delta)\,.
\end{align*}
\end{corollary}
\begin{proof}
We first observe that $\mathbb{E}_{\Omega_{\delta}}[\sigma_{0}\sigma_{2\delta}]=\mathbb{E}_{\Omega_{\delta}}[\sigma_{0}\sigma_{(1+i)\delta}\sigma_{(1+i)\delta}\sigma_{2\delta}]$
and 
\begin{align*}
\mathbb{E}_{\Omega_{\delta}}[\epsilon({a_{1}})\epsilon({a_{2}})]= & \frac{1}{2}+\mathbb{E}_{\Omega_{\delta}}[\sigma_{0}\sigma_{(1+i)\delta}\sigma_{(1+i)\delta}\sigma_{2\delta}] -\frac{\sqrt{2}}{2}(\mathbb{E}_{\Omega_{\delta}}[\sigma_{0}\sigma_{(1+i)\delta}]+\mathbb{E}_{\Omega_{\delta}}[\sigma_{(1+i)\delta}\sigma_{2\delta}])\,,
\end{align*}
where $a_{1}=\frac{1+i}{2},a_{2}=\frac{3+i}{2}$. By the first-order
Taylor expansion of \cite{hosm2013} (after rescaling) for the energy
density ($\mathbb{E}_{\Omega_{\delta}}[\epsilon({a_{1}})]=-\delta\cdot\frac{\sqrt{2}}{\pi}\cdot r_{\Omega}^{-1}(0)+o(\delta)$),
this implies that as $\delta\to0$,
\begin{align*}
\mathbb{E}_{\Omega_{\delta}}[\sigma_{0}\sigma_{2\delta}]= & \frac{1}{2}+\mathbb{E}_{\Omega_{\delta}}[\epsilon({a_{1}})\epsilon({a_{2}})]+\delta\cdot\frac{2}{\pi}\cdot r_{\Omega}^{-1}(0)+o(\delta)\,.
\end{align*}

From this it suffices to compute the first-order asymptotics of $\lim_{\delta\to0}\mathbb{E}_{\Omega_{\delta}}[\epsilon({a_{1}})\epsilon({a_{2}})]$.
In order to do so, we consider the antisymmetric matrix 
\begin{align*}
\mathbf{F}^{\left\{ e_{k}\right\} }= & \left(\begin{array}{cccc}
0 & F_{\mathbb{C}_{1}}^{a_{1}^{+}a_{2}^{+}} & F_{\mathbb{C}_{1}}^{a_{1}^{+}a_{2}^{-}} & F_{\mathbb{C}_{1}}^{\dagger a_{1}^{+}a_{1}^{-}}\\
 & 0 & F_{\mathbb{C}_{1}}^{\dagger a_{2}^{+}a_{2}^{-}} & F_{\mathbb{C}_{1}}^{a_{2}^{+}a_{1}^{-}}\\
 &  & 0 & F_{\mathbb{\mathbb{C}}_{1}}^{a_{2}^{-}a_{1}^{-}}\\
 &  &  & 0
\end{array}\right)\text{ }
= 
\left(\begin{array}{cccc}
0 & a & b & c\\
 & 0 & d & e\\
 &  & 0 & f\\
 &  &  & 0
\end{array}\right)
\end{align*}
where we choose orientations $o_{1}^{+}=e^{7\pi i/4}$ and
$o_{2}^{+}=e^{5\pi i/4}$ on $\mathbb{C}_{1}$.
Plugging in the explicit values from the observable defined in Eq.
(\ref{eq:fermion-fermion-explicit-1}) 
and using the definition $F_{\mathbb{C}_{1}}^{\alpha,\zeta}=\mbox{{Re}}[i\sqrt{p}H_{\mathbb{C}_{1}}^{\alpha,z}]$,
where we choose the principal branch of the square root, observe that
\begin{align*}
a=-\frac{1}{2\pi}+\frac{\sqrt{2}}{2\pi}+\frac{1}{4}\,,\qquad & b=-\frac{1}{4}+\frac{1}{2\pi}\,,\\
c=0\,,\qquad & d=0\,,\\
e=-\frac{1}{4}+\frac{1}{2\pi}\,,\qquad & f=\frac{1}{2\pi}+\frac{\sqrt{2}}{2\pi}-\frac{1}{4}\,.
\end{align*}
Combined with $\mbox{Pf}(\mathbf{F}^{\left\{ e_{k}\right\} })=af-be+cd$
and $\lim_{\delta\to0}\mathbb{E}_{\Omega_{\delta}}[\sigma_{0}\sigma_{2\delta}]=\frac{1}{2}+4\mbox{Pf}(\mathbf{F}^{\left\{ e_{k}\right\} })$
(where we applied Theorem~\ref{thm:spin-sym-energy} to $\mathbb{E}_{\Omega_{\delta}}[\epsilon({a_{1}})\epsilon({a_{2}})]$), 

\begin{align*}
\lim_{\delta\to0}\mathbb{E}_{\Omega_{\delta}}[\sigma_{0}\sigma_{2\delta}]=\mathbb{E}_{\mathbb{C}_{1}}[\sigma_{0}\sigma_{2\delta}]= & \frac{2}{\pi}\,.
\end{align*}
In order to compute the constant in the conformal correction, note
that the matrix $\mathbf{E}^{\left\{ e_{k}\right\} }$ is given by
$E_{\mathbb{C}_{1}}^{\alpha,\zeta}=\mbox{{Re}}\left[\frac{i\sqrt{p}\overline{\sqrt{o}}}{\sqrt{2}\pi}\right]$
so that, here,
\begin{align*}
\mathbf{E}^{\left\{ e_{k}\right\} }= & \left(\begin{array}{cccc}
0 & E_{\mathbb{C}_{1}}^{a_{1}^{+}a_{2}^{+}} & E_{\mathbb{C}_{1}}^{a_{1}^{+}a_{2}^{-}} & E_{\mathbb{C}_{1}}^{a_{1}^{+}a_{1}^{-}}\\
 & 0 & E_{\mathbb{C}_{1}}^{a_{2}^{+}a_{2}^{-}} & E_{\mathbb{C}_{1}}^{a_{2}^{+}a_{1}^{-}}\\
 &  & 0 & E_{\mathbb{\mathbb{C}}_{1}}^{a_{2}^{-}a_{1}^{-}}\\
 &  &  & 0
\end{array}\right)=\frac{1}{\sqrt{2}\pi}\left(\begin{array}{cccc}
0 & \frac{\sqrt{2}}{2} & \frac{\sqrt{2}}{2} & 1\\
 & 0 & 1 & \frac{\sqrt{2}}{2}\\
 &  & 0 & -\frac{\sqrt{2}}{2}\\
 &  &  & 0
\end{array}\right)\,.
\end{align*}
Inverting $\mathbf{F}^{\left\{ e_{k}\right\} }$ by hand, and using
the Pfaffian expansion formula $\mbox{Pf}(A+\delta B)=\mbox{Pf}(A)+\text{\ensuremath{\delta}}\mbox{Pf}(A)\mbox{Tr}(A^{-1}B)$,
 along with the above expressions, we see that, in fact,
$4D_{\mathbf{E}^{\left\{ e_{k}\right\} }}\mbox{Pf}(\mathbf{F}^{\left\{ e_{k}\right\} })=4\mbox{Pf}(\mathbf{F}^{\left\{ e_{k}\right\} })\mbox{Tr}((\mathbf{F}^{\left\{ e_{k}\right\} })^{-1}\mathbf{E}^{\left\{ e_{k}\right\} })= 0
$
which, combined with Theorem \ref{thm:spin-sym-energy} implies the
desired geometric correction.
\end{proof}

\begin{corollary}\label{cor:l-shaped}Consider the Ising model on $\Omega_\delta$  with plus boundary conditions and $0\in \mathcal F_{\Omega_\delta}$; then as $\delta \to 0$, 
$$\frac{\mathbb{E}_{\Omega_\delta}[\sigma_0\sigma_{(1+i)\delta}\sigma_{2\delta}]}{\mathbb{E}_{\Omega_\delta}[\sigma_0]}=2(\sqrt{2}-1)+\delta\cdot \frac{5\sqrt 2-7}{2}\cdot\re\left[ \partial_z \log r_\Omega(z)|_{z=0} \right]+o(\delta)
$$
\end{corollary}
\begin{proof}
First observe that the edge $e$ separating $\sigma_{(1+i)\delta}$ from $\sigma_{2\delta}$ has midpoint at $\delta a$ where $a = \frac 32 +\frac{i}2$. 
\begin{align*}
\frac{\mathbb E_{\Omega_\delta}[\sigma_0 \epsilon_{[0]}(\delta a)]}{\mathbb E_{\Omega_\delta} [\sigma_0]}= \frac{\mathbb E_{\Omega_\delta}[\sigma_0 \sigma_{(1+i)\delta} \sigma_{2\delta}]}{\mathbb E_{\Omega_\delta} [\sigma_0]} - \mu_{ a}\,.
\end{align*}
They by Eq.~\eqref{eq:mu-e-explicit}, picking an s-orientation $o$ on $a$, 
\begin{align*}
\mu_{ a} = \sqrt {o}[H_{[\mathbb C_1,0]}^{a^{o}}(a_{+}^{o}) + H_{[\mathbb C_1,0]}^{a^{o}}(a_{-}^{o})]
\end{align*}
To compute the front and back values of $H_{[\mathbb C_1,0]}^{a^o}(a_{\pm})$ we have implemented the recursion procedure for $H$ using Mathematica as outlined in Proposition~\ref{prop:real-line-full-plane} and Corollary \ref{cor:full-plane-observable-construction} (see also Fig. \ref{fig:s-holomorphic-propagation}). That yields that 
\begin{align*}
\mu_{3/2+i/2}=2\sqrt{2}-2\,.
\end{align*}
Now by Theorem~\ref{thm:spin-sens-energy}, we wish to compute $\mathcal P_{[0]}^{a}=(-2) \mbox{Pf} (\mathbf F_{[0]}^{a})$ but since $\mathbf F_{[0]}^a$ is a $2n \times 2n$ antisymmetric matrix that is zero on its anti-diagonal, $\mathcal  P_{[0]}^a=0$. On the other hand, that implies that $Q_{[0]}^a=\mathbf E_{[0]}^{a}$ whose entries are given by Definition~\ref{def:antisymmetric-matrix-entries} in terms of values of $G_{[\mathbb C_1,0]}$ and $\tilde G^{\pm}_{[\mathbb C_1,0]}$ evaluated on oppositely oriented $a^{o^\pm}$. Via the explicit construction of the slit-plane harmonic measure recursion procedure outlined in Remark~\ref{rem:g-explicit}, one can calculate
\begin{align*}
E_{\left[\mathbb{C}_{1},0\right]}^{a^{o^+},a^{o^-}}={5\sqrt 2 -7}\,.
\end{align*}
Putting these together and plugging in to the expansion given by Theorem~\ref{thm:spin-sens-energy} for the spin-weighted correlation, we obtain the desired geometric correction.
\end{proof}

\end{document}